\documentclass[aps,prb,showpacs,twocolumn,superscriptaddress]{revtex4-1}

\usepackage{bm,color,amsmath,amssymb,latexsym,graphicx,psfrag,accents,float}
\usepackage{amscd,dsfont,wasysym,mathrsfs,mathtools}
\usepackage{multirow}
\usepackage{dcolumn}
\usepackage{xcolor}
\usepackage{comment}
\newcolumntype{L}[1]{>{\raggedright\arraybackslash}p{#1}}
\newcolumntype{C}[1]{>{\centering\arraybackslash}p{#1}}
\newcolumntype{R}[1]{>{\raggedleft\arraybackslash}p{#1}}

			\newcommand{\e}[1]{\begin{align}{#1}\end{align}}	
		
		\newcommand{\f}[2]{\frac{#1}{#2}}
		\newcommand{\tf}[2]{\tfrac{#1}{#2}}
		
		\newcommand{\p}[2]{\frac{\partial #1}{\partial #2}}

		\newcommand{\la}[1]{\label{#1}}

		\newcommand{\q}[1]{Eq.\ (\ref{#1})}
		\newcommand{\qq}[2]{Eqs.\ (\ref{#1})-(\ref{#2})}
		\newcommand{\s}[1]{Sec.\ \ref{#1}}
		\newcommand{\fig}[1]{Fig.\ \ref{#1}}		
		\newcommand{\app}[1]{App.\ \ref{#1}}				
		\newcommand{\tab}[1]{Tab.\ \ref{#1}}
		\newcommand{\ocite}[1]{Ref.\ \onlinecite{#1}}

		\newcommand{\iand}{\ins{and}}

		\newcommand{\eq}{=&\;}

		\newcommand{\Z}{\mathbb{Z}}

\newcommand{\var}{\varepsilon}
\newcommand\as{\;\;\;\;}

\newcommand{\hbp}{\hat{\bp}}

\newcommand{\ba}{\boldsymbol{a}}
\newcommand{\bb}{\boldsymbol{b}}

\newcommand{\bk}{\boldsymbol{k}}

\newcommand{\bp}{\boldsymbol{p}}
\newcommand{\bq}{\boldsymbol{q}}
\newcommand{\br}{\boldsymbol{r}}

\newcommand{\bx}{\boldsymbol{x}}
\newcommand{\by}{\boldsymbol{y}}
\newcommand{\bz}{\boldsymbol{z}}

\newcommand{\bA}{\boldsymbol{A}}

\newcommand{\bG}{\boldsymbol{G}}

\newcommand{\bR}{\boldsymbol{R}}

\newcommand{\bcalf}{\boldsymbol{\calf}}

		\newcommand{\bhp}{\boldsymbol{\hat{p}}}

\newcommand{\W}{{\cal W}}

\newcommand{\mir}{\mathfrak{r}}

\newcommand{\ins}[1]{\;\;\;\;\text{#1}\;\;\;\;}

\newcommand{\cala}{{\cal A}}
\newcommand{\calb}{{\cal B}}
\newcommand{\calc}{{\cal C}}

\newcommand{\calf}{{\cal F}}

\newcommand{\cali}{{\cal I}}

\newcommand{\call}{{\cal L}}

\newcommand{\calo}{{\cal O}}
\newcommand{\calp}{{\cal P}}

\newcommand{\cals}{{\cal S}}

\newcommand{\noi}[1]{\noindent (#1)}
\newcommand{\imp}{\;\;\Rightarrow\;\;}
\newcommand{\mo}{\text{-}1}

\newcommand{\braket}[2]{\big\langle #1 \big| #2 \big\rangle}

\newcommand{\braopket}[3]{\big\langle #1 \big| #2 \big| #3 \big\rangle}
\newcommand{\bra}[1]{\big\langle#1\big|}
\newcommand{\ket}[1]{\big|#1\big\rangle}

\newcommand{\pdg}[1]{{#1}^{\phantom{\dagger}}}

\newcommand{\lin}{\notag \\}

\newcommand{\ab}{\alpha\beta}

\newcommand{\low}{L$\ddot{\text{o}}$wdin\;}

\newcommand{\bpm}{\begin{pmatrix}}
\newcommand{\epm}{\end{pmatrix}}

\newcommand{\bal}{\begin{align}}

\newcommand{\dg}[1]{#1^{\scriptstyle{\dagger}}}

\newcommand{\sma}[1]{\scriptscriptstyle{#1}}

\newcommand{\noc}{n_{\sma{{occ}}}}

\newcommand{\bkpar}{\bk_{\parallel}}
\newcommand{\hatbmx}{\hat{g}_x}
\newcommand{\qed}{\nobreak \ifvmode \relax \else
      \ifdim\lastskip<1.5em \hskip-\lastskip
      \hskip1.5em plus0em minus0.5em \fi \nobreak
      \vrule height0.75em width0.5em depth0.25em\fi}

\begin{document}

\title{ Glide-resolved photoemission spectroscopy:\\
 Measuring topological invariants in nonsymmorphic space groups
%remove Perfect spin-momentum locking in the photoemission of nonsymmorphic topological solids
}
 \author{A. Alexandradinata} \affiliation{Department of Physics and Institute for Condensed Matter Theory, University of Illinois at Urbana-Champaign, Urbana, Illinois 61801, USA}
  \author{Zhijun Wang}\affiliation{Beijing National Laboratory for Condensed Matter Physics,
and Institute of Physics, Chinese Academy of Sciences, Beijing 100190, China}\affiliation{University of Chinese Academy of Sciences, Beijing 100049, China}   
	\author{B. Andrei Bernevig} \affiliation{Department of Physics, Princeton University, Princeton NJ 08544, USA} 
	\affiliation{Physics Department, Freie Universitat Berlin, Arnimallee 14, 14195 Berlin, Germany}
	\affiliation{Max Planck Institute of Microstructure Physics, 06120 Halle, Germany}
  \author{Michael Zaletel}\affiliation{Department of Physics, Princeton University, Princeton NJ 08544, USA}   
  
%\homepage[]{Your web page}
%\thanks{}
%\altaffiliation{}

%\affiliation{Department of Physics, Princeton University, Princeton NJ 08544, USA}   \affiliation{Department of Physics, Yale University, New Haven, Connecticut 06520, USA} 

% 0123456789012345678901234567890123456789012345678901234567890123456789
% 0         1         2         3         4         5         6
\begin{abstract}
The two classes of 3D, time-reversal-invariant insulators are known to subdivide into four classes in the presence of  glide symmetry.\cite{Hourglass,Cohomological,Nonsymm_Shiozaki} Here, we extend this classification of insulators to include glide-symmetric Weyl metals, and find a finer $\Z_4\oplus \Z $ classification.  We further elucidate the smoking-gun experimental signature of each class in the photoemission spectroscopy of surface states. 
Measuring the $\Z_4$ topological invariant by photoemission relies on identifying the glide representation of the initial Bloch state before photo-excitation -- we show how this is accomplished with relativistic selection rules, combined with standard spectroscopic techniques to resolve both momentum and spin. Our method relies on a novel spin-momentum locking that is characteristic of all glide-symmetric solids (inclusive of insulators and metals in trivial and topological categories). As an orthogonal application, given a glide-symmetric solid with an ideally symmetric surface, we may utilize this spin-momentum locking  to generate a source of \emph{fully} spin-polarized photoelectrons, which have diverse applications in solid-state spectroscopy. Our ab-initio calculations predict Ba$_2$Pb, stressed Na$_3$Bi, and KHgSb to realize all three, nontrivial insulating phases in the $\Z_4$  classification.
\end{abstract}
\date{\today}

%\pacs{74.20.Mn, 74.20.Rp, 74.25.Jb, 74.72.Jb}
% insert suggested keywords - APS authors don't need to do this
%\keywords{}

\maketitle

%While topological invariants are used to distinguish different phases of matter having the same symmetry, not all topological invariants are directly measureable because they depend on the choice of coordinate system. We formulation a notion of coordinate-independent topological invariants, which is a subset of all the topological invariants that determine the topological classification.
%We exemplify our formulation with band systems having glide symmetry, for which we extend the previous classification for band insulators to also include Weyl metals. For both gapped and gapless systems, we show how the full set of topological invariants reduces to a smaller subset of coordinate-independent invariants, which we formulate physically as: (i) a Fermi-level criterion for surface-localized states, as well as (ii) the Berry-Zak phases acquired in bulk transport. These signatures are shown to be accessible by modern experimental techniques; in particular, we propose a method to measure glide eigenvalues from photoemission spectroscopy, which facilitates the identification of our Fermi-level criterion. We exemplify this discussion with three materials (Ba$_2$Pb, Na$_3$Bi, KHgSb) which realize all three nontrivial phases of in our proposed, coordinate-independent $\Z_4$ classification. 

%Two insulators with the same symmetry are considered equivalent if their Hamiltonians are continuously connected without closing of the energy gap between conduction and valence bands. 

\begin{figure}[ht]
\centering
\includegraphics[width=8.6 cm]{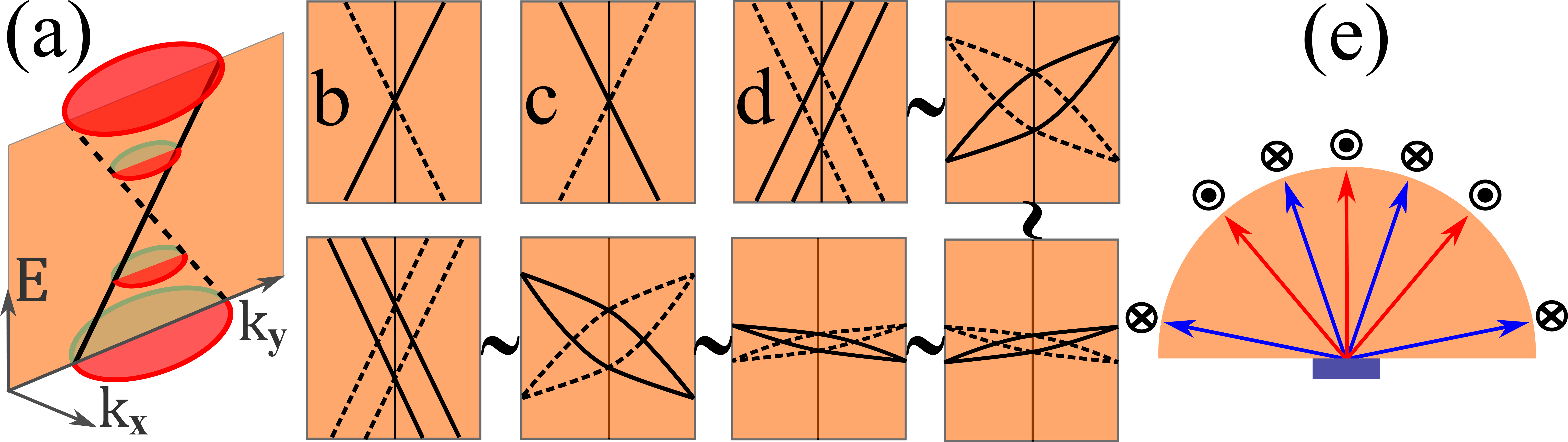}
\caption{(a)  Band dispersion of a Dirac fermion. The glide-invariant cross-section of a Dirac cone corresponds to a right- and left-moving mode, as indicated by two intersecting black lines. A solid (resp.\ dashed) line corresponds to the glide eigenvalue being $\Delta_+$ (resp.\ $\Delta_-$). (b-d) Glide-invariant cross sections of a variety of surface states. (e) A glide-invariant Bloch state (localized to the rectangular sample) absorbs a photon and is emitted as a superposition of plane waves travelling in several directions, as illustrated by the fan of arrows parallel to the glide-invariant plane (colored orange); for blue arrows, the photoelectron spin points into the board, and for red it points out.}\label{fig:front}
\end{figure}

The recent theoretical prediction\cite{Hourglass,Cohomological} and 
experimental 
discovery\cite{Ma_discoverhourglass} of 
hourglass-fermion surface states 
in 
%cancel the
KHgSb 
%material class
heralds a new class of topological solids  protected by nonsymmorphic crystalline symmetries\cite{Nonsymm_Shiozaki,unpinned,Shiozaki2015,ChaoxingNonsymm,ezawa_hourglasstb,Poyao_mobiuskondo,singlediraccone,bandcombinatorics_kruthoff,TQC,wieder_wallpaperfermion} 
-- symmetries that unavoidably translate space by a rational fraction of the lattice period.\cite{Lax} The two well-known classes\cite{fukanemele_3DTI,moore2007,Rahul_3DTI,Inversion_Fu} of 3D, time-reversal-invariant  insulators subdivide into four classes\cite{Nonsymm_Shiozaki,charlesAA_organizing} in the presence of  glide symmetry -- defined as the composition of a reflection symmetry with a half of a lattice translation. 
Indeed, while the $\mathbb{Z}_2$ classification in the absence of glide symmetry corresponds to the number (even vs. odd) of Dirac fermions on the surface of an insulator, glide symmetry further assigns to each Dirac fermion a ``chirality'' which enriches the classification to $\Z_4$. To appreciate this, consider a glide-invariant cross-section (in $\bk$-space) of a Dirac fermion, as illustrated in \fig{fig:front}(a); each Bloch state (with  wavevector in this cross-section) carries a glide eigenvalue which takes on one of two values (denoted as $\Delta_{\pm}$). The chirality of the Dirac fermion is defined  to be positive (resp.\ negative) if  the right-moving mode has eigenvalue $\Delta_+$ (resp.\ $\Delta_-$), as illustrated in \fig{fig:front}(b) [resp.\ (c)]. Two  fermions with positive chirality [first panel of \fig{fig:front}(d)] represent a nontrivial insulator  whose surface-band dispersion resembles an hourglass [second panel of \fig{fig:front}(d)].\cite{Hourglass,Cohomological} This same dispersion can be deformed to two  fermions with negative chirality [sequenced panels in \fig{fig:front}(d)] while preserving surface states at any energy in the  bulk  gap (as illustrated in third column of \fig{fig:wilson_insulators}); \footnote{This deformation argument was first presented in Ref.\ \onlinecite{Nonsymm_Shiozaki}}   
this provides a heuristic argument  for the $\Z_4$ classification of glide-symmetric insulators.

One of our aims is  to extend this classification to describe glide-symmetric solids -- inclusive of insulators and topological metals -- and to further elucidate the smoking-gun {experimental} signature of each class of solids. As described in \s{sec:solid}, the classification of topological solids is $\Z_4 {\oplus} \Z$, with $\Z$ corresponding to the net number of  Weyl points in a symmetry-reduced quadrant of the Brillouin zone. Each class of $\Z_4 {\oplus} \Z$ can be experimentally distinguished through (a) the holonomy of bulk Bloch functions over noncontractible loops of the Brillouin torus, as well as  through (b) the photoemission spectroscopy\cite{cardona_photoemissionbook,hufner_book} (PES) of surface states, as discussed in \s{sec:pesZ4}. (a) and (b) are related by the bulk-boundary correspondence\cite{fidkowski2011,ZhoushenHofstadter,Cohomological} of topological insulators and metals.  

We propose that our theory is materialized in \s{sec:material}  by Ba$_2$Pb, uniaxially-stressed Na$_3$Bi, and  KHgSb; they respectively   fall into the classes: ($\chi^+{\in}\Z_4,\calc{\in}\Z){=}(3,0$),  ($1,0$), and ($2,0$). For 
the Dirac semimetal
Na$_3$Bi, we consider a stress that preserves the glide symmetry but destabilizes the Dirac crossings between conduction and valence bands,\cite{zhijun_3DDirac} thus inducing a transition from a Dirac semimetal (with space group $D_{6h}^4$) to a $\chi^+{=}1$ topological insulator (with nonsymmorphic space group 63); 
%add
such a transition is deducible using the methods of Topological Quantum Chemistry.\cite{TQC} While it
%
%remove It
is known that  Ba$_2$Pb and gapped Na$_3$Bi  belong 
%cancel in
%add
to
the same nontrivial phase under the $\Z_2$ time-reversal-symmetric classification,\cite{zhijun_3DDirac,Sun_Ba2Pb} 
%remove Here, 
%add
here 
we propose that they are distinct  phases in the $\Z_4$ glide-symmetric classification, and may be distinguished by  photoemission spectroscopy.

%at high-symmetry wavevectors (away from Weyl points, if such exist at generic wavevectors)

Measuring the $\Z_4$ topological invariant through photoemission relies on identifying the glide eigenvalues ($\Delta_{\pm}$) of Bloch states before they are photo-excited [cf.\ \fig{fig:wilson_insulators}(c-h)]. By combining angle-resolved PES with dipole selection rules,\cite{gobeli_dipoleselection,HERMANSON_dipoleselection} it is known how to determine the integer-spin representation of glide for solids without spin-orbit coupling.\cite{PESCIA_glideresolve,PRINCE_glideresolve} However, this method is insufficient to  determine half-integer-spin representations of glide for spin-orbit-coupled solids, which are the subject of this work. Here, we show that spin- and angle-resolved PES, which was 
%remove unfeasible 
%add
not addressed
during the previous works,\cite{gobeli_dipoleselection,HERMANSON_dipoleselection,PESCIA_glideresolve,PRINCE_glideresolve,Borstel_relativisticdipoleselection} provides the missing ingredient to identify glide eigenvalues -- {and therefore the $\mathbb{Z}_4$ index} -- in spin-orbit-coupled solids. 

Our proposed method relies on photoexciting a glide-invariant Bloch state with linearly-polarized radiation. The excited photoelectron is emitted (into vacuum) as a quantum superposition of plane waves, with wavevectors differing only by reciprocal vectors of the solid (with a surface). The wavevectors  lying within the glide-invariant plane form a fan of rays that is illustrated in \fig{fig:front}(e). If the polarization vector of the incoming radiation lies orthogonal to the glide-invariant plane, then photoelectrons on any pair of adjacent rays are fully spin polarized in opposite directions -- normal and antinormal to the glide-invariant plane. As we will demonstrate in \s{sec:pes}, this perfect spin-momentum locking of the photoelectron is a \emph{general} manifestation of  spin-orbit coupling in all glide-symmetric solids (trivial or topological, insulating or metallic); the generalization to mirror-symmetric solids will also be discussed. As an orthogonal application of this locking, one may generate a fully spin-polarized photoelectronic current (photocurrent, in short) by isolating one of the rays in \fig{fig:front}(e) using standard angle-resolved PES techniques. The potential applications to solid-state spectroscopy are discussed in \s{sec:discussion}.

%As an orthogonal application of this locking, one may generate a fully spin-polarized photoelectronic current (photocurrent, in short) by isolating one of the rays in \fig{fig:front}(e) using standard angle-resolved PES techniques. Photoemission sources of spin-polarized electrons have diverse applications as spectroscopic probes of solid-state systems;\cite{kirschner_sources_detectors} such sources form the basis for spin-polarized bremsstrahlung isochromat spectroscopy,\cite{scheidt_spinbremsstrahlung} spin-polarized low-energy electron diffraction,\cite{KIRSCHNER_spinLEED} spin-polarized electron-energy-loss spectroscopy (e.g., for the investigation of Stoner excitations\cite{kirschner_spinEELS}), and spin-polarized appearance potential spectroscopy.\cite{KIRSCHNER_spinappearancepotential} While beam current densities of existing GaAs-based, photoemission sources  are satisfactory, their spin polarization is theoretically limited to 50\%, with experiments achieving just over 40\%;\cite{kirschner_sources_detectors}  in comparison, our proposed spin polarization can in principle be complete (100\%).

The reader who is solely interested in this spin-momentum locking (and how it is utilized to resolve glide eigenvalues in PES) may jump straight to \s{sec:pes}, which has been designed to be a self-contained exposition. In \s{sec:discussion}, we elaborate on our proposal to generate spin-polarized photocurrents, as well as compare it with existing theoretical proposals. We also summarize our main results, and discuss further experimental implications.

%The ability to resolve the glide eigenvalues of initial states is essential for the experimental determination of the $\Z_4{\otimes} \Z$ topological invariants ($\chi^{\sma{+}}{\in}\Z_4,\calc{\in}\Z$), which we will introduce below.  

%with glide and time-reversal symmetries are fall into four classes which are distinguished by a topological invariant $\chi^+ \in \{0,1,2,3\}=\Z_4$.

%Our procedure permits an \emph{absolute} measurement of the $\Z_4$ topological invariant, that does not rely on a comparative agreement with a first-principles calculation.

%a simple generalization of the procedure allows for an absolute measurement of mirror eigenvalues and therefore the mirror Chern number.   

%which of the two is forbidden changes across the BZ edge. In other words, the selection rules result in missing intensity peaks for half the bands in the first Brillouin zone, but this missing half is observable in the second Brillouin zone; this effective doubling of the Brillouin zone has been experimentally observed[Prince],

%Consider a glide-symmetric crystal in the orthorhombic crystal system (e.g., space group $Pnma$), 

%$(\chi^+{=}2,\calp_{01}{=}0)$ is exemplified by KHgSb, whose surface states display an unremovable `hourglass-flow' connectivity.

\section{Preliminaries on nonsymmorphic space-group representations}\la{sec:prelim}

Throughout this work, we focus on spin-orbit-coupled solids whose space groups contain (minimally) the operations of time reversal and glide. We adopt a Cartesian coordinate system $(x,y,z)$, with corresponding unit directional vectors $(\vec{\bx},\vec{\by},\vec{\bz})$, such that the glide symmetry (denoted as $g_x$) maps $(x, y, z) {\to} ({-}x,y{+}R_2/2, z)$, where $R_2$ is the lattice period in the $\vec{\by}$ direction. That is, $g_x$ is the composition of two commuting operations: a reflection ($r_x$) that inverts $x$, and a translation by half a lattice period in $\vec{\by}$. This implies $g_x^2$ is the product of a full lattice translation and $r_x^2$; the latter acts on spinor wavefunctions like a $2\pi$ rotation, i.e.,  it produces a ${-}1$ phase factor.

\begin{figure}[h]
\centering
\includegraphics[width=6 cm]{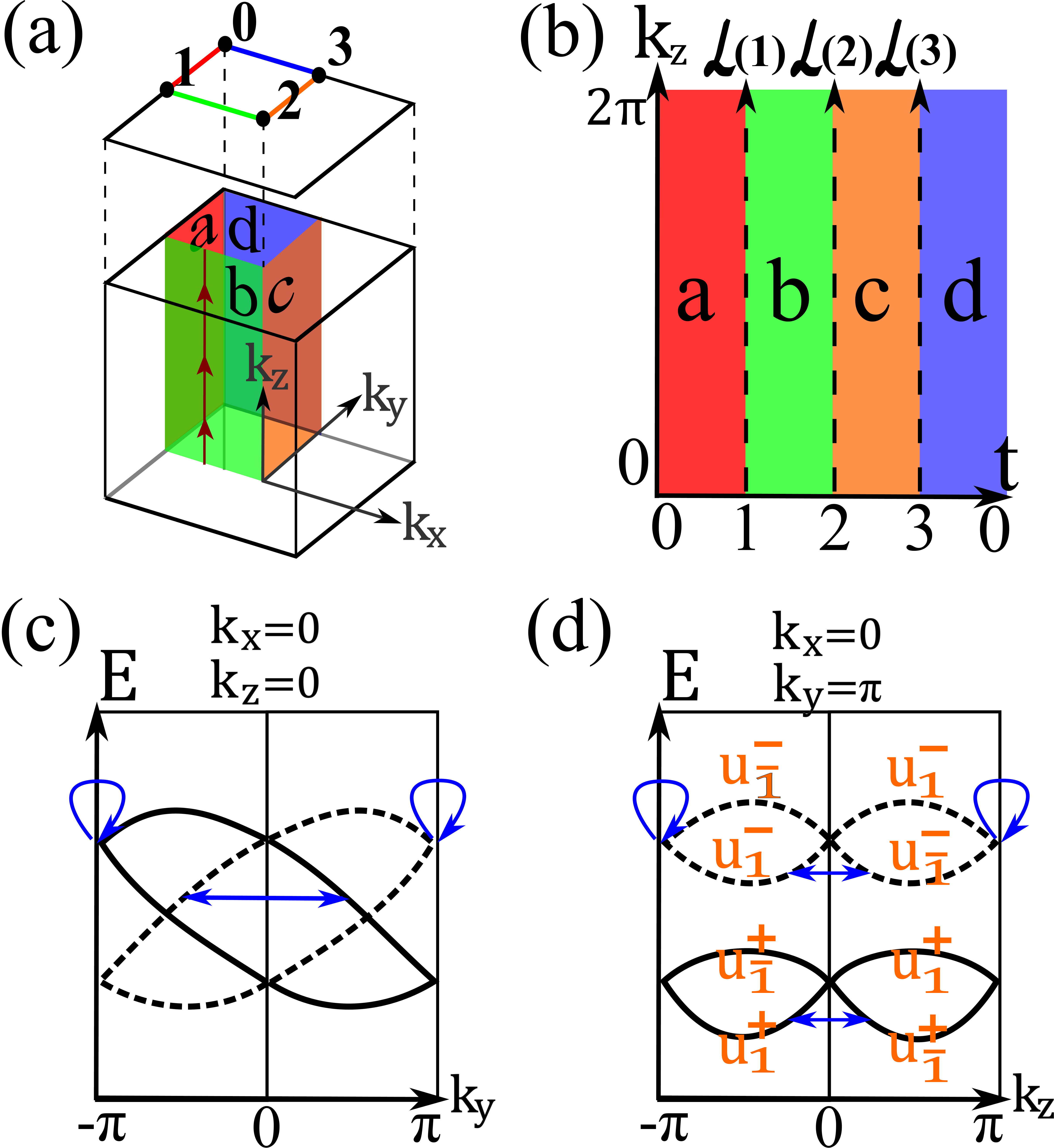}
\caption{(a) 3D Brillouin zone (BZ) of glide-symmetric solids. For certain nonsymmorphic space groups which contain glide symmetry (e.g., $D_{6h}^4$, the space group of KHgSb), their BZs are not cubic, and figure (a) should be understood as a modified BZ corresponding to a non-primitive real-space unit cell; further details may be found in \app{app:projectiveglide}.  %The 2D subregion is a combination of four colored faces, 
We will define topological invariants on  a 2D subregion that combines four colored faces,  which are labelled $abcd$ and parametrized in (b). Red ($a$) and orange ($c$) faces are glide-invariant. (c-d) For noncentrosymmetric space groups, we illustrate typical energy-band dispersions on two glide- and time-reversal-invariant  lines, the first  at fixed $k_x{=}k_z{=}0$, and the second at fixed $k_x{=}0,k_y{=}\pi$ ($R_2{=}1$). Solid and dashed lines  respectively indicate bands in the $\Delta_+$ and $\Delta_-$ representations, 
%new 
with corresponding glide eigenvalues $\Delta_{\pm}(k_y){=}{\pm} i \exp[{-}i k_y/2]$.  
%zadd
Arrows indicate states related by time reversal.}\label{fig:bz}
\end{figure}

Let us review the irreducible half-integer-spin representations of glide and discrete translational symmetries. The irreducible representations of translations are Bloch states labelled by a crystal wavevector $\bk{=}(k_x,k_y,k_z)$ in the first Brillouin zone (BZ). Since $g_x$ maps $\bk{\to}({-}k_x,k_y,k_z)$, the glide-invariant Bloch functions lie in two cuts of the BZ: the $k_x{=}0$ cut through the BZ center will be referred to as the central glide plane, and the $k_x{=}\pi/a_1$ plane (with $2\pi/a_1$ a reciprocal period in the $\vec{\bx}$ direction) will be referred to as the off-center glide plane. As illustrated in \fig{fig:bz}(a), the positive-$k_y$ halves of the central and off-center glide plane are labelled by $c$ and $a$ respectively.  

%remove merely
Let $\hatbmx$ be an operator representing $g_x$ on spinor wavefunctions. The action of $\hatbmx^2$ on a glide-invariant spinor Bloch function  produces a phase ${-}e^{{-}ik_yR_2}$, hence the possible eigenvalues of $\hatbmx$ fall into two branches of $\Delta_{\pm}(k_y) {:}{=} {\pm} i \exp[-i k_yR_2/2]$. A Bloch state with glide eigenvalue $\Delta_{\pm}(k_y)$ is said to be in the  $\Delta_{\pm}(k_y)$ representation; we will 
%remove often 
use `eigenvalue' and `representation' interchangeably. The typical energy band dispersions along two glide- and time-reversal-invariant lines are illustrated in \fig{fig:bz}(c-d); each solid black line (resp.\ dashed black line) indicates a band in the $\Delta_+$ (resp.\ $\Delta_-$) representation; this convention is adopted in all figures. The symmetry-enforced band connectivities in \fig{fig:bz}(c-d) are further explained in \app{sec:symm}.

\section{Classification of nonsymmorphic topological solids}\la{sec:solid}

\subsection{Zak-phase expression of the $\Z_4$ invariant} \la{sec:zakexpressZ4}

In \ocite{Nonsymm_Shiozaki}, a topological invariant  $\chi^+{\in}\Z_4{:}{=}\{0,1,2,3\}$ --  expressible as an integral of the Berry connection and curvature -- was introduced  to classify glide-invariant topological insulators. The same invariant provides a \emph{partial} classification of glide-invariant topological (semi)metals, so long as touchings  -- between conduction and valence bands -- occur  away from the bent, 2D subregion colored in \fig{fig:bz}(a). This subregion resembles the face of a rectangular pipe (with its ends  identified due to the periodicity of the BZ). The faces of the cylinder are denoted $a,b,c$ and $d$, with $c$ and $a$ belonging to the central and off-center glide planes respectively. 
%add: c-v, such, 
In the absence of additional point-group symmetry that might restrict conduction-valence touchings to $abcd$,\cite{chen_multiweyl} we may assume in the generic situation that such touchings occur elsewhere.

%The corresponding topological invariants are the Shiozaki-Sato-Gomi invariant $\chi^+{\in} \Z_4{=} \{0,1,2,3\}$ that is known from the classification of glide-symmetric insulators,\cite{Nonsymm_Shiozaki} as well as a  Chern number $\calc {\in} \Z$  which counts the net number of Weyl points enclosed by the colored subregion in \fig{fig:wilson_insulators}(a). 

\begin{figure}[ht]
\centering
\includegraphics[width=8.6 cm]{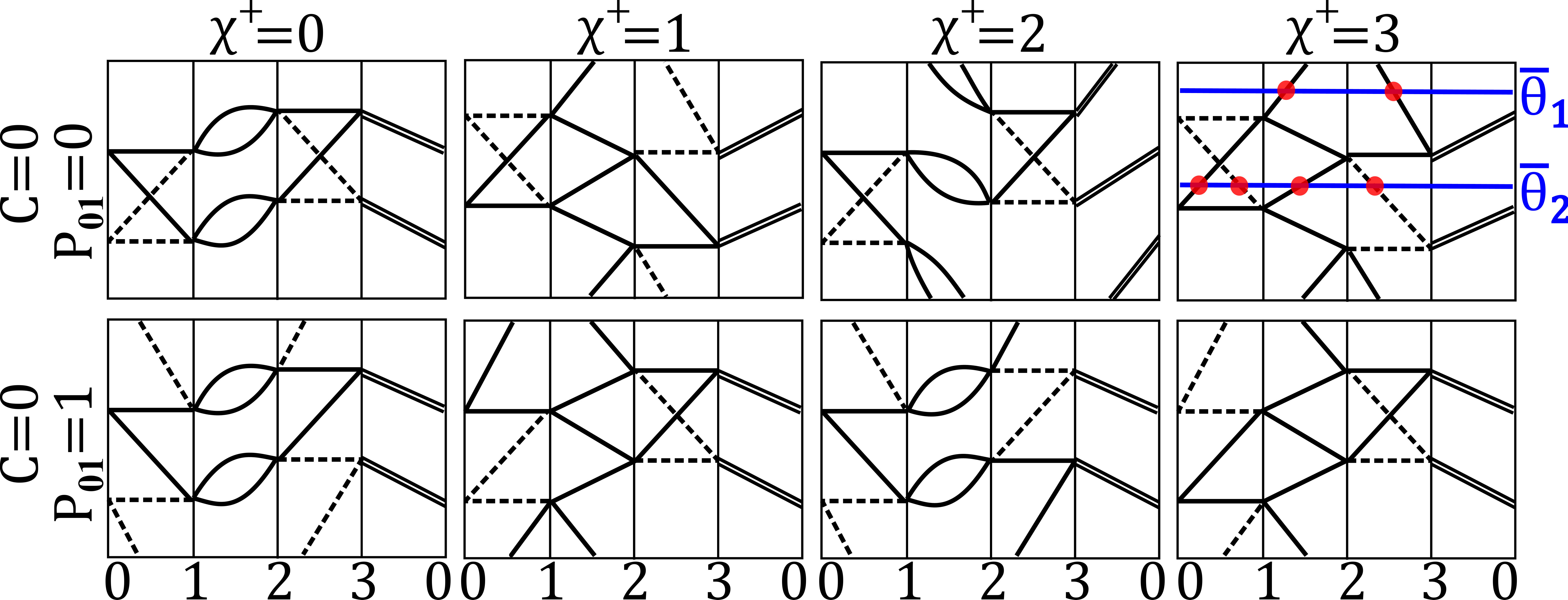}
\caption{ Classification of glide-symmetric insulators by a strong $\Z_4$ invariant ($\chi^+$) and a weak $\Z_2$ invariant ($\calp_{01}$). The vertical axis has the double interpretation as a Berry-Zak phase $\theta{\in}[0,2\pi]$, or as the energy of a surface-localized state (that interpolates between conduction and valence bands). Bands are doubly-degenerate along 30, and  glide-invariant along 01 and 23 only. }\label{fig:wilson_insulators}
\end{figure}

Let us present an equivalent reformulation of the $\Z_4$-invariant ($\chi^+$) through the matrix holonomy of multi-band Bloch functions over the Brillouin torus.     The comparative advantages of our formulation are that the eigenvalues of the holonomy matrix, as represented by the graphs in \fig{fig:wilson_insulators}: (i) are potentially measurable by interference experiments,\cite{atala2013,Tracy} (ii) are  directly relatable to surface states through the bulk-boundary correspondence,\cite{fidkowski2011,ZhoushenHofstadter,Cohomological} as will be elaborated below, and (iii)  are efficiently computed from tight-binding models and first-principles calculations.\cite{soluyanov2011,yu2011,AA2014} In this section, we will explain how the aforementioned graphs are attained, and describe an elementary method to identify $\chi^+$  from these graphs. The proof of equivalence between our holonomy-formulation of $\chi^+$ and the Shiozaki-Sato-Gomi formulation is postponed to \app{app:z4}.

To begin, let us consider the parallel transport of Bloch states in the z-direction, i.e., the wavenumber $k_z$ of a Bloch state is advanced by a reciprocal period, while the \emph{reduced wavevector} $\bk_{\parallel}{=}(k_x,k_y)$ is fixed. We consider a family of noncontractible loops within  $abcd$ [\fig{fig:bz}(a)]; this family is parameterized by $t {\in} [0,4]$ with $4{\equiv} 0$ [\fig{fig:bz}(b)].  A Bloch state that is parallel-transported over a loop does not necessarily return to its initial state; the mismatch between initial and final states  is represented by a holonomy matrix $\W$ in the space of occupied bands (numbering $\noc$). $\W$ is known as the Wilson loop of the non-abelian Berry gauge field,\cite{wilczek1984} and its unimodular eigenvalues $\{\exp[i\theta_j(t)]|j{=}1,2,\ldots,\noc; t{\in} [0,4]\}$ are the Zak phase factors. In analogy with energy bands, we may refer to $\theta_j(t)$ as the dispersion of a  `Zak band' with band index $j$. For  $t{\in} [0,1]$ and $[2,3]$ (which correspond to the glide-invariant faces $a$ and $c$), $\W$ block-diagonalizes into two $\noc/2$-by-$\noc/2$ blocks,\cite{TBO_JHAA} corresponding to the two representations ($\Delta_{\pm}$) of glide; we may therefore label the Zak bands as $\{\theta_j^{\pm}\}_{\sma{j{=}1}}^{\sma{\noc/2}}$.

The $\Z_4$ topological invariant is expressible as:      
 \e{ \chi^{\sma{\pm}} \eq \f1{\pi} \sum_{j=1}^{\noc/2} {\bigg[} \theta_j^{\sma{\pm}}\bigg|_{0} {-}\theta_j^{\sma{\pm}}\bigg|_{3} {+} \int_0^1  d\theta_j^{\sma{\pm}} {+} \int^3_2 d\theta_j^{\sma{\pm}} {\bigg]} \lin
&{+}\; \f1{2\pi} \sum_{j=1}^{\noc} \int_1^2 d\theta_j. \la{wilsonex}}
%mod four
For this expression to be well-defined modulo four, 
%a branch must be chosen for each of $\theta_j$ and $\theta_j^{\pm}$ (which would otherwise have a mod-$2\pi$ ambiguity) --
%We insist that our branch choice satisfies two properties: 
we choose that (i) $\theta_j$ is smooth with respect to $t$ over $[1,2]$, (ii) $\theta_j^{\pm}$ is smooth over $[0,1]$ and $[2,3]$, 
%Old: We further impose that $\theta_j^{\pm}(t)$ are pairwise degenerate at $t{=}0$ and $3$. 
%(ii) Due to the  degeneracy of Zak bands at $t{=}0$ and $3$, we impose 
and  (iii) $\theta_j^{\pm}(t)$ are pairwise degenerate at $t{=}0$ and $3$. To clarify (iii), for any $j\in \{1,\ldots,\noc/2\}$, there exists $j'\in \{1,\ldots,\noc/2\}$ such that $j\neq j'$, $\theta_j^{\pm}(0)=\theta_{j'}^{\pm}(0)$, and $\theta_j^{\pm}(3)=\theta_{j'}^{\pm}(3)$.
%That is, for any Zak band with phase $\theta^+_j(0)$, we will pick a branch for a distinct Zak band (labelled $j'$) such that $\theta^+_{j'}(0){=}\theta^+_j(0)$ (viewed as a strict equality, rather than an equivalence modulo $2\pi$).
%Consequently, ${\sum}_{\sma{j=1}}^{\sma{\noc/2}}\theta_{j}^{+}(0)$ is uniquely defined modulo $4\pi$. 
%old:In combination, the conditions of pairwise degeneracy and smoothness 
%(i-ii) together
%(i-ii) together ensure that $\chi^{\pm}$ is uniquely defined modulo $4$. 

%everywhere with the possible exception of two isolated points: $t{=}1$ and $2$.

%While each of $\exp[i\theta_j(t)]$ is invariant under $U(\noc)$ transformations in the space of occupied bands,

From \q{wilsonex}, we derive the simplest way to identify $\chi^{\pm}$ from the Zak-phase spectrum: for an arbitrarily chosen $\bar{\theta}$, draw a constant-$\bar{\theta}$ reference line (as illustrated in blue in the right-most column of \fig{fig:wilson_insulators}) and consider its intersections with Zak bands (indicated by red dots). For each intersection occurring at $t{\in}(1,2)$, we calculate the sign of the velocity $d\theta/dt$, and sum this quantity over all intersections
% new  [over $t{\in}(1,2)$]
[over $t{\in}(1,2)$] to obtain $\cals_{12}(\bar{\theta})$; for $t{\in}[0,1]$ and $[2,3]$, we consider only intersections with Zak bands in the $\Delta_{\pm}$ representation, and we similary sum over sgn[$d\theta/dt$] to obtain $\cals^{\pm}_{01}(\bar{\theta})$ and $\cals^{\pm}_{23}(\bar{\theta})$ respectively. The following \emph{weighted} sum of $\cals^{\pm}_{ij}$ and $\cals_{12}$, 
\e{\cals^{\pm}(\bar{\theta}) = 2\cals^{\pm}_{01}(\bar{\theta})+\cals_{12}(\bar{\theta})+2\cals^{\pm}_{23}(\bar{\theta}),\la{z4inv}}
satisfies that $(\cals^{\pm}(\bar{\theta}_1){-}\cals^{\pm}(\bar{\theta_2}))/4{\in}\Z$ for any two reference lines at constant $\bar{\theta}_1$ and $\bar{\theta}_2$, e.g., compare $\cals^{+}(\bar{\theta}_1){=}2(0){+}1{+}2({-}1){=}{-}1$ 
%new
[upper blue line in \fig{fig:wilson_insulators}] 
with $\cals^{+}(\bar{\theta}_2){=}2({+}1){+}1{+}2(0){=}3$ 
%new
[lower blue line]. 
%This equivalence modulo four is denoted as $\cals^{\pm}(\bar{\theta}_1) {\equiv} \cals^{\pm}(\bar{\theta}_2)$ 
Equivalently stated, if we henceforth view  $\cals^{\pm}(\bar{\theta})$ as an element in $\Z_4$, then this quantity becomes independent of $\bar{\theta}$. 
% Comparing \q{wilsonex} with \q{z4inv}, we further identify $\cals^{\pm} {\equiv} \chi^{\pm}$.   
By also viewing $\chi^{\pm}$ as a $\Z_4$ quantity, we may  identify $\cals^{\pm} {\equiv} \chi^{\pm}$ by comparing  \q{wilsonex} with \q{z4inv}. To clarify, $\equiv$ denotes an identity between two equivalence classes in $\Z_4$.

\subsection{Extended classification of glide-symmetric topological solids}

\begin{figure}[h]
\centering
\includegraphics[width=8 cm]{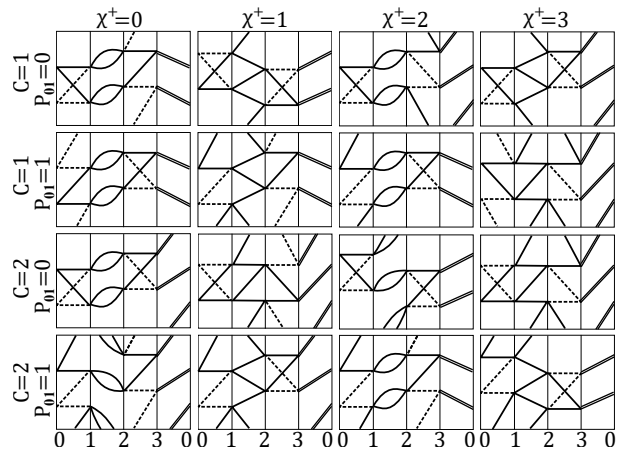}
\caption{Topological classification of glide-symmetric metals with $\calc {\in}\Z$ and $\chi^+{\in}\Z_4$.  A finer classification is possible with the introduction of $\calp_{01}{\in}\Z_2$ -- a weak topological invariant that is defined in \s{sec:material}.
%new
Note that $\chi^+{\in}\{0,1,2,3\}$ should be viewed as the mod-four equivalence class of the quantity defined in \q{wilsonex}, or equivalently in \q{z4inv}. }\label{fig:wilson_metals}
\end{figure}

We now demonstrate that $\chi^+ {\equiv}{-}\chi^-$ for insulators, while this is not necessarily true for Weyl metals. We are considering time-reversal- and glide-symmetric Weyl metals that occur only in non-centrosymmetric space groups.\cite{wan2010,halasz2012} Such metals may be characterized by counting the net number of Weyl nodes in the open Brillouin-zone quadrant $\calo$ surrounded by  (but not including) the faces $abcd$ [\fig{fig:wilson_insulators}(a)]. $\calo$ resembles the interior of a rectangular pipe, and its properties determine those of the other three quadrants owing to $g_x$ and time-reversal symmetry. Each Weyl node has a signed charge ($q$) corresponding to whether it is a source ($q{=}{+}1$) or sink ($q{=}{-}1$) of the Berry field strength; the net charge within $\calo$ is quantified by the bent Chern number ($\calc$),\cite{AAchen} which may be formulated as the net winding of $\theta(t)$ for $t{\in} [0,4]$, or equivalently as the summation of sign[$d\theta/dt$], over all intersections with a constant-$\bar{\theta}$ reference line. The sum is carried out over all bands indiscriminate of their symmetry representations, therefore
\e{\calc = \tf1{2}\big[\;\cals^{+}(\bar{\theta})+\cals^{-}(\bar{\theta})\;\big] + \cals_{30}(\bar{\theta}). \la{bentchern}}
%old: \cals_{30}$ here is the partial sum 
%new
To clarify, $\cals_{30}$ here is the summation of sign[$d\theta/dt$]
over the interval $t\in (3,4)$, which corresponds to the blue line $30$ in \fig{fig:bz}(a); 
$\cals_{30}$ must be even because Zak bands are doubly-degenerate due to $g_xT$ symmetry.\cite{Hourglass} While each of $\cals^{\pm}$ and $\cals_{30}$ 
% old: depends piecewise-continuously on $\bar{\theta}$, 
%new
may individually depend on the choice of reference line, 
their weighted sum ($\calc$) does not. Applying that $2S_{30}$ is an integer multiple of four, and the relation $\cals^{\pm} {\equiv} \chi^{\pm}$ from the previous paragraph, we derive
\e{ (\chi^+ +\chi^-) \equiv 2\calc \;\bmod\;4, \la{bentchern2}}
which implies a $\Z {\oplus} \Z_4$ classification of glide-symmetric solids, inclusive of metals and insulators. To recapitulate, $\Z$ counts the net number of Weyl points in a symmetry-reduced quadrant of the BZ. Representative examples for $\calc{=}1$ and $2$ are illustrated in \fig{fig:wilson_metals}.

\subsection{Surface states of nonsymmmorphic topological solids}

We now extend our discussion to the physics of surface states. We terminate the solid in the z-direction by introducing a surface that is symmetric under glide and discrete translations in the xy plane. We further assume that the surface is clean and does not undergo a symmetry-breaking reconstruction. So long as the above-stated symmetries are preserved, the exact termination of the surface (including relaxation effects) is not essential to our discussion -- we are concerned only with topological aspects of the surface states.

The translational symmetry implies the existence of a surface Brillouin zone ($sBZ$) that is parametrized by the wavevector $\bkpar$; recall that $(\bkpar,k_z)$ parametrizes the bulk Brillouin zone ($bBZ$) of a solid that is periodic in three directions. Energy bands whose wavefunction is localized to the surface shall be referred to as surface bands. Such surface bands can only exist at $\bkpar{\in}sBZ$ for which there is a bulk energy gap at the reduced wavevector $\bkpar$; in particular, they cannot exist at $\bkpar{\in}sBZ$ if a Weyl point  lies at  $(\bkpar,k_z){\in}bBZ$ for some $k_z$.  

Our previous discussion of Zak bands may be related to surface bands by the bulk-boundary correspondence. This correspondence states that the connectivity of Zak bands (over the reduced wavevector $\bkpar$) is topologically equivalent to the connectivity of surface bands (over the surface wavevector $\bkpar$).\cite{fidkowski2011,ZhoushenHofstadter,Cohomological} We shall only concern ourselves with the connectivity over on the high-symmetry lines  $01$, $12$, $23$, $30$ in $sBZ$ [see \fig{fig:wilson_insulators}(a)]; they are respectively the projections of the faces $a$, $b$, $c$ and $d$  in $bBZ$.  
Given our assumption that Weyl points (if they exist) lie away from $abcd$,  surface bands potentially exist  along $01230$, and their connectivity is then well-defined.

$\chi^{\pm}$ may  be identified by considering intersections between surface  bands (over $0123$) and a constant-energy reference line (e.g., the Fermi level). This reference line is chosen so as not to intersect any bulk bands; $\cals_{12}$ and $\cals_{ij}^{\pm}$ in  \q{z4inv} are defined analogously with the velocities ($d\var/dt$) of surface bands, instead of Zak bands. 
%old We may further relate $\cals$ to our argument of chiral Dirac fermions in the introduction: 
We are now ready to justify our heuristic  argument for the $\Z_4$ classification of glide-symmetric insulators, as formulated in the introductory paragraph: suppose our reference Fermi level lies above the Dirac node,  
each positive-chirality Dirac surface band  (centered at  $\bk_{\parallel}(2){=}\Gamma$) singly intersects  the reference line at each of $12$ and $23$; each therefore contributes to $\chi^+$ the quantity $2(0){+}({-}1){+}2(1)$; we may therefore interpret the deformation 
%old argument in \fig{fig:wilson_insulators}(e-h)
in \fig{fig:front}(d)
as the equivalence: $\chi^+{=}2{\equiv}{-}2$.

\section{Materialization of nonsymmorphic topological insulators}\la{sec:material}

In this section, we identify three  \emph{insulating} materials which realize all three nontrivial phases in the $\Z_4$  classification given by $\chi^+ {\equiv}{-}\chi^-$. This classification is characterized as strong, in the sense that any nontrivial phase (with $\chi^+{\neq}0$ mod 4) cannot be realized by layering lower-dimensional glide-symmetric topological insulators.\footnote{This is distinct from the noncrystalline, $\Z_2$ strong invariant\cite{fukanemele_3DTI,moore2007}} 

As is known from topological K theory,\cite{Nonsymm_Shiozaki} the full classification of glide-symmetric surface bands is $\Z_4{\oplus}\Z_2$, where the additional $\Z_2$ summand corresponds to a weak classification by a Kane-Mele\cite{kane2005A} invariant (denoted  $\calp_{01}{\in}\{0,1\}$) defined over the time-reversal- and glide-invariant plane containing the face $a$. $\calp_{01}$ may be determined by the connectivity of Zak/surface bands on the off-center glide line 01:\footnote{In principle we could consider the Kane-Mele invariants $\calp_{ij}{\in}\Z_2$ defined over four time-reversal-invariant planes, which project respectively to $ij{=}01,12,23,30$. However, given the strong invariant $\chi^+$, only one of the four $\calp_{ij}$ is independent. To appreciate this, note that the parity of $\chi^+$ uniquely determines $\Z_2$ strong invariant $\Gamma{\in}\Z_2$ of 3D time-reversal-symmetric insulators.\cite{charlesAA_organizing} Precisely, $\chi^+$ mod 2 $=\Gamma$, where $\Gamma{=}1$ corresponds to the nontrivial phase. Moreover, $\Gamma$ is uniquely determined by the Kane-Mele invariants on parallel planes: $\Gamma{=}\calp_{01}{+}\calp_{23}{=}\calp_{12}{+}\calp_{30}$.\cite{yu2011} Due to $g_x$ and time-reversal symmetries, $\calp_{30}{=}0$.\cite{charlesAA_organizing} Consequently, $\calp_{12}$ and $\calp_{23}$ is determined uniquely by $\chi^+$ and $\calp_{01}$.  }  
 $\calp_{01}{=}0$ corresponds to a gapped, 
%new
hourglass-type 
connectivity along $t{\in}[0,1]$ in the top row of \fig{fig:wilson_insulators}, and $\calp_{01}{=}1$ to  a zigzag 
%new
(quantum-spin-Hall)
connectivity\cite{QSHE_Rahul,fu2006} in the bottom row. 

\begin{figure}[h]
\centering
\includegraphics[width=8 cm]{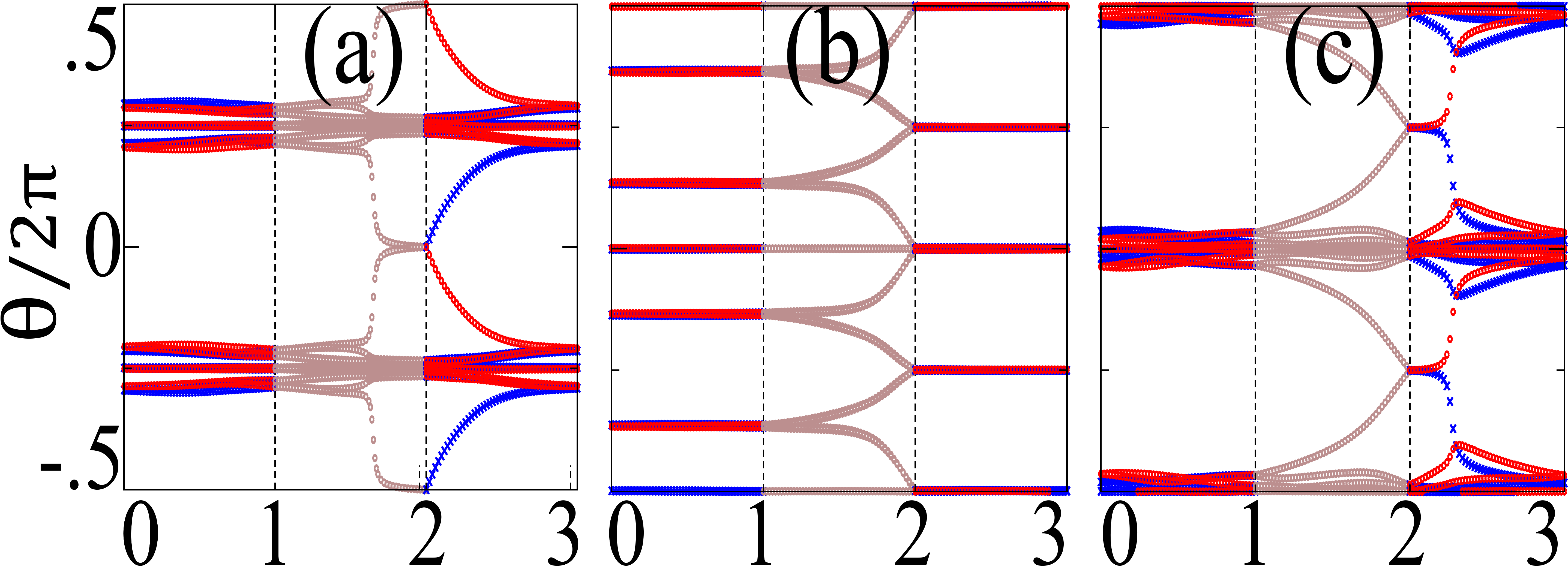}
\caption{Ab-initio-derived Zak phases of:  (a) Ba$_2$Pb, (b) KHgSb, and (c) uniaxially-stressed Na$_3$Bi. Along the glide-invariant lines $01$ and $23$, we decompose the Zak phases according to their glide representations: $\Delta_+$ is indicated by red circles,  and $\Delta_-$ by blue.}\label{fig:ex}
\end{figure}

Having described
%new
in \fig{fig:wilson_insulators}
 the connectivity in each 
%new
nontrivial
class of $\Z_4{\oplus}\Z_2$,  we are ready to identify Ba$_2$Pb, KHgSb, and uniaxially-stressed Na$_3$Bi  as corresponding to $(\chi^+,\calp_{01}){=}(3, 0), (2,0), (1,0),$ from their ab-initio-derived Zak-band connectivity in  \fig{fig:ex}(a-c).

The parity of $\chi^+$ being even (resp.\ odd)  is in one-to-one correspondence\cite{charlesAA_organizing} with the trivial (resp.\ nontrivial) phase in the time-reversal-symmetric, strong $\Z_2$ classification. We thus deduce that Ba$_2$Pb and uniaxially-stressed Na$_3$Bi  belong to the same phase in the time-reversal-symmetric classification (as was derived by other means in previous works\cite{zhijun_3DDirac,Sun_Ba2Pb} ), but belong to distinct $\Z_4$ phases in the presence of glide symmetry (a novel conclusion of this work). 
%new
This conclusion is further supported by our analysis of both compounds based on their elementary band representations\cite{Zak_bandrepresentations,Evarestov_bandrepresentations,Bacry_bandrepresentations,TQC,TBO_JHAA,nogo_AAJH} -- a perspective we develop in \app{app:materializations}.

% we find that the filled, low-energy bands of both compounds are not band representations, and moreover they do not differ by band representations. The first statement implies that both compounds are topologically nontrivial, and the second implies that they are topologically distinct. This perspective is elaborated in \app{app:materializations}

% the modern theory\cite{TQC,TBO_JHAA,nogo_AAJH} of elementary band representations (EBRs):\cite{Zak_bandrepresentations,Evarestov_bandrepresentations,Bacry_bandrepresentations} which we elaborate upon in  \app{app:materializations}.

In comparison,  KHgSb is trivial in the time-reversal-symmetric $\Z_2$ classification but nontrivial in the glide-symmetric $\Z_4$ classification; additional crystalline symmetries (beyond glide) in the space group ($D_{6h}^4$) of KHgSb are known to lead to an even finer classification.\cite{Cohomological} It was 
%old: first conjectured 
%new
argued
in \ocite{Nonsymm_Shiozaki} that KHgSb should belong to the $\chi^+{=}2$ class based on 
%old: a surface-state argument; 
the connectivity of its surface states;
\fig{fig:ex}(b) provides the first evidence based on an explicit calculation of the bulk topological invariant. We remark that a recent polarized Raman scattering study\cite{dchen_latticeinstabilityKHgSb} suggests of a low-temperature lattice instability in KHgSb; such an instability would not break glide symmetry, and we expect that  $\chi^+{=}2$ should remain valid.

In \app{app:materializations}, we detail the space groups and elementary band representations of these materials, and further describe the  stress
%old: in
%new
that should be applied to
Na$_3$Bi -- so that it becomes a topological insulator.

\section{Photoemission spectroscopy of $\Z_4$ invariant}\la{sec:pesZ4}

Let us describe how the $\Z_4$ invariant [cf.\ \q{z4inv}] is measurable from PES. The velocities ($d\var/dt$) 
%new
of surface states 
are measurable from angle-resolved PES using standard techniques.\cite{cardona_photoemissionbook,hufner_book} The counting of $\cals^{\pm}_{23}$ (resp.\ $\cals^{\pm}_{01}$) further requires that we identify the glide representation ($\Delta_{\pm}$) of the pre-excited Bloch state  on the glide line intersecting the surface-BZ center (resp.\ lying on the surface-BZ edge). We propose a spectroscopic method for identifying $\Delta_{\pm}$ on the central glide line 23 ($k_x{=}0$) in the next section [\s{sec:pes}]. 

This method cannot be applied to determine  $\Delta_{\pm}$ for the off-center glide line $01$, as explained at the end of \s{sec:pes}. However, we may anyway determine the $\Z_4$ invariant for materials with no Fermi-level surface states along $01$, in which case $\cals^{\pm}_{01}{=}0$. Indeed, there is no topological reason to expect surface states along $01$ for materials with a trivial weak index ($\calp_{01}{=}0$), as explained in \s{sec:material}.
Our calculations show that all three materials (proposed in \s{sec:material}) have  $\calp_{01}{=}0$, and have no Fermi-level surface states along $01$ for a perfect surface termination (i.e., ignoring surface relaxation or reconstruction). We remark that 
$\calp_{01}{=}0$ is guaranteed for certain space groups (including those of KHgSb and uniaxially-stressed Na$_3$Bi), owing to a symmetry of a discrete translation (in a direction oblique to the surface); this is elaborated in \app{app:projectiveglide}.

%, the triviality of $\calp_{01}$ implies a gapped
%add
%hourglass connectivity for the surface bands along $01$. 
%remove: The  generic situation is that there are no Fermi-level surface bands along $01$
%add
%It is possible that these hourglass surface bands are not intersected by the Fermi level, which 
%remove: implies trivially
%add
%would imply
%that $\cals^{\pm}_{01}{=}0$. 
%This generic situation describes the Fermi levels of
%add
%remove

%Hypothetically, even if surface states (along $01$) were to exist at the Fermi level,  one may utilize the invariance of $\chi^{\pm}$ (modulo four) under a change in the reference energy level (where the velocities and glide representations are measured in PES); this invariance has been demonstrated in \s{sec:solid}. Supposing that a reference energy exists (within the bulk energy gap) where there are no $01$ surface states,  
%old: we are again obviated of the 
%add
%there is no
%need to measure glide eigenvalues along $01$.  

%By  measuring $\chi^{\pm}$ at a reference energy  below the Fermi energy (but still within the bulk spectral gap), and where there are no intersections with surface states along $01$, we again release ourselves from the necessity of .

%if there are no surface-localized states along $01$, it follows trivially that $\cals^{\pm}_{01}{=}0$. This scenario describes the Fermi levels of  the three materials we have proposed in \s{sec:material}. 

Let us address one final subtlety about the identification of  $\chi^{\pm}$ (or $\calc$) from photoemission. $\chi^{\pm}$ and $\calc$ have been defined with respect to a fixed, Cartesian, right-handed coordinate system  parametrized by $(k_x,k_y,k_z)$. A spectroscopist who examines a solid  necessarily has to pick a coordinate system and measure the topological invariants with respect to this choice. Will two measurements of $\chi^{\pm}$  -- of the same solid but based on different coordinates chosen by the spectroscopist -- unambiguously agree?

% it remains to confirm if two spectroscopists measuring $\chi^{\pm}$ on the same solid would unambiguously obtain the same answer -- this is not apriori obvious, since $\chi^{\pm}$ is defined with respect to a  coordinate system [parametrized by $(k_x,k_y,k_z)$], but generically no coordinate system is unique. To put it bluntly, if two spectroscopist might 

The glide symmetry may be exploited to reduce this coordinate ambiguity:  we may always choose a right-handed, Cartesian coordinate system where $\vec{\bx}$ (resp.\ $\vec{\by}$) lies parallel to the reflection (resp.\ fractional translational) component of the glide, i.e., the glide  maps $(x,y,z) \rightarrow (-x,y\pm R_2/2,z)$;\footnote{Any solid that is symmetric under $g_x:(x,y,z) \rightarrow (-x,y+ R_2/2,z)$ would also be symmetric under $(x,y,z) \rightarrow (-x,y-R_2/2,z)$, since $(x,y,z) \rightarrow (x,y+R_2,z)$ is also a symmetry of the solid. 
} from the experimental perspective, this presupposes some knowledge about the crystallographic orientation of a sample, as discussed further in \app{app:lightsource}.
This prescription  does not uniquely fix the coordinate system: supposing $(x,y,z)$ satisfies the above condition, so would $(x',y',z')=(-x,-y,z)$, and more generally any coordinate system that is related to $(x,y,z)$ by two-fold rotations about $\vec{\bx},\vec{\by}$ or $\vec{\bz}$; such rotations, denoted as
$p\in \{C_{2x},C_{2y},C_{2z}\}$ respectively, preserve the orientation (or handedness) of the coordinate system. 

%remove: To restate our question precisely, suppose two spectroscopists are given an identical sample on which they 
%add
It follows from the above discussion that two spectroscopists, given an identical sample, may 
set down different coordinate systems parametrized by $(x,y,z)$ and  $(x',y',z')=p\circ (x,y,z)$ respectively; $p$ need not be a symmetry of the solid. Following identically the instructions of this work, the two spectroscopist would determine the $\Z_4$ and $\Z$ invariants based on their chosen coordinates; suppose the first  spectroscopist measures the numbers $(\chi^{\pm},\calc)$, and the second measures  $(p\circ \chi^{\pm},p\circ \calc)$. As proven in \app{sec:ambiguity}, for $p{\in}\{C_{2x},C_{2y}\}$,
$p {\circ} \chi^{\pm} {=}{-}\chi^{\mp}$ and $p{\circ} \calc{=}{-}\calc$. On the other hand, $C_{2z} {\circ} \chi^{\pm} {=}\chi^{\pm}$ and $C_{2z}{\circ} \calc{=}\calc$. In all cases, \q{bentchern2} is invariant. We may then draw the following conclusions depending on whether $\calc$ is even or odd: if even (which includes the insulating case), then $\chi^{\pm}{\equiv}{-}\chi^{\mp}$ according to \q{bentchern2}, and  two right-handed (or two left-handed)
spectroscopists always agree on their measured values for $\chi^{\pm}$. That is to say, $\chi^{\pm}=p\circ \chi^{\pm}$ for $p{\in}\{C_{2x},C_{2y},C_{2z}\}$.  If, however,  $\calc$ is odd, two right-handed spectroscopists are only guaranteed to agree  on the parity of $\chi^{\pm}$.  
%add 
Despite this ambiguity, once a convention for a coordinate system is fixed, the \emph{distinction} between phases is well-defined.

\section{Glide-resolved photoemission spectroscopy}\la{sec:pes}

This section is  a self-contained exposition on a spectroscopic method to identify the glide representation $\Delta_{\pm}(k_y)$ of initial Bloch states (i.e., Bloch states before photo-excitation). We assume only that the reader is familiar with basic notions in the representation theory of space groups, as reviewed briefly in  \s{sec:prelim}. 

Our method is applicable to surface or bulk photoemission. That is to say, our initial Bloch states may be localized to the surface (on which the radiation is incident) or delocalized throughout the bulk of the solid. In both  cases, we focus on  initial Bloch states with wavevectors on the glide-invariant line intersecting the surface-BZ center (the central glide line), as indicated by 23 in \fig{fig:bz}(a). Adopting our choice coordinates for real and quasimomentum spaces, this glide-invariant line lies at $k_x{=}0$, for a glide operation $g_x$ that maps $(x,y){\rightarrow}({-}x,y{+}R_2/2)$, with $R_2$ a primitive surface-lattice period. 

We will first describe the basic idea  in simple, intuitive terms in  \s{sec:heuristic}, where we specialize to normally incident, linearly polarized and monochromatic light. We shall assume  that the radiation gauge and dipole approximation are applicable to the electron-photon coupling; the dipole approximation is relaxed in the formal theory presented in \s{sec:rigorous}, where we also generalize to other incident angles and polarizations.

\subsection{Basic principle}\la{sec:heuristic}

\begin{figure}[h]
\centering
\includegraphics[width=8.6 cm]{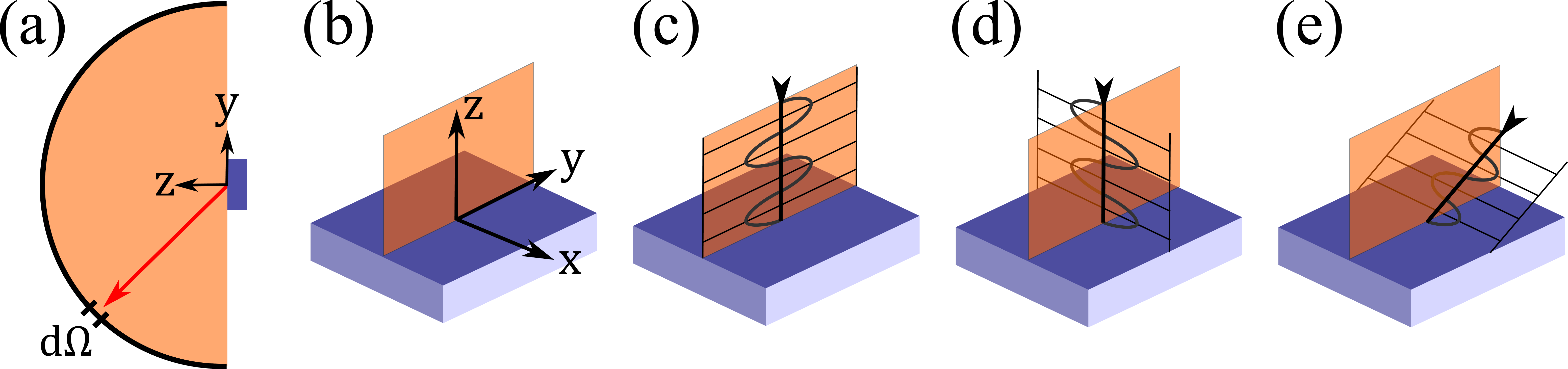}
\caption{(a) Photoemission setup: a sample (colored blue) is radiated and emits photoelectrons which are collected in a hemispherical cup. (b) Our choice coordinate system. (c-e) illustrate our favored incidence angles and polarizations. The spatial variation and directional vector of the electric field are indicated by the sinuisoidal lines. }\label{fig:pes}
\end{figure}

Suppose an electron -- with  Bloch wave function $\psi_i$, initial energy $\var_i$, and wavevector $\bkpar{=}(0,k_y)$ -- absorbs a photon and is excited to a photoelectronic state with energy $\var_p$. The electron-photon coupling is proportional to $\ba{\cdot} \hbp$ in the radiation gauge, where  $\ba{\propto}\vec{\epsilon}$ is the divergence-free electromagnetic vector potential, and the electromagnetic scalar potential is chosen to vanish. $\hbp$ above is the electronic momentum operator, which should be distinguished from the crystal wavevector $\bk$.  
%old: The  polarization vector $\vec{\epsilon}$ is chosen to lie parallel to the glide plane; 
%new
We choose normally-incident, linearly-polarized radiation with the polarization vector $\vec{\epsilon}$ lying parallel to the glide plane;
for the $g_x$-invariant yz-plane, $\vec{\epsilon}{=}\vec{\by}$ is the unit vector in the y-direction [cf.\ \fig{fig:pes}(c)]. In the dipole approximation, $\ba{\cdot} \hbp$ reduces to a spatially-homogeneous constant multiplied with $\hat{p}_y$. Since $\hat{p}_y$ is invariant under $g_x$ and surface-parallel translations, $\psi_i$ and the emitted photoelectron belong to the same representation of these symmetries; 
%add
we shall refer to this constraint as a selection rule.

This  selection rule has observable consequences for a photoelectron that is measured 
%add
at 
the detector. This photoelectron generically has a complicated wavefunction with a component in vacuum that extends toward the detector, and a separate component that penetrates the solid up to an escape depth.\cite{mahan_theoryphotoemission} 
%Crucially, the two components are not mutually related by any spacetime symmetry (e.g., $g_x$ and surface-parallel translations) of a surface-terminated solid. Consequently, 
Consider how a photoelectron transforms under any spacetime symmetry of a surface-terminated solid (in short, surface-preserving symmetry), as exemplified in this context by $g_x$ and surface-parallel translations. Such transformation is completely determined by the transformation of the photoelectron's component in vacuum, because a surface-preserving isometry never maps a point inside a solid to a point outside. 
%Crucially, the transformation of the entire photoelectron under such symmetries is completely determined by the transformation of the photoelectron's component in vacuum. 
Since vacuum is symmetric under continuous translations and $SU(2)$ spin rotations,\footnote{These are not symmetries of a spin-orbit-coupled solid.} the vacuum component  is simply  a linear combination of plane waves  with energy $\var_p{:}{=}(\hbar p)^2/2m$ and  wavevector $\bp$ (note $|\bp|{:}{=}p$);  for each $\bp$, there are two plane waves distinguished by the photoelectron spin. Due to the symmetry of discrete surface-parallel translations,  the surface-parallel component of $\bp$ must equal $\bkpar$ -- of the initial Bloch state -- modulo a surface-parallel reciprocal vector $\bG_{\parallel}$; each $\bG_{\parallel}$ corresponds to a different angle for photoelectrons to come out of the solid, as illustrated by the fan of arrows in \fig{fig:front}(e).

To understand the symmetry representation of the photoelectron, we must therefore analyze the symmetry properties of spin-polarized plane waves. 
Each $g_x$-invariant plane-wave state
%old  (denoted $\ket{\phi_{\bp,s}}{:}{=} \ket{\phi_{\bp}} {\otimes}\ket{s}$ )  have momentum $\bp$
%add
$\ket{\phi_{\bp,s}}$ is a tensor product ($\ket{\phi_{\bp}} {\otimes}\ket{s}$) of a spinless plane wave ($\braket{\br}{\phi_{\bp}}{=}e^{i\bp\cdot \br}$) and a spinor $\ket{s{=}{\pm} 1}$ in the eigenbasis of $S_x$. The momentum $\bp$ lies 
 parallel to the glide plane ($p_x{=}0$), and
%add
the
 spin orthogonal to the glide plane,
%add
such that
\e{ \hat{g}_x\ket{\phi_{\bp,s}}{=}{-}i\,s \, e^{-ip_yR_2/2}\ket{\phi_{\bp,s}};\;S_x\ket{s}{=}s \f{\hbar}{2} \ket{s}. \la{spinpolplanewave}}
The phase ${-}i\,s$ originates from reflecting $\ket{s}$  in the x-direction; after all, this reflection is just the composition of spatial inversion (which acts trivially on spin) and a two-fold rotation $e^{-i \pi (L_x+S_x)}$ about the x-axis. The phase  $e^{-ip_yR_2/2}$ in \q{spinpolplanewave} originates from translating $\phi_{\bp}$ by half a lattice period in the y-direction. We 
%remove may 
%add
can always 
express  $p_y{=}k_y{+}2\pi n/R_2$ such that $k_y$ lies in the first Brillouin zone (BZ) and $n{\in} \Z$. Recalling from \s{sec:prelim} that a Bloch state in the $\Delta_{\pm}(k_y)$ representation has glide eigenvalue ${\pm}ie^{-ik_yR_2/2}$, we conclude that $\phi_{\bp,s}$ transforms in the $\Delta_{-s}(k_y)$ representation if $n$ is even, and in the $\Delta_{+s}(k_y)$ representation if $n$ is odd. 

Combining this symmetry analysis with our selection rule, we find the following constraint for a photoelectron that is excited from an initial Bloch state ($\bkpar{=}(0,k_y)$) in the $\Delta_{+}(k_y)$ representation. Namely, 
%old: if a photoelectronic plane wave is detected with surface-parallel momentum $\bp_{\parallel}{=}(0,p_y)$,
%add
the photoelectronic plane wave ($\ket{\phi_{\bp,s}}$) that is detected must also belong in the $\Delta_{+}(k_y)$ representation; this implies that  
the spin of the photoelectron is nontrivially locked to its momentum: 
% remove: $(p_y{-}k_y)R_2/2\pi$ is odd (resp.\ even) if and only if $\langle S_x\rangle{=}\hbar/2$ (resp.\  ${-}\hbar/2$);
%new
expressing the surface-parallel component of $\bp$ as $\bp_{\parallel}{=}(0,p_y{=}k_y{+}2\pi n/R_2)$, then 
\bal
\langle S_x\rangle=s\f{\hbar}{2}=\begin{cases}+\f{\hbar}{2}, &\text{if}\; n\in 2\Z+1\\
                                                -\f{\hbar}{2}, &\text{if}\; n\in 2\Z\end{cases}
																										\la{spinmomlock}\end{align}
%$n$ is odd  (resp.\ even) if and only if $\langle S_x\rangle{=}s\hbar/2{=}\hbar/2$ (resp.\  ${-}\hbar/2$); 
If the initial Bloch state were in the $\Delta_-$ representation, then \q{spinmomlock} holds with the interchange of `odd' and `even'.  This spin-momentum locking manifests the glide symmetry of the spin-orbit interaction. As a consequence, each ray of the fan [in \fig{fig:front}(e)], corresponding to a unique value of $n$, is fully spin polarized; nearest-neighbor rays always have opposing polarizations. The angle of each ray is determined by energy conservation: 
\e{ \var_i+\hbar\omega=\f{\hbar^2 (k_y+2\pi n/R_2)^2 +\hbar^2p_z^2}{2m}. }
Tantalizingly, each ray may be isolated experimentally by standard spin- and angular-resolution techniques that measure $\langle S_x\rangle$ and $p_y$;\cite{hufner_book} this allows us to spectroscopically identify the glide representation of an initial state.

%\footnote{ We assume the relativistic mass correction is negligible for electrons bound to the solid, and also for the photoelectron. This assumption places an upper bound on the photon energy ($\hbar \omega{\ll}mc^2$) which is more than adequately satisfied in the visible and ultraviolet regime.}

\subsection{One-step theory of glide-resolved photoemission}\la{sec:rigorous}

 To justify this spin-momentum locking rigorously, we employ the steady-state scattering formulation\cite{lippmann_schwinger,gellmann_scattering,bethe_quantummechanics} of the one-step theory\cite{adawi_theoryphotoelectric,mahan_theoryphotoemission,feibelman_theoryphotoemission} of photoemission. 
%old: In the absence of radiation, the independent-electron Hamiltonian 
We begin with the component of the Hamiltonian that describes the solid in the absence of radiation; in the independent-electron approximation, this 
assumes the standard Pauli form: $H_e{=}(\hbar\hbp)^2/2m{+}{V}$, in the non-relativisitic limit\cite{foldy_wouthuysen,blount_extendsfoldy} of the Dirac Hamiltonian; $V$ includes a scalar potential, the spin-orbit coupling, and in principle also the Darwin term. Since $V$ encodes a mean-field interaction of a single electron with other electrons as well as the ionic lattice,  $V$ falls off to zero rapidly away from the solid.\cite{ashcroft_mermin} Here, we have adopted the usual electrostatic convention for the zero of energy -- as the %old: minimal energy of a plane wave 
%new:
energy of a zero-momentum plane wave
in free space (far away from the solid). 

%old: The aforementioned initial Bloch state $\psi_i$ is 
%new:
Suppose $\psi_i$, 
an eigenstate of $H_e$ with energy $\var_i$ below the Fermi level, 
%add
absorbs a single photon with energy $\hbar \omega$; $i$ here includes all quantum numbers of the eigenstate, including the band index and the crystal wavevector. The corresponding photoelectron has energy $\var_i{+}\hbar \omega{>}0$, and a spinor  wavefunction of the form:
\e{\Psi_{p,i}= G^{\sma{+}}(\var_p)H_{int}\psi_i, \as \var_i+\hbar \omega:=\var_p:=(\hbar p)^2/2m \la{scatteredwf}} 
to lowest order in the 
%old: fine-structure constant.
%new
electron charge.\footnote{$\Psi$ may be derived by a simple generalization of Adawi's calculation\cite{adawi_theoryphotoelectric} to include the effect of spin. The essential structure of the derivation is identical; one merely has to include a spin-orbit-coupling and Darwin terms to $V$ in Eq.\ (2.1) of \ocite{adawi_theoryphotoelectric}, and to interpret $\phi_0$ in Eq.\ (2.3) as a spinor state.} Here we have introduced the advanced/retarded Green's functions: $G^{\sma{\pm}}(\var){=}(\var{-}H_e{\pm}i\delta)^{\mo}$, with infinitesimal $\delta{>}0$.  The electron-photon coupling has the  form $H_{int}{=}|e| (\ba{\cdot} \hbp{+}\hbp {\cdot}\ba)/2mc$ in the temporal gauge, where the scalar potential vanishes; $\ba$ here is the screened\cite{feibelman_theoryphotoemission,feibelman_review} electromagnetic vector potential in the solid. The Zeeman interaction with the spin magnetic moment typically has a small effect relative to the $\ba {\cdot}\hbp$ term,\cite{feuchtwang_reviewspinpolII,feder_review} and is therefore neglected from $H_{int}$; a further evaluation of  the Zeeman interaction is provided in \s{sec:discussion}.

%\footnote{More quantitative comparative studies of the two terms seem desirable; in principle the Zeeman effect may reduce the full spin polarization  } 

Given that $\psi_i$ belongs to a certain glide representation, we would like that the photoelectron transforms in  a glide representation that is uniquely determined by the representation of $\psi_i$. Such a selection rule exists if the  electron-photon coupling $H_{int}$ transforms in a one-dimensional representation of glide symmetry, i.e., $\hatbmx H_{int}\hatbmx^{-1}$ equals $H_{int}$ up to a phase, with $\hatbmx$ the operator that implements glide reflection [cf.\ \q{spinpolplanewave}].

As shown in \app{app:lightsource}, the desired transformation of $H_{int}$ exists for a linearly-polarized light source, with wavevector  parallel to the glide-invariant yz plane, and with the polarization vector $\vec{\epsilon}$  either orthogonal [see \fig{fig:pes}(d-e)] or parallel [\fig{fig:pes}(c)] to the glide-invariant plane. In the standard convention, we identify the orthogonal alignment   as $s$ polarization, and the parallel alignment  as $p$ polarization, though such identifications are not meaningful for normal incidence.

In the case of normal incidence, the Fresnel equations inform us that the  light remains linearly polarized (with the same polarization vector $\vec{\epsilon}$) upon transmission into the solid; that is to say, the vector potential $\ba$ within the solid remains parallel to $\vec{\epsilon}$. In the orthogonal alignment, $H_{int}{\propto}\ba{\cdot}\hbp{\propto}\hat{p}_xe^{-i\omega z/c}$  anticommutes  with the glide operator $\hatbmx$; with the parallel alignment, $H_{int}{\propto}\hat{p}_ye^{-i\omega z/c}$ commutes\footnote{Because $[H_e,\hatbmx]{=}0$ and $\hatbmx$ is a unitary operator.} with $\hatbmx$. In the more general case of non-normal incident angles [see \fig{fig:pes}(e)], it is shown in \app{app:lightsource} that 
\e{\hatbmx H_{int}\hatbmx^{-1}=\pm e^{-iq_yR_2/2}H_{int},\la{onedimrepHint}} 
with the plus (resp.\ minus) sign applying to $p$ (resp.\ $s$) polarization, and the additional phase factor $e^{-iq_yR_2/2}$ originating from a half-lattice translation of the photon field (having wavenumber $q_y$ within the solid).

%so long as the dipole approximation is valid; the latter approximation is  well justified  in the visible or ultraviolet regime ($\hbar\omega {\lesssim}200$eV). Note that the dipole approximation is not needed in the aforementioned cases with normal incidence.

% with a polarization vector $\vec{\epsilon}$ that points either to $\vec{\bx}$  [for any incident angle, as illustrated in \fig{fig:pes}(d-e) ], or to  $\vec{\by}$ [for normal incidence \emph{only}, as illustrated in \fig{fig:pes}(c)]. 

Since $\hatbmx$ commutes with $G^{\sma{+}}(\var_p)$ [cf.\ \q{scatteredwf}], $\Psi_{p,i}$ and $H_{int}\psi_i$ transform in the same representation of $g_x$.  That is, if $\psi_i$ is a Bloch function ($\bkpar{=}(0,k_y)$) in the $\Delta_{\pm}(k_y)$ representation, then $\Psi_{p,i}$ belongs in the $\Delta_{\mp}(k_y+q_y)$ [resp.\ $\Delta_{\pm}(k_y+q_y)$] representation for $\vec{\epsilon}$ orthogonal (resp.\ parallel) to the glide-invariant plane; the addition of $q_y$ in the argument represents the absorption of the photon's momentum [cf.\ \q{onedimrepHint}]. Assuming the surface is clean and unreconstructed, $\Psi_{p,i}$ also transforms under discrete translations in the representation $\bkpar{=}(0,k_y+q_y)$.

Let us translate these selection rules to a spin-momentum-locking constraint on the measured photocurrent. We begin with an identity
%remove with the free-space Green's function $G_0^{\sma{\pm}}(\var){=}(\var{-}(\hbar\hbp)^2/2m{\pm}i\delta)^{\mo}$, 
%add
relating $G^{\sma{\pm}}$ to the  free-space Green's function $G_0^{\sma{\pm}}$:
\e{G^{\sma{\pm}}=G_0^{\sma{\pm}}+G_0^{\sma{\pm}}VG^{\sma{\pm}};\; G_0^{\sma{\pm}}(\var):=\big(\var-\tf{(\hbar\hbp)^2}{2m}\pm i\delta\big)^{\mo}.\la{greensiden}}
%remove whose asymptotic, spherical-wave form is well-known:
%add
The asymptotic, spherical-wave form of $G_0^{\sma{\pm}}$ is well-known:\cite{griffiths_introQM} 
for $r({:}{=}|\br|)$ and 
\e{ r \gg r',\;\bra{\br,s}G_0^{\sma{\pm}}(\var_p)\ket{\br',s'}\sim {-}\f{m}{\hbar^2}\f{e^{\pm ipr}}{2\pi r}e^{\mp i\bp\cdot \br'}{\delta_{ss'}},  \la{asymptoticfree}} 
where $\vec{\br}$ is the unit vector parallel to $\br$, $\bp{:}{=} p\vec{\br}{:}{=}(p_x,p_y,p_z)$, 
%new:
$\ket{\br,s}$ is an eigenstate of position and $S_x$ operators, and $\sim$ denotes the leading asymptotic form for large $r$.

Let us apply the identity \q{greensiden} and  the asymptotic form of $G_0^{\sma{+}}$ [\q{asymptoticfree}] to evaluate $\Psi_{p,i}$ defined in \q{scatteredwf}. 
%new
Combining \qq{scatteredwf}{greensiden}, we derive
\e{\Psi_{p,i}(\br,s)\eq \sum_{s'=\pm 1}\int d\br'\bra{\br,s}G_0^{\sma{+}}(\var_p)\ket{\br',s'}\lin
&\times \bra{\br',s'}\big(\,I+VG^{\sma{+}}(\var_p)\,\big )H_{int}\ket{\psi_i}. \la{forandrei}}
For the scattering geometry illustrated in \fig{fig:pes}(a), we take $\br$ to be a position on the hemispherical detector, and choose our spatial origin to lie within the solid.
%remove: $\br'$ in  \q{asymptoticfree} is an integration variable in $\int d\br' G_0^{\sma{+}}\ket{\br'}\bra{\br'}VG^{\sma{+}}$.
 Since $V$ vanishes rapidly away from the solid,\cite{ashcroft_mermin} the domain of integration (over $\br'$) may effectively be limited to a finite volume that is at most the order of the sample volume.\footnote{In practice, this domain may be even smaller due to a short  photoelectron escape depth, and possibly also the finite cross-sectional area of the photon beam.} Assuming that the detector-to-sample distance is much greater than the sample dimension (which is valid in most modern ARPES set-ups),  the  condition $r{\gg}r'$ is satisfied 
%new
for all $\br'$ in the domain of integration, 
hence we may utilize the asymptotic form of the free-space Green's function in \q{asymptoticfree}. Thus, combining \q{forandrei} with \q{asymptoticfree}, we derive
\begin{flalign}
\text{as}\; r \rightarrow\infty,&\as \Psi_{p,i}(\br,s)\sim -\f{m}{\hbar^2}\f{e^{ipr}}{2\pi r} \braopket{\Phi_{\bp,s}}{H_{int}}{\psi_i}, \la{asymptoticPsi}
\end{flalign}
where $\ket{\Phi_{\bp,s}}$ is defined as 
\e{\ket{\Phi_{\bp,s}} := \ket{\phi_{\bp,s}}+G^{\sma{-}}(\var_p)V\ket{\phi_{\bp,s}}.\la{lippPhi}}
We remind the reader that  $\ket{\phi_{\bp,s}}$ is a plane-wave state with momentum $\bp$ and spin eigenvalue $s\hbar/2$ under $S_x$ [cf.\ \q{spinpolplanewave}]. \q{lippPhi} may be identified as the Lippmann-Schwinger equation\cite{lippmann_schwinger} with the retarded Green's function; this informs us that $\ket{\Phi_{\bp,s}}$ is an eigenstate of $H_e$  with the boundary condition of an inverse low-energy electron diffraction (LEED) experiment.\footnote{ In an LEED experiment, an electron beam is directed to and diffracted off the surface of a solid.  It is well-known\cite{adawi_theoryphotoelectric,mahan_theoryphotoemission,feibelman_theoryphotoemission} that $\phi_{-\bp}{+}G^{\sma{+}}V\phi_{-\bp}$ is the spinless wavefunction for an LEED experiment in which the incident electron beam has momentum ${-}\bp$; this wavefunction shall be referred to as an LEED state. In the spinless theory, the inverse-LEED state can be defined as the time-reversed LEED state; time-reversal has the effect of inverting momentum ${-}\bp{\rightarrow} \bp$ and sending $G^{\sma{+}}{\rightarrow} G^{\sma{-}}$, thus producing the spinless analog of \q{lippPhi}. In our spinful theory, the LEED state ($\phi_{-\bp,s}{+}G^{\sma{+}}V\phi_{-\bp,s}$) describes an incoming, spin-polarized electron beam with momentum ${-}\bp$, and the inverse LEED state [right-hand side of \q{lippPhi}] describes an outgoing  beam with inverted momentum ($\bp$)  but  the \emph{same} spin polarization $s$.  } 

Let us evaluate the spin-resolved probability current  through a solid angle element $d\Omega$ centered at $\br$, as depicted in \fig{fig:pes}(a). The current contributed by $\psi_i$ is expressible as a Fermi golden rule:
\e{  \f{dI^i_{\bp,s}}{ d\Omega} {=} \f{r^2\hbar}{m} {\text{Im}}\left[ \Psi^*_{p,i}\partial_{r}\Psi_{p,i} \right]_{\br,s}{\sim} \f{2\pi}{\hbar}\rho_p |\braopket{\Phi_{\bp,s}}{H_{int}}{\psi_i}|^2, \la{fgr}}
where $\rho_p{:}{=}mp/(2\pi)^3\hbar^2$ is the density of plane-wave states per unit real-space volume and solid angle. The measured current at the detector is obtained by summing $dI^i_s$ over all initial states. 
%add
\qq{asymptoticPsi}{fgr} are the generalization of the inverse-LEED (or one-step) theory of photoemission (as originally formulated by Adawi\cite{adawi_theoryphotoelectric} and Mahan\cite{mahan_theoryphotoemission}) to include the effect of spin. Equivalent golden-rule formulae (for spin systems) have previously been derived\cite{ackermann_spin_onestep,ginatempo_spin_onestep,feder_review} based on a different formalism by  Pendry.\cite{pendry_onestep}

%One aspect of the spinful theory differs from the spinless: our inverse-LEED state is not equivalent to a time-reversed LEED state, in the sense that an inverse-LEED state does not have its spin inverted relative to the LEED state.\footnote{This is explained in the previous footnote.} 

Let us  consider the subgroup $\cala$ of \emph{spatial} symmetries that are preserved in the presence of the surface, i.e., they are the (subset of) symmetries of $H_e$ that do not involve time reversal.\footnote{We are describing the physically-motivated scattering geometry in \fig{fig:pes}(a), where the sample is finite in all three directions.  Rigorously, surface-parallel translations and glide cannot be symmetries of any surface with finite area. Practically, if the sample area is macroscopic, the effect of corners and edges are negligible to a state whose wavefunction is extended in $\vec{\bx}$ and $\vec{\by}$. Assuming that $\psi_i$ is such an extended state, it may be characterized to a good approximation by the symmetries of a semi-infinite solid (with a single surface); the unitary subgroup of such symmetries is denoted $\cala$.} A nonvanishing $dI^i_{\bp,s}$ requires that $\bra{\Phi_{\bp,s}}H_{int}\ket{\psi_i}{\neq}0$; according to the Wigner-Eckhart theorem, this further requires that 
\e{\Gamma^*_{\Phi_{\bp,s}}\otimes\Gamma_{H_{int}}\otimes\Gamma_{\psi_i}=E\oplus \ldots,\la{interresult}}
where $\Gamma_{\alpha}$ is the representation of $\alpha$ under $\cala$, $\Gamma^*$ denotes the complex-conjugate representation, and $E$ is the trivial representation. Since spatial symmetries are  represented unitarily, each element in $\cala$ commutes with both $G^{\sma{-}}$ and $V$. Therefore, we deduce from \q{lippPhi} that $\Phi_{\bp,s}$ and $\phi_{\bp,s}$ belong to the same representation of $A$. In combination, 
\e{dI_{\bp,s}^i\neq 0 \imp \Gamma^*_{\phi_{\bp,s}}\otimes\Gamma_{H_{int}}\otimes\Gamma_{\psi_i}=E\oplus \ldots,\la{keyresult}}
summarizes  a key result of this work: it states that the spin-resolved photocurrent satisfies selection rules based on the overlap of $H_{int}\psi_i$ with a spin-polarized plane wave $\phi_{\bp,s}$. The full generality of this result is explored in \s{sec:discussion}, but for now we content ourselves with the application at hand.

Applying \q{keyresult} to the representation of discrete surface-parallel  translations, we derive the well-known result that if $\psi_i$ has crystal wavevector $\bkpar$ and the photon has wavevector $\bq_{\parallel}$ within the solid, then $dI_s^i$ is only nonvanishing for $\bp_{\parallel}{=}\bkpar{+}\bq_{\parallel}$ modulo a surface reciprocal vector.

Applying \q{keyresult} to the representation of glide symmetry ($g_x$), and to plane waves propagating parallel to the glide plane (i.e., $p_x{=}0$), we derive that $dI_s^i$ is only nonvanishing for one spin component $s$; which component depends on the magnitude of $p_y$ and the glide representation of $H_{int}\psi_i$, as has been explained in  \s{sec:heuristic}
%add
[cf.\ \q{spinmomlock}].

%old: In contrast, such a full correlation between spin and momentum does \emph{not} occur for  plane waves that propagate in a direction non-parallel to the glide plane (i.e., $p_x{\neq}0$). Two scenarios for this are worth considering: (i) $\psi_i$ belongs to the central glide line (23, $k_x{=}0$), and the corresponding photoelectron is emitted with wavevector $p_x{=}2\pi m/a_1$ with $m$ a nonzero integer and $a_1$ the primitive surface-lattice period, and (ii) $\psi_i$ belongs to the off-center glide line (01, $k_x{=}\pi/a_1$), and the corresponding photoelectron is emitted with wavevector $p_x{=}\pi(2n{+}1)/a_1$ with $n{\in}\Z$.  In either scenario, the plane-wave state $\phi_{\bp,s}$ (with $p_x{\neq}0$) may always be decomposed as $\phi_{\bp,s}(x){=}$cos$(p_x x)\ket{\phi_{p_y,s}}{+}i$sin$(p_x x)\ket{\phi_{p_y,s}}$; these two components belong to distinct representations of $g_x$, because the cosine (resp.\ sine) is even (resp.\ odd) under the reflection: $x{\rightarrow}{-}x$. Consequently, no matter the glide representation of $\psi_i$ and no matter the magnitude of $p_y$, glide symmetry does not enforce $dI_{\bp,s}^i$  to vanish for any spin component ($s$) of $S_x$. Restating this conclusion, the photocurrent is not expected to be spin-polarized in the $x$ direction on grounds of glide symmetry. An implication is that the spin-momentum-locking technique cannot be used to determine the glide representation of initial Bloch states on the off-center glide line 01. 

%remove: In contrast,
%add
For the \emph{same}, glide-invariant initial state $\psi_i$ (with $k_x{=}0$), 
such a full correlation between spin and momentum does \emph{not} occur for 
%add
photoelectronic 
plane waves that propagate in a direction non-parallel to the glide plane (i.e., 
%remove: $p_x{\neq}0$). Two scenarios for this are worth considering: (i) $\psi_i$ belongs to the central glide line (23, $k_x{=}0$), and the corresponding photoelectron is emitted with wavevector $p_x{=}2\pi m/a_1$ with $m$ a nonzero integer and $a_1$ the primitive surface-lattice period, and (ii) $\psi_i$ belongs to the off-center glide line (01, $k_x{=}\pi/a_1$), and the corresponding photoelectron is emitted with wavevector $p_x{=}\pi(2n{+}1)/a_1$ with $n{\in}\Z$.  In either scenario, the
%add
$p_x{=}2\pi m/a_1$ with $m$ a nonzero integer and $a_1$ the primitive surface-lattice period). To explain this, consider that a one-dimensional plane wave $e^{ip_xx}$ (with $p_x\neq 0$) is a sum of two components [cos$(p_x x)+i $sin$(p_x x)$] that transform in even and odd  representations of the reflection: $x{\rightarrow}{-}x$; likewise, $\phi_{\bp,s}$ is the sum of two components belonging     to distinct representations of $g_x$.
%cos$(p_x x)\ket{\phi_{p_y,s}}{+}i$sin$(p_x x)\ket{\phi_{p_y,s}}$   
%plane-wave state $\phi_{\bp,s}$ (with $p_x{\neq}0$) may always be decomposed as
%$\phi_{\bp,s}(x){=}$cos$(p_x x)\ket{\phi_{p_y,s}}{+}i$sin$(p_x x)\ket{\phi_{p_y,s}}$;
 %these two components belong to distinct representations of $g_x$, because the cosine (resp.\ sine) is even (resp.\ odd) under the reflection: $x{\rightarrow}{-}x$. 
Consequently, no matter the glide representation of $\psi_i$ and no matter the magnitude of $p_y$, glide symmetry does not enforce $dI_{\bp,s}^i$  to vanish for any spin eigenvalue ($s$) of $S_x$. Restating this conclusion, the photocurrent is not expected to be spin-polarized in the $x$ direction on grounds of glide symmetry.

%insert
Finally, let us consider a glide-invariant initial state $\psi_i$ belonging to the off-center glide line (01, $k_x{=}\pi/a_1$). The corresponding photoelectron must be emitted with \emph{nonzero} wavenumber $p_x{=}q_x{+}\pi(2n{+}1)/a_1$ with $n{\in}\Z$. By the same argument (given in the previous paragraph), we may conclude that the photocurrent will not be spin-polarized. To recapitulate, 
the spin-momentum-locking technique cannot be used to determine the glide representation of initial Bloch states on the off-center glide line 01.

\section{Discussion and summary}\la{sec:discussion}

 We have proposed a $\Z_4{\oplus}\Z$ strong classification of glide-symmetric solids (inclusive of band insulators and metals); for each nontrivial class of the $\Z_4$ classification, we have proposed a materialization in KHgSb, Ba$_2$Pb and stressed Na$_3$Bi. The smoking-gun signatures of each phase are described
%remove: for the parallel transport of bulk states, as well as 
in the photoemission of surface states.  To facilitate the identification of $\chi^{\pm}{\in} \Z_4$, we have proposed a  method to measure initial-state glide eigenvalues in photoemission spectroscopy. It is further shown that any two spectroscopists -- employing distinct spatial coordinate systems but with the same orientation -- will agree on: (a) $\chi^{\pm}$ modulo four, if $\calc$ is even, and (b) $\chi^{\pm}$ modulo two, if $\calc$ is odd. The implications of (a) for Ba$_2$Pb ($\chi^+{=}{-}1$) and stressed Na$_3$Bi ($\chi^+{=}{+}1$) is that they may be sharply distinguished through glide-resolved photoemission.

Our method to resolve glide eigenvalues exploits  a spin-momentum locking that  characterizes the photoemission of \emph{any} glide-symmetric solid. In more detail, a photoelectron is emitted 
%remove: into a fan of rays illustrated in \fig{fig:front}(e), with alternating rays having opposite spin orientations.
%new:
into vacuum as a superposition of plane waves, whose wavevectors are  illustrated in \fig{fig:front}(e) as a fan of rays; distinct rays differ by a surface reciprocal vector, and every adjacent pair of rays has opposite spin orientations. 

As an orthogonal application of this locking, one may generate a photocurrent  with near-perfect spin polarization by isolating one of these rays, using standard angle-resolved PES techniques. 
%add:
Photoemission sources of spin-polarized electrons have diverse applications as spectroscopic probes of solid-state systems;\cite{kirschner_sources_detectors} such sources form the basis for spin-polarized bremsstrahlung isochromat spectroscopy,\cite{scheidt_spinbremsstrahlung} spin-polarized low-energy electron diffraction,\cite{KIRSCHNER_spinLEED} spin-polarized electron-energy-loss spectroscopy (e.g., for the investigation of Stoner excitations\cite{kirschner_spinEELS}), and spin-polarized appearance potential spectroscopy.\cite{KIRSCHNER_spinappearancepotential} While beam current densities of existing GaAs-based, photoemission sources  are satisfactory, their spin polarization is theoretically limited to 50\%, with experiments achieving just over 40\%;\cite{kirschner_sources_detectors}  in comparison, our proposed spin polarization can in principle be complete (100\%), assuming the surface-terminated solid perfectly maintains glide symmetry.

For the above applications, spin-orbit-split energy bands are desirable; otherwise,  distinct glide representations would be energy-degenerate at each crystal wavevector,\cite{Cohomological} and their combined photoemission would result in cancelling spin polarizations. 
%add
Practically, the spin-orbit splitting should be larger than the energy resolution of the detector in PES.
Two types of spin-orbit-split energy bands may be utilized: (a) for bulk bands (whose wavefunctions extend over the entire solid), it is necessary 
%add: 
(but not sufficient\footnote{Even in noncentrosymmetric space groups, certain other point-group symmetries might result in energy-degenerate initial states that have the same reduced wavevector ($k_x,k_y$) but belong in distinct glide representations. For example, a reflection symmetry $M_z$ (that inverts $z$) anticommutes with glide $g_x$ in the spinor representation; hence $M_z$ relates two energy-degenerate states [at $(k_x,k_y,\pm k_z)$] that belong to distinct representations of $g_x$.}) 
that the space group is noncentrosymmetric; otherwise distinct glide representations would remain energy-degenerate (at each crystal wavevector\cite{Cohomological}) even in the presence of spin-orbit coupling. (b)  For spin-orbit-split surface bands, no such restriction on the space group is needed, because spatial inversion symmetry is anyway broken at a surface.

%remove: Our derivation of 100\% spin polarization  relies on three main approximations: 
%add
Our derivation of 100\% spin polarization is based on a model of the first-order photoelectric effect with the following approximations: 
(i) the independent-electron approximation,  (ii) a classical, Maxwell-based approximation to the electromagnetic wave in the solid,  (iii) the neglect of  the Zeeman 
%remove: magnetic moment 
interaction 
%add
(with the magnetic field of the radiation)
relative to the minimal coupling $\propto \ba {\cdot} \hbp$ [cf.\ the discussion in \s{sec:rigorous}], and (iv) a surface termination which perfectly respects glide symmetry. One effect of many-body interactions in Fermi liquids is to add a continuous background to the photoemission intensity, which may reduce (but not eliminate) the full spin polarization associated to a sharp peak. (ii) is a good approximation for radiation of certain polarization and incidence angles, as explained in \s{sec:rigorous} and \app{app:lightsource}. (iii) is widely believed to be valid\cite{feder_review} and has been substantiated by model calculations;\cite{feuchtwang_reviewspinpolII} however, further quantitative studies are desirable. 
%add
(iv) Our prediction of 100\% spin polarization also assumes that the surface of a glide-symmetric solid is also glide-symmetric. That is to say, if any surface relaxation or reconstruction occurs, we assume it  preserves the glide symmetry; this assumption should be checked for any candidate material. In principle, glide-asymmetric surface defects may also reduce the spin polarization.  We briefly remark on  the spin polarization of  the second-order photocurrent, which is induced by the absorption of two photons: for simplicity we consider normally-incident light with polarization vector parallel or orthogonal to the glide plane; in both cases, the second-order photocurrent is fully spin-polarized; in the former (resp.\ latter) case, the second-order spin polarization is parallel (resp.\ anti-parallel) to the first-order spin polarization.\footnote{This may be derived from a simple generalization of our theory. Second-order processes are either quadratic in $\ba\cdot\bhp$ or linear in $\ba^2$, with $\ba=a_0\vec{y}e^{-i\omega z/c}$ for the parallel alignment and $\ba=a_0\vec{x}e^{-i\omega z/c}$ for the anti-parallel alignment. For either alignment, $(\ba\cdot\bhp)^2$ and $\ba^2$ are invariant under the glide operation, hence the photoelectron has the same glide representation as the initial state.  }

A comparison  with existing proposals for spin-polarized photocurrents is instructive. It is not surprising that photoemission from a spin-polarized groundstate  would be spin-polarized;\cite{feder_polarizedelectrons_book} such groundstates  exhibit either long-ranged magnetic order or spontaneously-broken spatial symmetries leading to a spin-split Fermi surface.\cite{hirsch_spinsplit,congjun_dynamicSOC,congjun_fermiliquidinstabilities,chubukov_spinsplit,rashba_spinsplit_AAJH}
For groundstates without spontaneous ordering,  only partially spin-polarized photocurrents have been realized practically, and typically  only with circularly-polarized light.\cite{kirschner_sources_detectors} We highlight an existing theoretical proposal which relies on neglecting the $\ba{\cdot}\hbp$ interaction in favor of the Zeeman interaction: a fully spin-polarized photocurrent may then  be generated by radiating a solid (having \emph{negligible} spin-orbit coupling) with {circularly}-polarized light. The neglect of the $\ba{\cdot}\hbp$ interaction is valid only for special geometric configurations,\cite{feuchtwang_reviewspinpolII} and even so the Zeeman-induced photocurrent is expected to be  weak.\cite{feder_review} 
%remove: For these reasons, we believe our complementary proposal (with linearly-polarized light and spin-orbit-coupled solids) to be practically more effective.

While we have focused on glide-symmetric solids throughout this work, we highlight a result that is generally applicable to the photoemission of any spin-orbit-coupled solid, no matter its space group. 
Our result is that the spin-resolved photocurrent (contributed by an initial Bloch state $\psi_i$) satisfies a Wigner-Eckhart-type selection rule based on the overlap of $H_{int}\psi_i$ with a spin-polarized plane-wave state, as summarized in \q{keyresult}. Here, our selection rules are based only on spatial symmetries that are preserved in the presence of a surface; these symmetries are determined by the exact conditions of the surface, including possible relaxation or reconstruction effects. $H_{int}$ here is the electron-photon coupling, and may in principle include either or both of the  $\ba {\cdot}\hbp$ and Zeeman interactions. It should be emphasized that \q{keyresult} has been derived \emph{without} the dipole and Born approximations. In the Born approximation, the final state of photoemission [cf.\ \q{lippPhi}] is approximated as  a plane wave;\cite{adawi_theoryphotoelectric} this approximation is certainly invalid at lower photon energies.\cite{feder_review} Also, we remark that \q{keyresult} has been derived within the one-step theory, which is more accurate\cite{goldmann_onestep_Cu} and more generally applicable\cite{mahan_theoryphotoemission,feibelman_theoryphotoemission} than the three-step theory\cite{spicer_threestep,berglund_spicer_threestep}  -- only the one-step theory  can describe surface photoemission.

%Our result may be intuitively rationalized with the Born approximation, where one approximates the final state of photoemission with a plane wave.\cite{adawi_theoryphotoelectric} However, it should be emphasized that our result has been derived \emph{without} the Born approximation, and is therefore applicable to lower photon energies where the Born approximation  

%We also remark that \q{keyresult} has been derived without the Born approximation, i.e., by the replacement of the final state in photoemission by a plane wave, as is commonly done in the literature.[wang] While it remains questionable if the Born approximation is quantitatively accurate (especially in the visible and UV regime), we point out that the group-theoretic selection rule of \q{keyresult} has been derived without the Born 

For a final illustration, we apply \q{keyresult} to solids with a reflection (or mirror) symmetry that is \emph{not} a glide symmetry.  For simplicity, we consider normally- incident light with a polarization vector lying parallel to the 
%add
mirror-invariant plane. The associated photocurrent would also spread out in a fully-spin-polarized fan illustrated in \fig{fig:front}(e), except the direction of spin polarization would \emph{not} alternate between adjacent rays. This alternation is a fundamental property of glide symmetry,
%add
which is special for having a momentum-dependent  eigenvalue ${\propto} e^{-ik_y/2}$. 
This provides a sharp distinction between the photoemission of mirror- and glide-symmetric solids. This distinction exists for both insulators and metals, in both trivial and topological categories. In particular, one may compare the surface photoemission of the mirror-symmetric topological insulator SnTe\cite{Hsieh_SnTe} with any of the glide-symmetric topological insulators that have been proposed in this work.

%add
In the late stages of this work, Ryoo et al. have independently formulated\cite{Ryoo_glideARPES} the glide selection rule  that is one result of this work. While their selection rule is derived assuming the dipole approximation (which is generally invalid for 
surface photoemission induced by $p$-polarized light\cite{feibelman_surfacereflectivity,levinson_surfacephotonfield,feibelman_review}), the pedagogical derivation presented in \s{sec:rigorous} does not rely on the dipole approximation.  \\

%While some conclusions overlap, we point out that our formal derivation of the selection rule [in \s{sec:rigorous}] rigorously justifies the heuristic arguments presented in \ocite{Ryoo_glideARPES}. Moreover, we disagree on the validity of the dipole selection rule: while \ocite{Ryoo_glideARPES} claims that the selection rule is applicable with any $p$-polarized light, past works\cite{feibelman_surfacereflectivity,levinson_surfacephotonfield,feibelman_review,goldmann_onestep_Cu,feder_review} have demonstrated that the dipole approximation (for surface photoemission) is generically invalid for $p$-polarized light (except for normally-incident $p$-polarized light) [cf.\ \app{app:lightsource}]. One superficial discrepancy (in the formulation of the selection rule) stems from a semantic error in their use of `second Brillouin zone': they have claimed that the spin polarization of a photoelectron emitted in the first Brillouin zone is opposite to  that emitted in the second Brillouin zone;\cite{Ryoo_glideARPES} the correct statement is that the spin polarizations are opposite for two photoelectrons whose wavevectors differ by a primitive surface reciprocal vector (parallel to the fractional translation of the glide operation) [cf.\  \q{spinmomlock} in this work and also Eq.\ (1-5) in \ocite{Ryoo_glideARPES}].   \\   

\begin{acknowledgments}
We are grateful to Ken Shiozaki and Masatoshi Sato for informative discussions that linked this work to their K-theoretic classification. Ji Hoon Ryoo, Ilya Belospolski, Ilya Drozdov and Peter Feibelbaum helped to clarify the discussion on photoemission spectroscopy. We especially thank Ji Hoon Ryoo, Ken Shiozaki and Judith H\"oller for a critical reading of the manuscript.  AA was supported by the Yale Postdoctoral Prize Fellowship and the Gordon and Betty Moore Foundation EPiQS Initiative through Grant No. GBMF4305 at the University of Illinois. Z. W. was supported by the CAS Pioneer Hundred Talents Program. BAB acknowledges support from the Department of Energy de-sc0016239, Simons Investigator Award, the Packard Foundation, the Schmidt Fund for Innovative Research, NSF EAGER grant DMR-1643312, ONR - N00014-14-1-0330, ARO MURI W911NF-12-1-0461, and NSF-MRSEC DMR-1420541.
\end{acknowledgments}

\begin{widetext}
\appendix

\section*{Appendix}

The appendices are organized as follows:\\

\noindent (\ref{sec:general}) We briefly review symmetries in the tight-binding method and establish notation that would be used throughout the appendix.\\

\noindent (\ref{app:z4}) We show the equivalence between the $\Z_4$ invariant defined by Shiozaki-Sato-Gomi,\cite{Nonsymm_Shiozaki} and the Zak-phase expression in \q{wilsonex}.\\

\noindent (\ref{app:materializations}) We detail the space groups and elementary band representations of Ba$_2$Pb, stressed Na$_3$Bi, and KHgSb,
%add 
so as to provide a complementary perspective on their topological nontriviality.\\

\noindent (\ref{app:projectiveglide}) We introduce two symmetry classes of solids with glide symmetry; the two classes are distinguished by the representation of glide symmetry in the 3D Brillouin zone (BZ). In one of the two classes, the weak $\Z_2$ invariant is trivial, and a non-primitive unit cell must be chosen to compute the strong $\Z_4$ invariant. \\

\noindent(\ref{sec:ambiguity}) We show if and how the topological invariants defined in the main text depend on the choice of coordinate system.\\

%add
\noindent(\ref{app:lightsource}) We discuss properties of the photoemission light source that allow us to utilize the  selection rule (derived in \s{sec:pes}).

\section{Review of symmetries in the tight-binding method} \label{sec:general}

%We will borrow certain results which are more pedagogically derived in Appendix A of \ocite{Cohomological}; \s{sec:symm} clarifies how glide and time-reversal symmetries are represented in glide planes. 

\subsection{Review of the tight-binding method}

In the tight-binding method, the Hilbert space is reduced to a finite number of atomic \low orbitals $\varphi_{\boldsymbol{R},\alpha}$, for each unit cell labelled by the Bravais lattice (BL) vector $\boldsymbol{R}$.\cite{slater1954,goringe1997,lowdin1950} In Hamiltonians with discrete translational symmetry, our basis vectors are
\begin{align} \label{basisvec}
\phi_{\boldsymbol{k}, \alpha}(\boldsymbol{r}) = \tfrac{1}{\sqrt{N}} \sum_{\boldsymbol{R}} e^{i\boldsymbol{k} \cdot (\boldsymbol{R}+\boldsymbol{r_{\alpha}})} \pdg{\varphi}_{\boldsymbol{R},\alpha}(\boldsymbol{r}-\boldsymbol{R}-\boldsymbol{r_{\alpha}}),
\end{align}
where $\alpha=1,\ldots, n_{tot}$, $\boldsymbol{k}$ is a crystal momentum, $N$ is the number of unit cells, $\alpha$ labels the \low orbital, and $\boldsymbol{r_{\alpha}}$ is the continuum spatial coordinate of the orbital $\alpha$ as measured from the origin in each unit cell. The tight-binding Hamiltonian is defined as 
\begin{align}
H(\boldsymbol{k})_{\alpha \beta} = \int d^dr\,\phi_{\boldsymbol{k},\alpha}(\boldsymbol{r})^* \,\hat{H} \,\phi_{\boldsymbol{k},\beta}(\boldsymbol{r}), 
\end{align}
where $\hat{H}$ is the single-particle Hamiltonian; $\hat{H}$ is a sum of the kinetic term, a scalar, $\br$-periodic potential (which accounts for the ionic lattice and a mean-field approximation of electron-electron interactions), as well as the spin-orbit interaction. The energy eigenstates are labelled by a band index $n$, and defined as $\psi_{n,\boldsymbol{k}}(\boldsymbol{r}) = \sum_{\alpha=1}^{n_{tot}}  \,u_{n,\boldsymbol{k}}(\alpha)\,\phi_{\boldsymbol{k}, \alpha}(\boldsymbol{r})$, where  
\begin{align}
\sum_{\beta=1}^{n_{tot}} H(\boldsymbol{k})_{\ab} \,u_{n,\boldsymbol{k}}(\beta)  = \varepsilon_{n,\boldsymbol{k}}\,u_{n,\boldsymbol{k}}(\alpha).
\end{align}
We employ the braket notation and rewrite the above equation as
\begin{align} \label{energyeigen}
H(\boldsymbol{k})\,\ket{u_{n,\boldsymbol{k}}} = \varepsilon_{n,\boldsymbol{k}}\,\ket{u_{n,\boldsymbol{k}}}.
\end{align}
Due to the spatial embedding of the orbitals, the basis vectors $\phi_{\boldsymbol{k},\alpha}$ are generally not periodic under $\boldsymbol{k} \rightarrow \boldsymbol{k}+\boldsymbol{G}$ for a reciprocal vector $\boldsymbol{G}$; indeed, by substituting $\bk$ with $\bk+\bG$ in \q{basisvec}, each summand acquires a phase factor $e^{i\bG\cdot \br_{\alpha}}$ which is generally not unity. This implies that the tight-binding Hamiltonian satisfies a condition we shall refer to as `Bloch-periodic':
\begin{align} \label{aperiodic}
H(\boldsymbol{k}+\boldsymbol{G}) = V(\boldsymbol{G})^{\mo}\,H(\boldsymbol{k})\,V(\boldsymbol{G}),
\end{align}
where $V(\boldsymbol{G})$ is a unitary matrix with elements: $[V(\boldsymbol{G})]_{\ab} = \delta_{\ab}\,e^{i\boldsymbol{G}\cdot \boldsymbol{r_{\alpha}}}$. Throughout this appendix, we shall describe any matrix-valued function of $\bk$ as `Bloch-periodic' if $f(\bk+\bG)=V(\bG)^{-1}f(\bk)V(\bG)$.\\ 

In the context of insulators, we are interested in Hamiltonians with a spectral gap that is finite for all $\bk$, such that we can distinguish occupied from empty bands;  the former are projected by
\begin{align} \label{periodP}
P(&\bk) = \sum_{n =1}^{\noc} \ket{u_{n,\bk}}\bra{u_{n,\bk}} \lin
\eq V(\bG)\,P(\bk+\bG)\,V(\bG)^{\mo},
\end{align}
where the last equality follows directly from Eq.\ (\ref{aperiodic}).

\subsection{Symmetries in glide-invariant planes} \la{sec:symm}

Consider a time-reversal-invariant insulator that is symmetric under the glide $g_x$, which is a composition of a reflection (in the x coordinate) and a translation by half a Bravais lattice vector in the y direction.  We explain in this section  how  time-reversal and glide symmetries constrain the projection $P(\bk)$ to filled bands, with $\bk$ lying in a glide plane; the restriction of $\bk$ to the plane will be denoted $\bk^r:=(k_y,k_z)$.  In this section  (and for the formulation of the topological invariants $\chi^{\pm}$), we shall concern ourselves only with   glide planes wherein  each wavevector is mapped to itself under glide; these glide planes are labelled \emph{ordinary}. 
%add
For example, any glide plane that includes the Brillouin-zone center is always ordinary;
non-ordinary glide planes only occur away from the zone center, and only for certain space groups, as elaborated in \app{app:projectiveglide}.\\

%In all space groups with glide symmetry, there are two inequivalent and parallel glide planes, one of which includes the Brillouin-zone center. A unit cell may always be chosen such that both glide planes are ordinary, though in certain space groups this unit cell is necessarily not the primitive choice. 

%the glide plane through the Brillouin center is ordinary. In certain space groups, for a Brillouin zone that is dual to a primitive, real-space unit cell, the glide plane off the Brillouin center is projective, i.e., not ordinary; however, a non-primitive unit cell can always be chosen so that both glide planes are ordinary. This section applies to all cases, just described, in which the glide plane is ordinary.\\

Let us parametrize the ordinary glide plane by $\bk^r:=(k_y,k_z)$, which we define to lie in the first Brillouin zone (BZ). Assuming that $\bG_y=2\pi \vec{\by}/R_2$ is a reciprocal vector, $k_y\in [-\pi,\pi]$ in units where $R_2=1$. $\hat{T}$ is defined as the antiunitary representation of time reversal in this plane, and $\hatbmx(k_y)$ as the unitary, wavevector-dependent representation of $g_x$; $\hatbmx(k_y)$ is the product of $\exp{(-ik_y/2)}$ and a momentum-independent matrix $U_{g_x}$ which commutes with $\hat{T}$, as shown in Appendix A1 of Ref.\ \onlinecite{Cohomological}. It follows that 
\e{ \hat{T}\hatbmx(k_y) = \hatbmx(-k_y)\hat{T}, \la{glidetrs}}
which we will shortly find to be useful. $P(\bk^r)$, as defined in \q{periodP}, projects to a $\noc$-dimensional vector space, with $\noc$ a multiple of four owing to glide and time-reversal symmetries, as proven in Appendix C of Ref.\ \onlinecite{Cohomological}. This vector space splits into two subspaces of equal dimension, which transform in the two representations of glide: $\Delta_{\pm}(k_y)$. That is, $\noc/2$ number of vectors in the $\Delta_+(k_y)$ representation have the glide eigenvalue $+i \exp[-ik_y/2]$ under the operation $\hatbmx(k_y)$; the other $\noc/2$ vectors have glide eigenvalue $-i \exp[-ik_y/2]$. The glide symmetry constrains the projection as
\e{ \hatbmx(k_y) P(\bk^r)\hatbmx(k_y)^{-1}=P(\bk^r),}
and time-reversal symmetry constrains as
\e{&\hat{T}P(\bk^r)\hat{T}^{-1} = P(-\bk^r) = V(\pm \bG_y) P(\pm\bG_y{-}\bk^r) V(\mp \bG_y), \la{trs0} \\
&\Rightarrow  \hat{T}_{\pm} P(\bk^r) \hat{T}_{\pm}^{-1} = P(\pm \bG_y-\bk^r),\;\text{with}\; \hat{T}_{\pm} \equiv V(\mp\bG_y)\hat{T} \iand \hat{T}^{\mo}_{\pm} = V(\mp\bG_y)\hat{T}^{\mo}.\la{trs}}
We have applied \q{periodP} in the second equality of \q{trs0}. From \q{glidetrs} and \q{trs0}, we deduce that time-reversed partner states at  $\pm \bk^r$ belong to orthogonal representations of $\Delta_{\pm}(k_y)$, as illustrated by the double-headed arrow in \fig{fig:bz}(c). Indeed,
\e{  \ins{if} \hatbmx(k_y)\ket{u_{\bk^r}} = \pm i e^{-ik_y/2}\ket{u_{\bk^r}},  \ins{then} \hatbmx(-k_y)\hat{T}\ket{u_{\bk^r}} = \hat{T}\hatbmx(k_y)\ket{u_{\bk^r}}= \mp i e^{ik_y/2}\hat{T}\ket{u_{\bk^r}}.} 
On the other hand, time reversal imposes a different constraint on the glide representations at the $k_y{=}{\pm}\pi$ edges of the glide plane:  $\hat{T}_{\pm}$ maps $\bk^r=(\pm \pi,k_z) \rightarrow (\pm \pi,-k_z)$, and $\hat{T}_{\pm}$-related states belong to the same glide representation, as illustrated by curved arrows in \fig{fig:bz}(c) and double-headed arrows in \fig{fig:bz}(d). This result follows from
\e{\hatbmx(\pm \pi)\hat{T}_{\pm}=\hatbmx(\pm \pi) V(\mp\bG_y)\hat{T}= e^{-i\bG_y \cdot \vec{\by}/2}V(\mp\bG_y)\hatbmx(\pm \pi)\hat{T}= e^{-i\pi}V(\mp\bG_y)\hat{T}\hatbmx(\mp \pi)=\hat{T}_{\pm}\hatbmx(\pm \pi), \la{glidetrans}}
and the reality of the eigenvalues of $\hatbmx(\pm \pi)$. The second equality in \q{glidetrans} follows from Eq.\ (A24) in Ref.\ \onlinecite{Cohomological}. \\

To restate the above result in slightly different words, within an ordinary glide plane, any time-reversed partner states which lie 
%remove: at distinct $k_y$  
%add
at $k_y$ and $-k_y$ 
belong in opposite glide representations;
%  add
this statement applies to $k_y{=}0$.
%remove: However, time-reversed partner states with the same $k_y$ (which occurs only for $k_y=\pm \pi$)  
%add
In comparison, time-reversed states with equal wavenumber ($k_y{=}\pi$) belong in the same glide representation; note at $k_y=\pi$ that the glide eigenvalue is real. This will be helpful in formulating the $\Z_4$ invariant in \s{app:z4}.

%To recapitulate, we have found the antiunitary representation ($\hat{T}_{\pm}$) of time reversal on the glide-invariant line on the BZ edge ($k_y=\pi$). Time-reversed partner states in the BZ edge belong in the same glide representation, while time-reversed partner states in the This will be helpful in formulating the $\Z_4$ invariant in \s{app:z4}.    

\section{ Zak-phase expression of strong $\Z_4$ invariant}\la{app:z4}

We show the equivalence between the $\Z_4$ invariant defined by Shiozaki et. al.,\cite{Nonsymm_Shiozaki} and the Zak-phase expression \q{wilsonex}.\\ 

Consider the bent quasimomentum region ($abc$) drawn in \fig{fig:wilson_insulators}(a), which is the union of three faces $a$ (red), $b$ (green) and $c$ (orange): $a$ and $c$ are each half of a glide plane, and $b$ is a half-plane orthogonal to both $a$ and $c$; due to the periodicity of the Brillouin torus, $abc$ has the topology of an open cylinder and is parametrized by orthogonal coordinates $\bk=(t,k_z)$, with $t \in [0,3]$ and $k_z \in [0,2\pi)$; $k_z=0$ is identified with $k_z=2\pi$. We define $\call(t)$ as constant-$t$ circles in $abc$, as illustrated by oriented dashed lines in 
%remove Fig. 3(b)
%add
\fig{fig:bz}(b); 
the sign of $\pm \call(t)$ indicates its orientation, and $abc$ is bounded by $\call(0) -\call(3)$. \\ 

In the half-plane $b$ [$t\in [1,2]$, corresponding to $k_x$ varying in the interval $(-\pi,0)$], we define the connection and curvature as
\e{ \text{for}\;\;\bk =(t,k_z) \in b,\;\; &\text{Tr}[\bA(\boldsymbol{k})] = \sum_{i=1}^{\noc} \langle {u_{i,\boldsymbol{k}}} |{\nabla_{\boldsymbol{k}}u_{i,\boldsymbol{k}}} \rangle, \lin
& F(\bk) \equiv \partial_{t}\text{Tr}[\bA_z(\bk)] - \partial_{z}\text{Tr}[\bA_t(\bk)].}
Here, $\bA = (A_t,A_z)$ with $A_t =\sum_i \langle {u_{i,\boldsymbol{k}}} |\partial_t u_{i,\boldsymbol{k}} \rangle$ and $A_z =\sum_i \langle {u_{i,\boldsymbol{k}}} |\partial_{k_z} u_{i,\boldsymbol{k}} \rangle$. Choosing wavefunctions in $a$ and $c$ to be eigenstates of the glide operation, they divide into two equally-numbered sets according to their glide eigenvalues, which fall into either branch of $\Delta_{\eta}(k_y)=\eta \,i\,$exp$(-ik_y/2)$, with $\eta = {\pm} 1$. We distinguish between these two sets by modifying  our wavefunction labels to $\{u_{n,\bk}^{\eta}|n{=}1,\ldots,\noc/2\}$. We then define the glide-projected, Berry connection, and its corresponding curvature as
\e{ \text{for}\;\;\bk  \in a \cup c,\;\; &\text{Tr}[\bA^{\eta}(\boldsymbol{k})] = \sum_{i=1}^{\noc/2} \langle {u^{\eta}_{i,\boldsymbol{k}}} |{\nabla_{\boldsymbol{k}}u^{\eta}_{i,\boldsymbol{k}}} \rangle, \lin
& F^{\eta}(\bk) \equiv \partial_{t}\text{Tr}[A^{\eta}_z(\bk)] - \partial_{z}\text{Tr}[A^{\eta}_t(\bk)].} 
Shiozaki et. al. defined a $\Z_4$ invariant by
\e{ \chi^{\eta} = 2 \calp^{\eta}(0) - 2\calp^{\eta}(3) + \f{i}{2\pi}\left[ 2\int_a F^{\eta} d^2\bk  + 2 \int_c F^{\eta}d^2\bk + \int_b F\,d^2\bk \right], \la{ssg}}
with the polarization (in the $\Delta_{\eta}$ representation) defined as
\e{ \calp^{\eta}(t) \equiv \f{i}{2\pi} \int_{\call(t)}\text{Tr}[\bA^{\eta}(\bk)] \cdot d\bk.\la{polarizationeta}}
Included in Shiozaki's definition is a gauge constraint for the wavefunctions on the boundary ($\call(0) -\call(3)$).   

Before defining this constraint in complete generality, let us develop some intuition by considering a specific realization. For noncentrosymmetric space groups, energy bands in each glide subspace are two-fold connected along $\call(3)$ [also true for $\call(0)$], due to the Kramers-degenerate points at  $k_z=0$ and $\pi$, as illustrated in \fig{fig:bz}(d); note here that the glide eigenvalue is real, hence time-reversal related states belong to the same glide representation.  For each pair of energy bands (within one glide subspace), one energy band may be denoted $u^{\sma{\eta}}_{\sma{\alpha,\bk}}$  and the other $u^{\sma{\eta}}_{\sma{\bar{\alpha},\bk}}$, as illustrated in \fig{fig:bz}(d). As is well known, any energy eigenfunction of a Hamiltonian is only well-defined up to a phase (which here can depend on $\bk$). Here, it is possible to choose this phase (or gauge) such that 
\e{\text{for}\;\;t  \in \{0,3\},\;\;& \ket{u^{\eta}_{\alpha,t,-k_z}} = T_t\ket{u^{\eta}_{\bar{\alpha},t,k_z}},\lin
 &\ket{u^{\eta}_{\bar{\alpha},t,-k_z}} = -T_t\ket{u^{\eta}_{\alpha,t,k_z}}, \la{trsgauge}}
with $T_{0}$ [$T_3$] the antiunitary representation of time reversal at the time-reversal invariant line $\call(0)$ [$\call(3)$]. We have shown in \s{sec:symm} that time-reversed partner states at $k_y=\pi$ belong in the same glide representation (here the glide eigenvalues is real); we may directly identify $T_3=\hat{T}_+$ in \q{trs}. By imposing \q{trsgauge} on the wavefunction, the invariant defined by Shiozaki becomes well-defined modulo four.\cite{Nonsymm_Shiozaki} More generally, \q{ssg} is well-defined with the following gauge constraint: decompose each   glide subspace  (within the filled-band subspace) into $\noc/4$ pairs of bands [labelled by $\{u^{\sma{\eta}}_{\sma{\alpha,\bk}},u^{\sma{\eta}}_{\sma{\bar{\alpha},\bk}} | \bk \in \call(0)-\call(3),\;\alpha=1,\ldots, \noc/4\}$ ], such that each of $u^{\sma{\eta}}_{\sma{\alpha,\bk}}$ and $u^{\sma{\eta}}_{\sma{\bar{\alpha},\bk}}$ is first-order differentiable in $k_z$, and together satisfy   \q{trsgauge}.

Calculating the $\Z_4$ invariant through \q{ssg} requires that we find glide-projected wavefunctions that are both first-order differentiable along
%add
the boundary of abc
($\call(0){-}\call(3)$) and constrained as in \q{trsgauge}. In the rest of this section, we reformulate \q{ssg} as an index [\q{wilsonex}] that is extractable from the  Wilson loop of the Berry gauge field, whose basic properties we review in \s{app:revwilson}. One advantage of a Zak-phase calculation is that it may be done without fixing a gauge; in comparison, the necessity of imposing a differentiable gauge [satisfying  \q{trsgauge}] makes \q{ssg} difficult to compute in practice. Our reformulation is a generalization of Ref.\ [\onlinecite{yu2011,alexey2011}] for insulators with glide symmetry. To organize this appendix section, we divide $\chi^{\eta}$ into two additive contributions: $\chi^{\eta}_{ac}$ from the glide-invariant faces $a$ and $c$, and $\chi_b$ from the 
%add
glide-variant
face $b$.
\e{ \chi^{\eta} \eq \chi_b+\chi^{\eta}_{ac} \lin
 \chi_b \eq \f{i}{2\pi}\int_b F\,d^2\bk  \\
\chi^{\eta}_{ac} \eq 2 \big[\,\calp^{\eta}(0) - \calp^{\eta}(3)\,\big] + \f{i}{\pi}\left[ \int_a F^{\eta} d^2\bk  +  \int_c F^{\eta}d^2\bk \right]. \la{ssg2}}
We tackle $\chi_b$ in \s{app:wilsoncurv}, and $\chi^{\eta}_{ac}$ in \s{app:chiac}.

%\subsection{Mod-$4$ invariance of $\chi^{\eta}$}
%In the first step, we apply the boundary constraint to express \q{ssg} as
%\e{ -i2\pi \chi^{\eta} = 4 \int_{l(0)} A^{\eta}_I - 4\int_{\call(3)} A^{\eta}_I + 2 \int_a F^{\eta}  + 2 \int_c F^{\eta} + \int_b F, \la{ssg}}
%where $\int A$ is shorthand for a line integral and $\int F$ for an area integral;  

\subsection{Review of Wilson loops}\la{app:revwilson}

\subsubsection{Basic definition}

We consider the parallel transport of occupied Bloch waves around a momentum loop $\call$, where at each $\bk  \in  \call$ a spectral gap separates a set of lower-energy, occupied states (numbering $\noc$) from a higher-energy, unoccupied subspace. The $\noc$-by-$\noc$ matrix representing such parallel transport is known as the Wilson loop,\cite{wilczek1984} and it may be expressed as the path-ordered exponential (denoted by $\overline{\text{exp}}$) of the Berry-Wilczek-Zee non-Abelian connection:\cite{wilczek1984,berry1984} 
\begin{align} \la{wilson}
\W[\call] \equiv \overline{\text{exp}}\,\left[{-\int_{\call}} d\bk \cdot \boldsymbol{A}(\boldsymbol{k})\,\right], \as{\bA}_{ij}(\boldsymbol{k}) =\langle {u_{i,\boldsymbol{k}}} |{\nabla_{\boldsymbol{k}}u_{j,\boldsymbol{k}}} \rangle,\;\; i,j=1,\ldots,\noc,
\end{align}
where $|{u_{\sma{j,\bk}}}\rangle$ belongs to the filled-band subspace of the tight-binding Hamiltonian in \q{energyeigen}; in keeping with the generality of this review, we will not adopt the specific gauge choice in \q{trsgauge}.  Henceforth, we consider only noncontractible loops  within the 2D subregion $abc$ [cf.\ \fig{fig:bz}(a)] parametrized by $t\in [0,3]$ and $k_z\in[-\pi,\pi]$; each loop  is oriented parallel to $\vec{\bz}$ and lies at fixed $t$, as illustrated by the triple-headed arrow in \fig{fig:bz}(a); we streamline our notation from $\W[\call(t)]$ to
\begin{align} \la{wilsonsim}
\W(t) \equiv \overline{\text{exp}}\,\left[{-\int_{\call(t)}} dk_z \,{A_z}(t,k_z)\,\right].
\end{align}
To calculate $\W(t)$ from this expression, it is implicit from the definition of $A_z$ that a first-order differentiable basis for $u_{i,\bk}$ is needed. Moreover, to uniquely define the eigenspectrum of $\W(t)$, we insist that this basis further satisfies the condition:\cite{AA2014} 
\e{ \forall \;i \in \{1,\ldots,\noc\},\as \ket{u_{i,\bk+\bG_z}} = V(-\bG_z)\ket{u_{i,\bk}}. \la{semiperiodic}}
That such a basis can be found follows from the Bloch-periodicity of the Bloch Hamiltonian in \q{aperiodic}, and so we shall refer to \q{semiperiodic} as the Bloch-periodic gauge. 

\subsubsection{The gauge-independent Wilson loop}

It is advantageous to equivalently formulate the $\W$-eigenvalues as the unimodular eigenvalues of a related operator that is gauge-independent. Following our treatment in Ref.\ [\onlinecite{AA2014}] with slightly different notation, we define an operator that effects parallel transport in the L$\ddot{\text{o}}$wdin-orbital basis as
\e{\hat{W}_{k_{2},k_{1}}(t) =  \prod_{k_z}^{k_2 \leftarrow k_1} P(t,k_z). \la{line}}
Here, we uniquely specify the path from $(t,k_1){\rightarrow}(t,k_2)$ by choosing $k_z{\in}[k_2,k_1]$ to always lie in $[-\pi,\pi]$;  the right-hand-side of \q{line} indicates a path-ordered product of projections [defined in \q{periodP}] where $k_z$ assumes any discrete value $2\pi m/ N_z$ between $k_1$ and $k_2$, for integral $m$.  $\hat{W}_{\sma{k_{2},k_{1}}}(t)$ defines a map from  $\calb(t,k_1)$ to $\calb(t,k_2)$, where $\calb(\bk)$ is the $\noc$-dimensional vector space spanned by the filled bands ($\{u_{j,\bk}\}$) at $\bk$. The Bloch-periodicity of the Bloch Hamiltonian [\q{aperiodic}] implies that $V(\bG_z)$ (with $\bG_z{\equiv}2\pi \vec{\bz}$) is a map from $\calb(t,k_z)$ to $\calb(t,k_z{-}2\pi)$, and therefore the composition of $V(\bG_z)$ and $\hat{W}$ (defined with a curly $\hat{\W}$, which is to be distinguished from $\hat{W}$) is a map:
\e{ \hat{\W}(t) \equiv &V(\bG_z)\hat{W}_{\pi,-\pi}(t) : \;\calb(t,-\pi) \rightarrow \calb(t,-\pi).\la{gaugeindwilson}}
In the limit $N_z{\rightarrow}\infty$, $k_z$ becomes a continuous variable, and we may identify $\W$ in \q{wilson} as a matrix representation of curly $\hat{\W}$ in a basis of $\calb(t,{-}\pi)$ (the filled-band subspace at the base point  of the loop):\cite{AA2014}
\begin{align} \label{con3}
[\W(t)]_{ij} = \bra{u_{i,(t,-\pi)}}\,\hat{\W}(t)\,\ket{u_{j,(t,-\pi)}}.
\end{align}
Here, $i=1,\ldots,\noc$ labels the basis vector, and need not label an energy band.
We therefore refer to curly $\hat{\W}$ as the gauge-independent Wilson loop. The full eigenspectrum of $\W$ comprises the unimodular eigenvalues of $\hat{\W}$, which we label by exp$[i\theta_{\sma{n},t}]$ with $n{=}1,\ldots, \noc$. The form of $\W$ in \q{con3} manifests the gauge-invariance of its eigenspectrum, since if
\begin{align}
\ket{u_{j,(t,-\pi)}} &\rightarrow \sum_{i=1}^{\noc} \ket{u_{i,(t,-\pi)}} S_{ij}, \;\text{with}\; S \in U(\noc),  \lin
&\text{then}\;\;\W \rightarrow \dg{S}\W S.
\end{align}
We remark that the $\W$-eigenvalues are also independent of the base point of the loop;\cite{AA2014} our choice of $(t,-\pi)$ as the base point merely renders certain symmetries transparent, as will be made evident in \app{app:chiac}.  

\subsubsection{Relation of the Wilson loop to polarization}

%Implicit in this equality is a specific wavefunction basis which selects a unique branch for the multivalued logarithm.

It is useful to relate the Wilson loop to the polarization,\cite{kingsmith1993} defined as the line integral of the $U(1)$ Berry connection:
\e{ \calp(t) := \f{i}{2\pi} \int_{\call(t)}\text{Tr}[\bA(\bk)] \cdot d\bk.\la{polarization}}
We caution that $\calp$ is the expectation value of a discrete position operator (taking only discrete values corresponding to the centers of localized, tight-binding basis vectors),\cite{ZhoushenHofstadter,AA2014} rather than that of the usual continuum position operator.\cite{kingsmith1993} Implicit in the definition of the Wilson loop [\q{wilson}] is that wavefunctions are first-order differentiable in $k_z$ and Bloch-periodic in $\bG_z$ -- this would also imply that the polarization quantity in \q{polarization} is well-defined. The polarization is related\cite{AA2014}  to the $U(\noc)$ Wilson loop through:
\e{\calp(t) \equiv -\f{i}{2\pi}\text{ln}\,\text{det}\big[\,\W(t)\,\big].  \la{polwilson}}
Throughout this section, $\equiv$ denotes an equivalence up to addition or subtraction of an integer. As with all polarization quantities, this integer ambiguity\cite{kingsmith1993}  reflects the discrete translational symmetry in $\vec{\bz}$.  Defining $\{\text{exp}[i\theta_j(t)]|\}_{j=1}^{\noc}$ as the eigenvalues of $\W(t)$, \q{polwilson} is expressible as
\e{\calp(t) \equiv \f{1}{2\pi} \sum_{j=1}^{\noc}\theta_{j}(t). \la{polwilson2}}

%where each of $\theta_j/2\pi$ is the polarization of $\tilde{u}_{j,\bk}$. That is, the polarization defined by

To prove the equivalence of $\Z_4$ invariants, it is useful (as an intermediate step) to work in a special basis (denoted $\{\tilde{u}_{j,\bk}\}_{j=1}^{\noc}$) of the filled-band subspace spanned by $\{ {u}_{j,\bk}\}_{j=1}^{\noc}$. The new basis is defined to 
satisfy two (related) propertries: (i) 
for each $j$,
\e{ \tilde{\calp}_j(t) := \f{i}{2\pi} \int_{\call(t)}\braket{\tilde{u}_{j,\boldsymbol{k}}}{\nabla_{\boldsymbol{k}}\tilde{u}_{j,\boldsymbol{k}}} \cdot d\bk \equiv \f{\theta_j(t)}{2\pi}. \la{polMLWF}}
%remove the tight-binding wavefunction $\tilde{\psi}_{j,\bk}(\alpha):=e^{i\bk\cdot (\bR+\br_{\alpha})}\tilde{u}_{j,\bk}$ is as smooth as possible (with respect to $k_z$) while maintaining periodicity in  $k_z \rightarrow k_z+2\pi$; 
(ii) The Fourier transform of $\tilde{\psi}_{j,\bk}(\alpha):=e^{i\bk\cdot (\bR+\br_{\alpha})}\tilde{u}_{j,\bk}$ with respect to $k_z$ is a hybrid Wannier function\cite{AA2014,Maryam2014}that is an eigenstate of the  $z$ position operator projected to the filled-band subspace; such eigenstates are always maximally-localized\cite{marzari1997} in the $z$ direction. We refer to $\{\tilde{u}_{j,\bk}\}_{j=1}^{\noc}$ as the \emph{maximally-localized basis/gauge}. 
Due to their nice localization properties in real space, the maximally-localized basis has found applications in several contexts;\cite{AA2014,alexey_smoothgauge,benalcazar_quantizedmultipole} we briefly review how this basis is constructed.\\

%equivalently stated, each of $\{\theta_j|j =1,\ldots, \noc \}$ corresponds to a specific branch choice in $\{\theta_j{+}2\pi a |a {\in}\Z\}$. An absolute determination of each branch is not necessary for the eventual calculation of the $\Z_4$ invariant, as we will demonstrate. \\

% We caution that $\{\tilde{u}_n\}$ do not generically correspond to an energy eigenstate (denoted previously by  $\{u_n\}$ without the tilde) of the Bloch Hamiltonian in \q{energyeigen}; however, $\{\tilde{u}_n\}$ and $\{u_n\}$ span the same vector space at each $\bk$.

\noindent \emph{Review} To construct this special basis, we first diagonalize the gauge-independent Wilson loop [\q{gaugeindwilson}] at the base point ($k_z=-\pi$) as 
\e{ \hat{\W}(t)\ket{\tilde{u}_{n,(t,-\pi)}} = e^{i\theta_{n,t}}\ket{\tilde{u}_{n,(t,-\pi)}}.\la{diagw}}  
We remind the reader that $\hat{W}$ is an $n_{tot}\times n_{tot}$ matrix operator with only $\noc$ unimodular eigenvalues (the rest being zero).   Basis vectors away from the base point are then constructed by parallel transport, composed with a multiplicative phase factor:\cite{AA2014,alexey2011,ZhoushenHofstadter} 
\e{ \ket{\tilde{u}_{n,(t,k_z)}} = e^{-i(k_z+\pi)\theta_n/2\pi}\hat{W}_{k_z,-\pi}(t)\ket{ \tilde{u}_{n,(t,-\pi)}}. \la{pttwist}}
% add
Note that $\tilde{u}_{n,(t,k_z)}$ diagonalizes the gauge-independent Wilson loop with base point $k_z$. Owing in part to
the phase factor 
%remove ensures the Bloch-periodicity condition:
%add
in  \q{pttwist}, $\tilde{u}_{n,(t,k_z)}$ satisfies the Bloch-periodicity condition:
\e{ \ket{\tilde{u}_{n,(t,\pi)}} =e^{-i\theta_n} V(-\bG_z)V(\bG_z) \hat{W}_{\pi,-\pi}(t)\ket{\tilde{u}_{n,(t,-\pi)}} =e^{-i\theta_n} V(-\bG_z) \hat{\W}(t)\ket{\tilde{u}_{n,(t,-\pi)}} = V(-\bG_z)\ket{\tilde{u}_{n,(t,-\pi)}}; \la{semip}}
in the last equality, we utilized that $\tilde{u}$ is an eigenstate of the gauge-independent Wilson loop [cf.\ \q{diagw}].
%remove while minimally deviating from the parallel-transport basis. 
We remark that the  Berry connection evaluated with  $\tilde{u}_{n,(t,k_z)}$ 
%add
equals 
\e{  \braket{\tilde{u}_{m,(t,k_z)}}{ \p{\tilde{u}_{n,(t,k_z)}}{k_z}}=-i\delta_{mn} \f{\theta_m}{2\pi},} 
%add 
which 
generically does not  vanish. It is instructive to demonstrate that these basis functions are orthonormal away from the base point, assuming such is true for the base point. Dropping the constant label $t$ in this demonstration,
\e{\braket{\tilde{u}_{m,k_z}}{\tilde{u}_{n,k_z}}\eq \braopket{\tilde{u}_{m,-\pi}}{\hat{W}_{-\pi,k_z}\hat{W}_{k_z,-\pi}}{ \tilde{u}_{n,-\pi}} \lin
\eq \braket{\tilde{u}_{m,-\pi}}{ \tilde{u}_{n,-\pi}} = \delta_{m,n}.\la{demons}}    
In the second equality, we applied that parallel transport within the valence bands is unitary, and therefore $\hat{W}_{-\pi,k_z}\hat{W}_{k_z,-\pi}$ acts on any state in $\calb({-}\pi)$ as the identity operator.

\subsubsection{Relation of the Wilson loop to the integral of the curvature} \la{app:wilsoncurv}

%with branches chosen for $\calp(t_2)$ and $\calp(t_1)$, such that the wavefunction is smooth in $\bk$ within the area in question.

Let us consider the area integral of the Berry curvature over faces $a$, $b$ or $c$; any of these faces is parametrized by $k_z \in [0,2\pi)$ and $t\in [t_1,t_2]$ with $t_2>t_1$. We can always choose the wavefunction (in a face) to be smooth with respect to $t$ and $k_z$.\cite{nogo_AAJH}  We may then utilize Stoke's theorem to convert the area integral to a line integral of the Berry connection over the face's boundary; in the Bloch-periodic gauge of \q{semiperiodic}, the line integral over the line segments orthogonal to $\vec{\bz}$ cancel, and what remains is:    
\e{ \f{i}{2\pi}\int F(\bk)d^2k = \calp(t_2)-\calp(t_1).\la{curv}} 
 We will find it useful to evaluate the area integral with the maximally-localized basis defined in \q{diagw}, (\ref{pttwist}) and (\ref{semip}); then applying \q{polwilson2} to \q{curv}, we obtain
\e{ \f{i}{2\pi}\int F(\bk)d^2k = \f{1}{2\pi} \sum_{j=1}^{\noc}\left[\theta_{j}(t_2)-\theta_j(t_1)\right].\la{curv}} 
By our assumption that basis vectors are smooth in $t$, we must choose a branch for $\theta_j(t)$ that is differentiable in $t$ for $t\in[t_1,t_2]$, and therefore,
\e{ \int F(\bk)d^2k = i\sum_{j=1}^{\noc}\int_{t_2}^{t_1} \f{d\theta_{j}}{dt}dt :=i\sum_{j=1}^{\noc}\int_{t_2}^{t_1} d\theta_{j}.\la{curv2}} 
An immediate implication is that
\e{ \chi_b = \f{i}{2\pi}\int_b F(\bk)d^2k=\f1{2\pi}\sum_{j=1}^{\noc} \int_1^2 d\theta_j. \la{wilsonchib}}

\subsection{Expressing $\chi^{\eta}_{ac}$ with the Wilson loop}\la{app:chiac}

In this subsection, we restrict our discussion to the glide-invariant half-planes $a$ and $c$, as illustrated in \fig{fig:wilson_insulators}(a-b). The component of $\chi^{\eta}$ [recall \q{ssg}] contributed by $a$ and $c$ has been defined as $\chi^{\eta}_{ac}$ in \q{ssg2}. It is known from \ocite{Nonsymm_Shiozaki} that $\chi^{\eta}_{ac}$ is well-defined modulo $4$, if we insist, at $\bar{t} \in \{0,3\}$, that the wavefunctions satisfy the time-reversal constraint in \q{trsgauge}. \\

The goal of this section is to express $\chi^{\eta}_{ac}$ [as defined in \q{ssg2}]  equivalently as
\e{ \chi^{\eta}_{ac} =  \f1{\pi} \sum_{j=1}^{\noc/2} \left[ \theta_j^{\eta}(0) -\theta_j^{\eta}(3) + \int_0^1  d\theta_j^{\eta} + \int^3_2 d\theta_j^{\eta} \right], \la{chiac}}
where $\theta_j^{\eta}(t)$ is the phase of the $j$'th eigenvalue of the Wilson loop [$\W^{\eta}(t)$] projected to the $\Delta_{\eta}$ glide representation. To clarify, if we begin at the base point of $\call(t)$ [$t\in [0,1]$ or $[2,3]$] with a Bloch state in the $\Delta_{\eta}(k_y)$ representation, such a Bloch state remains in the  $\Delta_{\eta}(k_y)$ representation as it is parallel-transported in the z direction.\cite{TBO_JHAA} Consequently, the $\noc {\times} \noc$ Wilson loop diagonalizes into two $(\noc/2){\times}(\noc/2)$ blocks, which we define as $\W^{\eta}(t)$; the superscript $\eta{\in}{\pm}$ distinguishes between the two glide representations. 
For \q{chiac} to be a well-defined modulo four, we impose that $\theta_j^{\eta}$ is first-order differentiable with respect to $t$, and that $\theta_j^{\pm}(t)$ are \emph{pairwise degenerate} at $t{=}0$ and $3$. To clarify `pairwise degeneracy', we mean that for any Zak band with phase $\theta^+_j(0)$, we pick a branch for a distinct Zak band (labelled $j'$) such that $\theta^+_{j'}(0){=}\theta^+_j(0)$ (viewed as a strict equality, not an equivalence modulo $2\pi$), so that ${\sum}_{\sma{j=1}}^{\sma{\noc/2}}\theta_{j}^{+}(0)$ is uniquely defined modulo $4\pi$.\\

To prove the equivalence of \q{ssg2} with \q{chiac}, we adopt the following strategy. Beginning with the filled-band subspace in each glide representation, we pick a  basis that is maximally-localized in the z direction [cf.\ \qq{diagw}{semip}] and simultaneously satisfies the time-reversal-symmetric gauge constraint [\q{trsgauge}]. If such a basis (denoted $\tilde{u}^{\eta}_{\alpha,\bk},\tilde{u}^{\eta}_{\bar{\alpha},\bk}$) can be found, then we may evaluate all terms in \q{chiac} and \q{ssg2} in this special basis and see straightforwardly that they are identical. By `evaluating ... in this special basis', we mean that we can
 express all Zak phases in \q{chiac} as 
\e{\theta_{\alpha}^{\eta}= i\int_{\call}\braket{\tilde{u}^{\eta}_{\alpha,\boldsymbol{k}}}{\nabla_{\boldsymbol{k}}\tilde{u}^{\eta}_{\alpha,\boldsymbol{k}}} \cdot d\bk}
(and an identical expression with $\alpha \rightarrow \bar{\alpha}$);
we can express the quantities occurring in \q{ssg2} as
\e{ \calp^{\eta}(t) = \f{1}{2\pi}\left( \sum_{\alpha=1}^{\noc/2}\theta^{\eta}_{\alpha}(t)+\sum_{\bar{\alpha}=1}^{\noc/2}\theta^{\eta}_{\bar{\alpha}}(t)\right) , \as t=0,3 \lin
\f{i}{2\pi}\int_a F^{\eta}(\bk)d^2k=\f1{2\pi}\left(\sum_{\alpha=1}^{\noc/2} \int_0^1 d\theta^{\eta}_{\alpha}+\sum_{\bar{\alpha}=1}^{\noc/2} \int_0^1 d\theta^{\eta}_{\bar{\alpha}}  \right), \la{speccialbasis}}
(and an identical expression with $a \rightarrow c$ and $\int_0^1\rightarrow \int_2^3$). \q{speccialbasis} follows from our previously-derived 
\q{polwilson2} and \q{curv2}. \\

Let us now prove that, indeed, such a basis can be found. While we have demonstrated how to construct the maximally-localized basis in \qq{diagw}{semip}, we have not shown that the time-reversal constraint can be simultaneously and consistently imposed. 
Specifically, we would show that our maximally-localized basis vectors $\{\tilde{u}^{\sma{\eta}}_{\sma{j,\bk}}\}_{j=1}^{\noc/2}$ can be relabelled as pairs of $\{\tilde{u}^{\sma{\eta}}_{\sma{\alpha,\bk}},\tilde{u}^{\sma{\eta}}_{\sma{\bar{\alpha},\bk}} \}_{\sma{\alpha{=}1}}^{\sma{\noc/4}}$, such that each pair $(\alpha,\bar{\alpha})$ satisfies \q{trsgauge} with $u {\rightarrow}\tilde{u}$.\\

\noindent \emph{Proof:} 
%remove: It is sufficient to consider
%add
Let us focus on the glide- and time-reversal-invariant lines $\call(0)$ and $\call(3)$.  The proof is essentially identical for either line, so let us just focus on $\call(3)$. We begin by defining $\tilde{u}^{\sma{\eta}}_{\sma{n,(t=3,k_z)}}$ as a basis vector in $\calb^{\eta}(3,k_z)$ [the filled-band subspace in the $\eta$ glide representation] satisfying three maximally-localized conditions \q{diagw}-(\ref{semip}). Our proof is eased by equivalently expressing two of these three conditions [\q{diagw} and (\ref{pttwist})] as
\e{&V(\bG_z)\hat{W}_{2\pi,0}(3)\ket{\tilde{u}^{\eta}_{n,(3,0)}} =  e^{i\theta^{\eta}_{n,3}} \ket{\tilde{u}^{\eta}_{n,(3,0)}},\lin
&\ket{\tilde{u}^{\eta}_{n,(3,k_z)}} = e^{-ik_z\theta^{\eta}_n/2\pi}\hat{W}_{k_z,0}(3)\ket{ \tilde{u}^{\eta}_{n,(3,0)}}. \la{pttwist2}} 
In the above equations, we have, for analytic convenience, shifted the base point of the loop from $k_z=-\pi$ to $k_z=0$, and $\hat{W}_{k_2,k_1}$ has been defined in \q{line}; note that $\{e^{i\theta^{\eta}_n}\}$ is invariant under changes of the base point.\cite{AA2014} 
$\tilde{u}^{\eta}_{n,(3,k_z)}$ occurring in the second line of \q{pttwist2} is an eigenstate of the gauge-independent Wilson loop with base point $k_z=0$.
The 
%remove: above eigenvalue equation 
%add
first line of \q{pttwist2}
leads equivalently to the inverse-eigenvalue equation:
\e{\hat{W}_{0,2\pi}(3)V(-\bG_z)\ket{\tilde{u}^{\eta}_{n,(3,0)}} =  e^{-i\theta^{\eta}_{n,3}} \ket{\tilde{u}^{\eta}_{n,(3,0)}},\la{inverseeig}}
which follows from $\hat{W}_{0,2\pi}(3)V(-\bG_z)V(\bG_z)\hat{W}_{2\pi,0}(3)$ acting as the identity map in $\calb(3,0)$, the filled-band subspace.

Following our discussion in \s{sec:symm}, we would generate a basis vector in $\calb^{\eta}(3,-k_z)$ by time-reversing $\tilde{u}^{\sma{\eta}}_{\sma{\alpha,(3,k_z)}}$. The operator representation of time-reversal in $\call(3)$ (where $k_y{=}\pi$) is $\hat{T}_+$, as defined in \q{trs}; we remind the reader that  any 
%add
$\hat{T}_+$-related
pair of Bloch states 
%remove (related by $\hat{T}_+$) 
%add 
(at $k_y{=}\pi$)
belong  in the \emph{same} glide representation $\Delta_{\eta}$. From \q{line} and \q{trs}, we deduce the effect of time-reversing the Wilson-line operators:
\e{ \hat{T}_+\hat{W}_{k_2,k_1}(3)\hat{T}_+^{-1} = \hat{W}_{-k_2,-k_1}(3), \la{trwilson}}
and also the Wilson-loop operator:
\e{\hat{T}_+ V(\bG_z)\hat{W}_{2\pi,0}(3)\hat{T}_+^{-1} \eq V(-\bG_z)\hat{W}_{-2\pi,0}(3) \lin
\eq \hat{W}_{0,2\pi}(3)V(-\bG_z).  \la{trloop}}
To simplify our notation, we henceforth drop the constant labels for the glide index $\eta$ and the quasimomentum parameter $t=3$ [e.g., $\tilde{u}^{\sma{\eta}}_{\sma{\alpha,(3,k_z)}} {\rightarrow} \tilde{u}_{\sma{\alpha,k_z}}$, $\calb^{\sma{\eta}}(3,k_z){\rightarrow}\calb(k_z)$]. Since $\hat{T}_+$ is antiunitary and squares to ${-}1$,  $\hat{T}_+|\tilde{u}^{\sma{\eta}}_{\sma{\alpha,0}}\rangle {\in} \calb(0)$ is orthogonal to $|\tilde{u}_{\sma{\alpha,0}}\rangle$. We would further show that $\hat{T}_+|\tilde{u}_{\sma{\alpha,0}}\rangle$ diagonalizes the gauge-independent Wilson-loop with the same eigenvalue as $|\tilde{u}_{\sma{\alpha,0}}\rangle$:
\e{&V(\bG_z)\hat{W}_{2\pi,0} \hat{T}_+\ket{\tilde{u}_{n,0}} = \hat{T}_+\hat{W}_{0,2\pi}V(-\bG_z)\ket{\tilde{u}_{n,0}} \lin
\eq \hat{T}_+ e^{-i\theta_n}\ket{\tilde{u}_{n,0}} =e^{i\theta_n} \hat{T}_+\ket{\tilde{u}_{n,0}}.\la{trdiag}}
In the second equality, we applied \q{trloop}, and in the third \q{inverseeig}. Applying \q{pttwist2} and (\ref{trwilson}),
\e{ \hat{T}_+\ket{\tilde{u}_{n,k_z}} = e^{ik_z\theta_n/2\pi}\hat{W}_{-k_z,0}\hat{T}_+\ket{ \tilde{u}_{\alpha,0}}. \la{int}}
Thus if we relabel
\e{ & \ket{\tilde{u}_{\alpha,k_z}} := \ket{\tilde{u}_{n,k_z}}, \lin
&\ket{\tilde{u}_{\bar{\alpha},-k_z}} := \hat{T}_+\ket{\tilde{u}_{n,k_z}},\lin
&\,e^{i\theta_{\alpha}}=e^{i\theta_{\bar{\alpha}}} := e^{i\theta_n}, \la{relabel}}
\q{trdiag} and (\ref{int}) may be expressed as two of the three maximally-localized conditions:
\e{ &V(\bG_z)\hat{W}_{2\pi,0} \ket{\tilde{u}_{\bar{\alpha},0}} = e^{i\theta_{\bar{\alpha}}}\ket{\tilde{u}_{\bar{\alpha},0}}, \lin
&\ket{\tilde{u}_{\bar{\alpha},k_z}}  = e^{-ik_z\theta_{\bar{\alpha}}/2\pi}\hat{W}_{k_z,0}\ket{ \tilde{u}_{\bar{\alpha},0}}, } 
and the third condition (Bloch-periodicity) is simple to show. By assumption, $\tilde{u}_{{\alpha},k_z}$ is also maximally-localized. By construction, each pair of $\{ \tilde{u}_{{\alpha},k_z},\tilde{u}_{\bar{\alpha},k_z}\}$ satisfies the time-reversal constraint [\q{trsgauge}].  $\blacksquare$\\

% two conditions (that ensured \q{ssg2} and \q{chiac} are  $\Z_4$ quantities) are in fact equivalent in this special basis. 

It is instructive to compare the respective gauge conditions that have been imposed to ensure that \q{ssg2} and \q{chiac} are well-defined $\Z_4$ quantities. The time-reversal condition of \q{trsgauge} implies 
\e{ i\braket{\tilde{u}_{\alpha,\boldsymbol{k}}}{\nabla_{\boldsymbol{k}}\tilde{u}_{\alpha,\boldsymbol{k}}}=i\braket{\tilde{u}_{\bar{\alpha},\boldsymbol{k}}}{\nabla_{\boldsymbol{k}}\tilde{u}_{\bar{\alpha},\boldsymbol{k}}}\big|_{\bk\rightarrow -\bk},}
which ensures the pairwise-degeneracy condition on \q{chiac}:
\e{ i\int_{\call}\braket{\tilde{u}_{\alpha,\boldsymbol{k}}}{\nabla_{\boldsymbol{k}}\tilde{u}_{\alpha,\boldsymbol{k}}} \cdot d\bk :=  \theta_{\alpha}=\theta_{\bar{\alpha}} := i\int_{\call}\braket{\tilde{u}_{\bar{\alpha},\boldsymbol{k}}}{\nabla_{\boldsymbol{k}}\tilde{u}_{\bar{\alpha},\boldsymbol{k}}} \cdot d\bk. \la{pwdeg}}
The above equality is strict, and is a stronger condition than the equivalence modulo $2\pi$ [which was proven earlier in \q{relabel}]. \\

%We remark that the time-reversal constraint [\q{trsgauge}] manifests as a constraint on the choice of branches for the Wilson-loop phases, i.e., we must always pick branches that are doubly-degenerate at $\bar{t} \in \{0,3\}$: $\theta_{\alpha}(\bar{t})=\theta_{\bar{\alpha}}(\bar{t})$, as proven in \q{relabel}. This ensures that the quantity $\chi^{\eta}_{ac}$, if identified with $\chi^{\eta}_{ac}+4$, is uniquely defined.\\    

Combining the results of this section with \q{wilsonchib}, we finally complete the proof 
%remove the
%add
of equivalence between \q{wilsonex} and \q{ssg}.  Having proven this equivalence in the maximally-localized and time-reversal-symmetric gauge, we emphasize that the computation of the Zak phase factors $\{e^{i\theta_{n}^{\eta}}\}$ is manifestly gauge-invariant; these phase factors are obtained from diagonalizing  the gauge-independent Wilson loop in \q{gaugeindwilson}. 

%, \emph{practically}, we do not need to construct maximally-localized wavefunctions in the computation of $\chi^{\eta}$. Every variable on the right-hand side of \q{wilsonex} can be obtained from diagonalizing the gauge-independent Wilson loop in \q{gaugeindwilson}.

\section{Two symmetry classes of solids with glide symmetry}\la{app:projectiveglide}

We introduce here two symmetry classes (labelled I and II) of solids with glide symmetry. The practical value of distinguishing these classes is that  in class II, the weak $\Z_2$ invariant is always trivial; while the strong $\Z_4$ classification holds for both classes, in class II a non-primitive unit cell must be chosen to compute the strong $\Z_4$ invariant.\\

The  two classes are distinguished by the representation of glide symmetry in the  Brillouin zone (BZ), which is  defined standardly as the Wigner-Seitz cell of the reciprocal lattice. Glide-invariant planes in the BZ are of two types: in an \emph{ordinary glide plane}, each wavevector $\bk$ is mapped to itself by glide. In a \emph{projective glide plane}, each $\bk$ is mapped by glide to a distinct wavevector ($g_x{\circ} \bk$) on said plane, such that $g_x{\circ} \bk$ is translated from $\bk$ by half a reciprocal vector. This is analogous to a nonsymmorphic symmetry whose fractional translation (traditionally defined in real space) now acts in $\bk$ space; this analogy is elaborated precisely in Ref.\ \onlinecite{Cohomological}.\\

Class-I glide-symmetric solids are defined to have two ordinary glide planes in the BZ, as exemplified by Ba$_2$Pb (space group 62). For a glide symmetry $g_x$ that inverts the wavenumber $k_x$, the two planes lie at $k_x=0$ and $k_x=\pi/R_1$, where $2\pi \vec{x}/R_1$ is a primitive reciprocal vector. In this class, the strong  ($\chi^+ {\in} \Z_4$) and weak ($\calp_{01} {\in} \Z_2$) invariants may independently assume any values, as representatively illustrated in \fig{fig:wilson_insulators}; this is consistent with a K-theoretic classification of surface states in Ref.\ \onlinecite{Nonsymm_Shiozaki}. We remind the reader that $\calp_{01}$ is a Kane-Mele invariant defined over the off-center glide plane. Ba$_2$Pb  falls into the $(\chi^+,\calp_{01}){=}(3, 0)$ class, as may be verified by its Zak phases in \fig{fig:ex}(a). \\

%We might ask if a finer distinction of class-I insulators arises, owing to  the analogous Kane-Mele invariant ($\calp_{23}$) in the glide plane that projects to 23. Given $(\chi^+,\calp_{01})$, $\calp_{23}$ is not an independent invariant. Indeed, $\calc{=}0$ imposes $\calp_{01}{=}\calp_{23}$ ($\calp_{01} {\neq} \calp_{23}$) if $\chi^+$ is even (odd), as we argue pictorially in \fig{fig:wilson_insulators}.  

Class-II solids are defined to have only a  single ordinary glide plane (containing the BZ center) in the BZ; an off-center glide plane exists but is  projective.   For a glide symmetry $g_x$ that inverts the wavenumber $k_x$, though an off-center glide plane exists at $k_x=\pi/R_1$, $2\pi \vec{x}/R_1$ is a \emph{not} primitive reciprocal vector; however, the existence of primitive vectors $2\pi \vec{x}/R_1+\pi\vec{z}/R_3$ and $2\pi\vec{z}/R_3$ ensure that glide-related states in the plane are separated by half a reciprocal vector $(\pi\vec{z}/R_3)$. Consequently, the Kane-Mele invariant for the off-center glide plane is always trivial ($\calp_{01}{=}0$), as was proven in the appendix of \ocite{Hourglass}; see also the \emph{reductio ad absurdum} argument through Wilson-loop connectivities in \ocite{Cohomological}. \\

There remains for class-II solids a $\Z_4$ strong classification, as exemplified by KHgSb [SG $D_{6h}^4$; $\chi^+{=}2$; \fig{fig:ex}(b)], and uniaxially stressed Na$_3$Bi [$\chi^+{=}1$; \fig{fig:ex}(c)].   The $\Z_4$ invariant [cf.\ \q{wilsonex}] is only well-defined for $\bk$ in a modified BZ (denoted BZ') wherein both glide planes are ordinary. To appreciate this, consider that a Bloch state with wavevector $\bk$ in a projective glide plane does not transform in either of the glide representations $\Delta_{\pm}$ (due to the glide-related states lying at inequivalent wavevectors). The simplest choice for BZ' would correspond to a non-primitive real-space unit cell that is consistent with a glide-symmetric surface termination, as exemplified 
% add
(for KHgSb) by the orange rectangle
in \fig{fig:nonprimitive}(a).
We remind the reader that  a non-primitive cell has larger volume than the primitive cell; it is a region that, when translated through  a \emph{subset} of vectors of the Bravais lattice, just fills all of space without overlapping itself or leaving voids;\cite{ashcroft_mermin} 
%add
the subset of vectors in our example is generated by $\bR_1'$ and $\bR_2'$ [ \fig{fig:nonprimitive}(a)]. 
This subset of vectors form a reduced Bravais lattice (denoted BL') that is distinct from the original. BZ' would then be the Wigner-Seitz cell of the reciprocal lattice dual to BL';
%add
both BZ and BZ' of KHgSb are illustrated respectively as the hexagon and orange rectangle in \fig{fig:nonprimitive}(b).
This prescription of enlarging the unit cell was first suggested in \ocite{Nonsymm_Shiozaki} to establish a connection between their K-theoretic classification and the material class of KHgSb. 
%add 
The utility of BZ' is that the $\Z_4$ invariant may  be calculated by diagonalizing a family of Wilson loops (over the nontrivial cycles of BZ'), as was described in \s{sec:zakexpressZ4}; an example of such a Wilson loop is illustrated with triple arrows in \q{fig:nonprimitive}(b). The result of this calculation for KHgSb has been shown in \fig{fig:ex}(b), from which we conclude $\chi^+{=}2$.

\begin{figure}[h]
\centering
\includegraphics[width=8.6 cm]{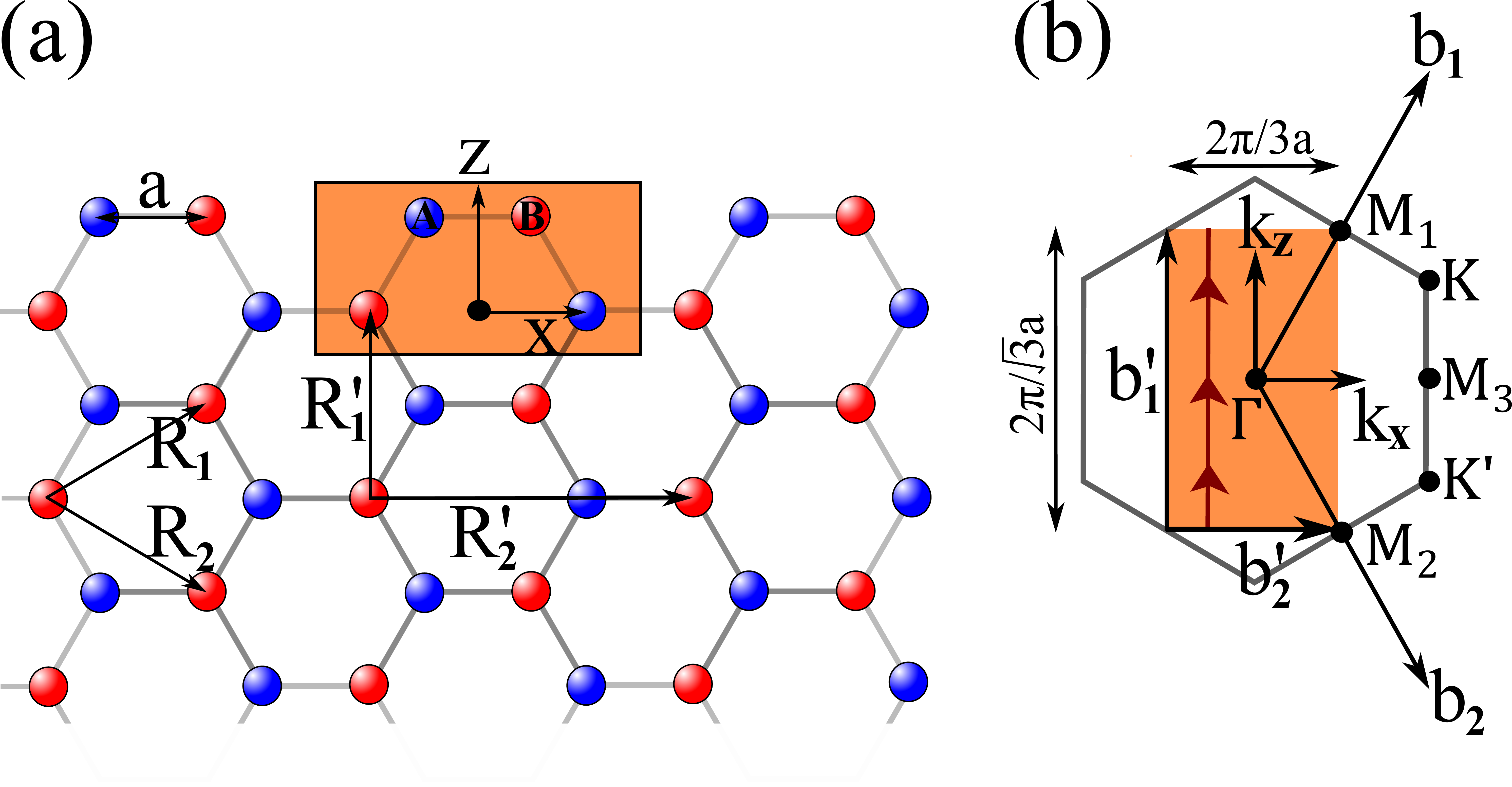}
\caption{(a) Constant-y cross section of the crystal structure of KHgSb, with bulk Bravais lattice vectors $\bR_1$ and $\bR_2$. The top armchair edge is the cross-section of a glide-symmetric surface. A nonprimitive unit cell consistent with a glide-symmetric surface contains four atoms in both A and B sublattices (colored red and blue respectively); 
%add
this nonprimitive cell, when translated by vectors $\bR'_1$ and $\bR'_2$, covers the entire xz-plane. 
Note that this  nonprimitive unit cell has twice the volume of the primitive cell.    (b) The hexagon illustrates the constant-$k_z$ cross-section of the BZ; $\bb_1$ and $\bb_2$ are primitive reciprocal vectors
%add 
dual to $\bR_1$ and $\bR_2$; 
the orange rectangle inscribed in the hexagon illustrates BZ', which is the Wigner-Seitz cell 
%remove: corresponding to the nonprimitive unit cell in (a).
%add
of a modified reciprocal lattice with basis vectors $\bb_1'$ and $\bb_2'$ (which are dual to $\bR'_1$ and $\bR'_2$). }\label{fig:nonprimitive}
\end{figure}

\section{Material analysis: space groups and elementary band representations }\la{app:materializations}

\subsection{Ba$_2$Pb}

The space group of Ba$_2$Pb is  SG62 ($Pnma$), which has an orthorhombic lattice. The spatial symmetries include: an inversion ($\cal I$), three screws ($\{C_{2x}|\frac{1}{2}\frac{1}{2}\frac{1}{2}\}$, $\{C_{2y}|0\frac{1}{2}0\}$ and $\{C_{2z}|\frac{1}{2}0\frac{1}{2}\}$), two glide ($g_x\equiv \{r_{x}|\frac{1}{2}\frac{1}{2}\frac{1}{2}\}$ and $g_z\equiv\{r_{z}|\frac{1}{2}0\frac{1}{2}\}$) and one mirror ($\{r_{y}|0\frac{1}{2}0\}$). 
%add
Note $r_j$ is a mirror operation that inverts the single coordinate $j$.\\

For the calculations
% add
of topological invariants, 
we redefine the lattice vectors as $\vec a'=2\vec a+\vec b$, $\vec b'=b$ and $\vec c'=c$, which are orthogonal. We can then set $a',b'$ and $c'$ as the $x,y,z$ axes. With respect to these new lattice vectors, the glide symmetry is represented by $g_x\equiv\{r_{x}|00\frac{1}{2}\}$.\\

%cannot be represented by locally-symmetric Wannier functions. By `locally-symmetric', we mean that for any spatial point $\bvarpi$, all Wannier functions centered at $\bvarpi$ form a representation of all point-group symmetries that preserve $\bvarpi$.\cite{nogo_AAJH} Equivalently stated,  the groundstate of Ba$_2$Pb 

Beside exhibiting a nontrivial connectivity of the Zak phases [cf.\ \fig{fig:ex}(a)], another manifestation\cite{TQC,nogo_AAJH} of the nontriviality of Ba$_2$Pb is that its groundstate is not a direct sum of elementary band representations.\cite{TQC,nogo_AAJH}  To prove this, it is sufficient to compare the irreducible representations (irreps) at high-symmetry wavevectors.\cite{TQC,elcoro_EBRinBilbao} By inspection, the irreps of Ba$_2$Pb (\tab{tab:irrepBa2Pb}) cannot be decomposed into a direct sum of irreps of the elementary band representations, as obtained from the Bilbao crystallographic server (reproduced in \tab{tab:irrep62}).

\begin{table}[H]
\centering
\caption{Elementary band representations\cite{TQC,elcoro_EBRinBilbao} for SG62. \la{tab:irrep62}}
\begin{tabular}{c|ccccc}
Wyckoff pos.  & 4a                 & 4a                 &   4b                     &    4b                    &  4c                 \\
Band-Rep.     &$A_g A_g$           &$A_u A_u$           &  $A_g A_g$               &   $A_u A_u$              & $^1E ^2E$           \\
\hline                                                                                                                              
$\Gamma$      & 4x5                & 4x6                &   4x6                    &    4x6                   &  4x6                \\
 R            & (3+3)$\oplus$(4+4)& (3+3)$\oplus$(4+4)&   (3+3)$\oplus$(4+4)    &    (3+3)$\oplus$(4+4)   &  (3+3)$\oplus$(4+4)\\
 S            & (3+3)$\oplus$(4+4)& (3+3)$\oplus$(4+4)&   (3+3)$\oplus$(4+4)    &    (3+3)$\oplus$(4+4)   &  (3+3)$\oplus$(4+4)\\
 T            & 2x(3+4)            & 2x(3+4)            &   2x(3+4)                &    2x(3+4)               &  2x(3+4)            \\
 U            & 2x(5+5)            & 2x(6+6)            &   2x(6+6)                &    2x(5+5)               &  (5+5)$\oplus$(6+6)\\
 X            & 2x(3+4)            & 2x(3+4)            &   2x(3+4)                &    2x(3+4)               &  2x(3+4)            \\
 Y            & 2x(3+4)            & 2x(3+4)            &   2x(3+4)                &    2x(3+4)               &  2x(3+4)            \\
 Z            & 2x(3+4)            & 2x(3+4)            &   2x(3+4)                &    2x(3+4)               &  2x(3+4)            \\
\end{tabular}
\end{table}

\begin{table}[H]
\centering
\caption{
%remove Computed 
Irreducible representations for Ba2Pb,
%add
as computed by VASP. \la{tab:irrepBa2Pb}}
\begin{tabular}{c|c}
   & \text{Valence bands}          \\
\hline
$\Gamma$& 6;5;6;5;5;6;5;6;6;5;6;6; \\
 R      & 4+4;3+3;3+3;4+4;3+3;4+4; \\
 S      & 4+4;3+3;3+3;4+4;4+4;3+3; \\
 T      & 3+4;3+4;3+4;3+4;3+4;3+4; \\
 U      & 5+5;6+6;6+6;5+5;6+6;5+5; \\
 X      & 3+4;3+4;3+4;3+4;3+4;3+4; \\
 Y      & 3+4;3+4;3+4;3+4;3+4;3+4; \\
 Z      & 3+4;3+4;3+4;3+4;3+4;3+4; \\
\end{tabular}
\end{table}

\subsection{Stressed Na$_3$Bi}\la{sec:na3bi}

For Na$_3$Bi that is stressed in the $x$ direction, the space group falls into $Cmcm$ (SG 63), which is a body-center structure. The conventional lattices are redefined as $\vec a'=0.98(\vec a+\vec b)$ where
%add
the factor
 0.98 is due to 
%remove: the
%add
a hypothetical
compression in the x direction, $\vec b'=b$ and $\vec c'=c$, where $a,b,c$ are the primitive lattice vectors in the original structure(SG 194). $\chi^+$ is calculated with the conventional (non-primitive) lattices. The glide symmetry is represented by $g_x\equiv\{r_{x}|00\frac{1}{2}\}$.

By comparing the irreps of all elementary band representations [in SG63; see \tab{ebr63}] with the irreps of stressed Na$_3$Bi [cf.\ \tab{irrepNa3bi}], we conclude that the groundstate of stressed Na$_3$Bi is not band-representable.

\begin{table}[H]
\centering
\caption{Elementary band representations for SG63. \la{ebr63}}
\begin{tabular}{c|ccccc}
Wyckoff pos.& 4a & 4a& 4b & 4b & 4c\\
Band-Rep.   & $^1E_g ^2E_g$ & $^1E_u ^2E_u$& $^1E_g ^2E_g$ & $^1E_u ^2E_u$ & E\\
\hline          
 $\Gamma$ & 2x5   & 2x6    & 2x5   & 2x6   &5+6 \\
  R       & 2+2   & 2+2    & 2+2   & 2+2   &2+2 \\
  S       &2x(3+4)&2x(5+6) &2x(5+6)&2x(3+4)&(3+4)$\oplus$(5+6) \\
  T       & 3+4   &    3+4 & 3+4   &    3+4&3+4 \\
  Y       & 2x5   &    2x6 & 2x5   &    2x6&5+6 \\
  Z       & 3+4   &    3+4 & 3+4   &    3+4&3+4 \\
\end{tabular}
\end{table}

\begin{table}[H]
\centering
\caption{
%remove Computed 
Irreducible representations for stressed Na3Bi,
%add
as computed by VASP. \la{irrepNa3bi}}
\begin{tabular}{c|c}
 & Valence bands \\
\hline          
 $\Gamma$ & 5;6;5;5;5+6;\\
  R       & 8; 12; 11; 9; 8; 12; \\
  S       & 5+6; 3+4; 3+4; 5+6; 3+4; 5+6; \\
  T       &  3+4; 3+4; 3+4; \\
  Y       &  5; 6; 6; 5; 6; 5;   \\
  Z       &  3+4; 3+4; 3+4;  \\
\end{tabular}
\end{table}

\subsection{KHgSb}

The space group of KHgSb is $D_{6h}^4$ or SG194; further details about its crystallographic structure may be found in \ocite{Hourglass}. By comparing the irreps of all elementary band representations [in SG194; see \tab{ebr194}] with the irreps of KHgSb [cf.\ \tab{irrepkhgsb}], we conclude that the groundstate of KHgSb is not band-representable.

\begin{table}[H]
\centering
\caption{Elementary band representations for SG194 \la{ebr194}}
\begin{tabular}{c|ccccccccccccc}
Wyckoff pos.& 2a& 2a& 2a& 2a & 2b& 2b& 2b& 2c&2c&2c &2d&2d&2d\\
Band-Rep.   &$^1E_g^2E_g$&$^1E_u^2E_u$&$E_{1g}$&$E_{1u}$ &$E_1$&$E_2$&$E_3$&$E_1$&$E_2$&$E_3$&$E_1$&$E_2$&$E_3$ \\
\hline          
 A          & (4+5)& (4+5)& 6& 6  & 6 &6&4+5&6&6&4+5&6&6&4+5\\
$\Gamma$    &2x7 & 2x10 & 8$\oplus$9 & 11$\oplus$12 &9$\oplus$11&8$\oplus$12&7$\oplus$10&9$\oplus$11&8$\oplus$12&7$\oplus$10&9$\oplus$11&8$\oplus$12&7$\oplus$10 \\
 H          &(4+5)$\oplus$(6+7)  &(4+5)$\oplus$(6+7) &8$\oplus$9 & 8$\oplus$9 &8$\oplus$9&8$\oplus$9&(4+5)$\oplus$(6+7)&(4+5)$\oplus$9&(6+7)$\oplus$8&8$\oplus$9&(4+5)$\oplus$9&(6+7)$\oplus$8&8$\oplus$9 \\
 K          & 2x7 & 2x7 & 8$\oplus$9 & 8$\oplus$9 &2x9&2x8&2x7&7$\oplus$8&7$\oplus$9&8$\oplus$9&7$\oplus$8&7$\oplus$9&8$\oplus$9\\
 L          &3+4  & 3+4 & 3+4 & 3+4 &3+4&3+4&3+4&3+4&3+4&3+4&3+4&3+4&3+4\\
 M          & 2x5 & 2x6 & 2x5 & 2x6 &5+6&5+6&5+6&5+6&5+6&5+6&5+6&5+6&5+6\\
\end{tabular}
\end{table}

\begin{table}[H]
\centering
\caption{
%remove Computed 
Irreducible representations for KHgSb,
%add
as computed by VASP.
\la{irrepkhgsb}}
\begin{tabular}{c|c}
 & Valence bands\\
\hline          
A & 6;6;6; \\
$\Gamma$ & 8;12; 11; 9; 8; 12; \\
H & 6+7; 8; 9; 8; 6+7; 8; \\
K & 7; 8; 9; 9; 7; 9;\\
\end{tabular}
\end{table}

\section{Ambiguity in the choice of coordinate systems}\la{sec:ambiguity}

This appendix addresses a question posed at the end of \s{sec:pesZ4}, which we will briefly recapitulate.  Suppose we choose a right-handed, Cartesian coordinate system where  where $\vec{\bx}$ (resp.\ $\vec{\by}$) lies parallel to the reflection (resp.\ fractional translational) component of the glide, i.e., the glide  maps $(x,y,z) \rightarrow (-x,y\pm R_2/2,z)$. Such a coordinate system would be called glide-symmetric. Would  the topological invariants $\chi^{+}$ (or $\calc$) differ if measured in distinct glide-symmetric coordinates? \\

As argued in \s{sec:pesZ4}, there are three glide-symmetric coordinates which are related to each other by two-fold rotations $C_{2j}$ about the directional axes $\vec{j}$ ($j=x,y,z$); we shall only concern ourselves with proper point-group transformations that preserve the orientation (or  handedness) of the coordinate system. We will refer to one glide-symmetric, right-handed (but otherwise arbitrarily chosen) coordinate system -- in $\bk$-space -- as the \emph{reference coordinate system}; all other coordinate systems are related to the reference by $\bk'=p \circ \bk$, with $p$ a point-group transformation (e.g., $C_{2x}\circ \bk:=(k_x,-k_y,-k_z)$ etc). It should be emphasized that $p$ is \emph{not necessarily} a symmetry of the solid (i.e., not an element of the space group), but merely reflects an ambiguity in the choice of coordinates. \\

To establish notation, a map between points: $\bk\rightarrow p \circ \bk$ induces naturally a map between  subregions of the Brillouin torus (e.g., lines denoted as $l$, or faces denoted as $a,b,c,d$.); we shall denote this as $l \rightarrow p \circ l$ etc; several examples are illustrated in \fig{fig:coordinate_invariance_chern}. It is useful (as an intermediate step in the following computations) to decompose $C_{2x}$ as the product of two reflections $r_y$ and $r_z$, such that each $r_j$ inverts only the $j$'th coordinate ($j=x,y,z$). We will also consider coordinate transformations induced by the inversion $\cali: (x,y,z)\rightarrow(-x,-y,-z)$, though inversion symmetry need not belong in the space group.

\begin{figure}[H]
\centering
\includegraphics[width=10 cm]{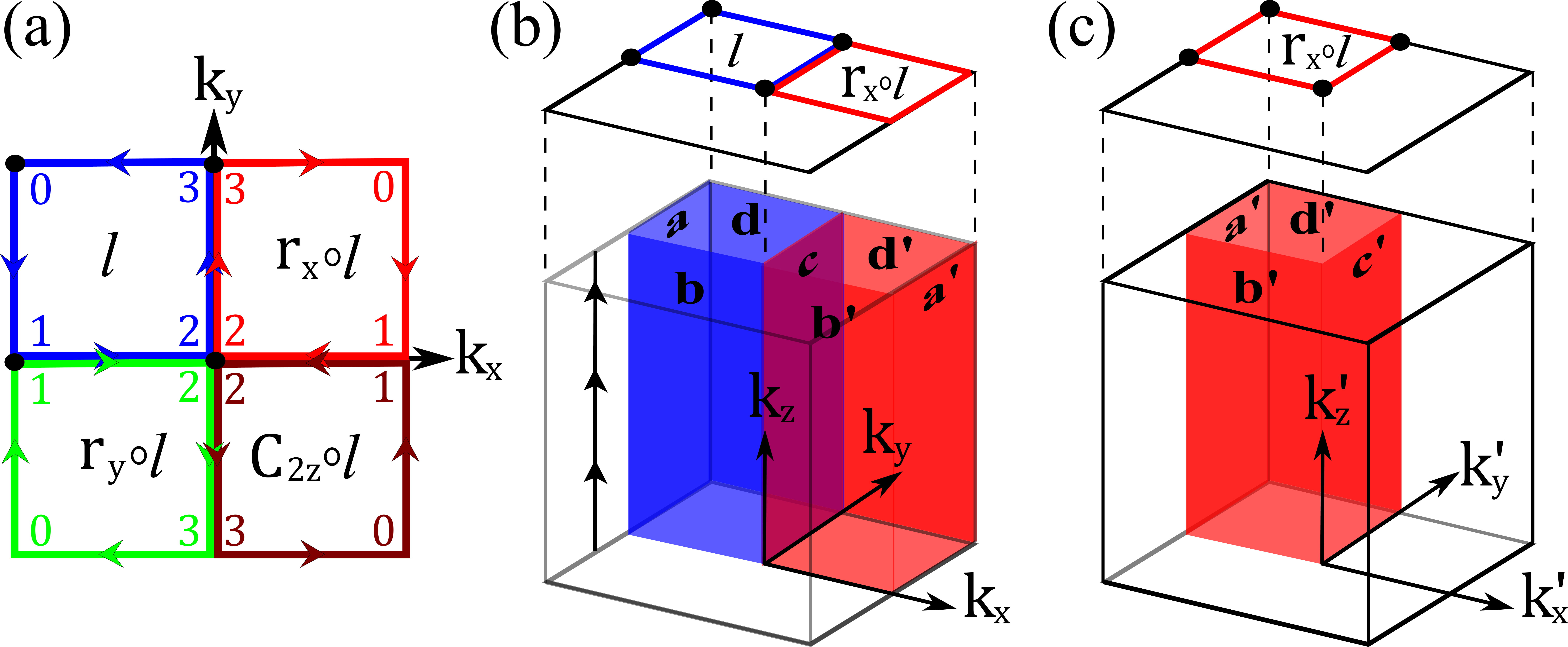}
\caption{(a) Illustration of $p \circ l$ in the surface Brillouin torus. (b) Bent manifolds $abcd$ and  $a'b'c'd'=r_x \circ abcd$ in the $\bk$ coordinates; note that $r_x\circ c$ and $c$ differ only in orientation. (c) $r_x \circ abcd$ in the reflected coordinates $(k_x',k_y',k_z')=(-k_x,k_y,k_z).$  }\label{fig:coordinate_invariance_chern}
\end{figure}

%For example, if $\bk$ represented a glide-symmetric choice, so certainly would $\bk'=(-k_x,-k_y,k_z)$ or $(-k_x,k_y,-k_z)$. (To appreciate this ambiguity, 

We separately analyze the coordinate dependence of $\calc$ and $\chi^{\pm}$ in \s{app:coordinateC} and \s{app:coordinatechi} respectively.

\subsection{Coordinate dependence of the bent Chern number $\calc$} \la{app:coordinateC}

We begin by defining the Berry curvature as a pseudovector field $\bcalf=(\calf_x,\calf_y,\calf_z)$, with components
\e{ \calf_a = i\epsilon_{abc}\sum_{n=1}^{\noc}\braket{\partial_bu_n}{\partial_cu_n};\la{field}}
$\partial_j$ is shorthand for the derivative with respect to $k_j$, $\epsilon_{abc}$ is the Levi-Cevita tensor, repeated indices (e.g., $b,c$ above) are summed over the Cartesian directions $x,y,z$. The bent Chern number is defined as the integral of the Berry curvature 
\e{\calc= -\int_{|a|}\calf_x dk_ydk_z -\int_{|b|}\calf_y dk_xdk_z+\int_{|c|}\calf_xdk_ydk_z+\int_{|d|}\calf_y dk_xdk_z; \la{absolute}}
where $|f|$ in the subscript of $\int_{|f|}$ denotes the face $f$ without its orientation. The $\pm$ signs in front of each integral reflects our convention that $\calc$ measures the outgoing Berry `flux', or equivalently the net charge of the Berry monopoles within the quadrant enclosed by $abcd$. An equivalent and useful expression is
\e{ \calc = \f{i}{2\pi}\int_0^4  \f{dt}{4} \int_0^{2\pi}\f{dk_z}{2\pi} \left[ \braket{\partial_tu}{\partial_{k_z}u}-\braket{\partial_{k_z}u}{\partial_{t}u}\right],}
where $t\in [0,4]$ (with $4\equiv 0$) parametrizes the loop $l$ on which $abcd$ projects in the z direction, as illustrated in \fig{fig:coordinate_invariance_chern}(a) [see also \fig{fig:bz}(b)]. $l$ is anticlockwise-oriented [as indicated by arrows in \fig{fig:coordinate_invariance_chern}(a)], and $t$ increases in the direction of the orientation  loop $l$.

Let $\calc$ be the Chern number defined over $abcd$ in the reference coordinate system (parametrized by $\bk$). We define $p\circ \calc$ as the same Chern number in a different coordinate system parametrized by $\bk'=p\circ \bk$; that is, $p\circ \calc$ is defined exactly as in \q{absolute} but with $\bk$ replaced by $\bk'$.  For the same Hamiltonian, we would prove that
\e{ \calc = r_x \circ \calc =r_y \circ \calc =-r_z \circ \calc. \la{bcherntr}}

To prove the first equality, consider that $r_x\circ \calc$ is the Chern number defined over $a'b'c'd'
%add
{=}r_x\circ abcd$ 
in the $\bk'=(-k_x,k_y,k_z)$ coordinates, as illustrated in \fig{fig:coordinate_invariance_chern}(c).   In the reference coordinates, $a'b'c'd'$ is comparatively illustrated with $abcd$ in  \fig{fig:coordinate_invariance_chern}(b). Since $a'b'c'd'$ and $abcd$ are related by the reflection $r_x$, they enclose different quadrants of the BZ (colored red and blue respectively). To deduce that $\calc = r_x \circ \calc$, we will rely on two observations: (i) While $r_x \circ \calc$ is defined to measure the outgoing Berry flux in the $\bk'$ coordinates, it measures the incoming Berry flux in the reference coordinates $\bk$; this may be deduced by the $r_x\circ l$ having an opposite orientation relative to  $l$, as illustrated in \fig{fig:coordinate_invariance_chern}
%add
(a-
b). (ii) Since the curvature transforms like pseudovector, we expect that glide-related Berry monopoles having opposite charge -- therefore the net monopole charge in the blue quadrant is negative the monopole charge in the red quadrant. In combination, (i-ii) produces the desired result. \\

%add
$\calc = r_y \circ \calc$ 
[the second equality in \q{bcherntr}]  may be derived by a simple generalization of the above argument. Now the two quadrants
%add
(enclosed by $abcd$ and $r_y\circ abcd$)
 are related by a composition ($Tg_x$) of time-reversal and glide symmetry. (i') $r_y \circ \calc$ also measures the incoming Berry flux
%add
in the reference coordinates,
and (ii') $Tg_x$-related monopoles have opposite charge.
%add
(Note that $r_y$ is not assumed be a symmetry in the space group, but if it were, we would similarly conclude that $r_y$-related monopoles have opposite charge.)\\ 

%add
$\calc = -r_z \circ \calc$
[the last equality in \q{bcherntr}] may be derived from the following argument. 
%remove: $a'b'c'd'$ in the reference coordinates, exactly overlaps with $abcd$; however the two subregions have opposite orientations because $k_z$ is inverted. 
%add
When both $abcd$ and $a'b'c'd'=r_z\circ abcd$ are viewed in the reference coordinates, the two surfaces occupy the same area (in $\bk$-space) and differ only in their orientations;
this difference in orientations originates from the reversal of $k_z$. 
%remove: In other words,
%add 
This implies that 
$r_z\circ \calc$ measures the \emph{incoming} Berry flux through $abcd$.\\

From \q{bcherntr} and $C_{2x}=r_yr_z$ etc., we derive that the bent Chern numbers -- for two coordinate parametrizations of the same Hamiltonian -- are related as 
\bal
 p \circ \calc = \begin{cases}-\calc, & p\in \{C_{2x},C_{2y}\},\\
                                    \calc, & p=C_{2z}.\end{cases}\la{pcirccalc}
\end{align}

\subsection{Coordinate dependence of topological invariant $\chi^{\pm}$} \la{app:coordinatechi}

Let us define $\chi^{\pm}$ as $\Z_4$ invariants defined with respect to a reference coordinate system parametrized by $\bk$; analogously, $p\circ \chi^{\pm}$ are defined as the $\Z_4$ invariants defined with respect to a distinct coordinate system with $\bk'=p\circ \bk$.
 For the same Hamiltonian, we will show that 
\bal
 p \circ \chi^{\pm} = \begin{cases}-\chi^{\mp}, & p\in \{C_{2x},C_{2y}\},\\
                                    \chi^{\pm}, & p=C_{2z}.\end{cases}\la{gchi}
\end{align}
This would imply, in combination with \q{pcirccalc}, that $\chi^++\chi^-\equiv 2\calc$ mod $4$ [cf.\ \q{bentchern2}] is invariant under proper coordinate transformations -- a result applicable to both band insulators and Weyl metals.\\

\begin{figure}[H]
\centering
\includegraphics[width=16 cm]{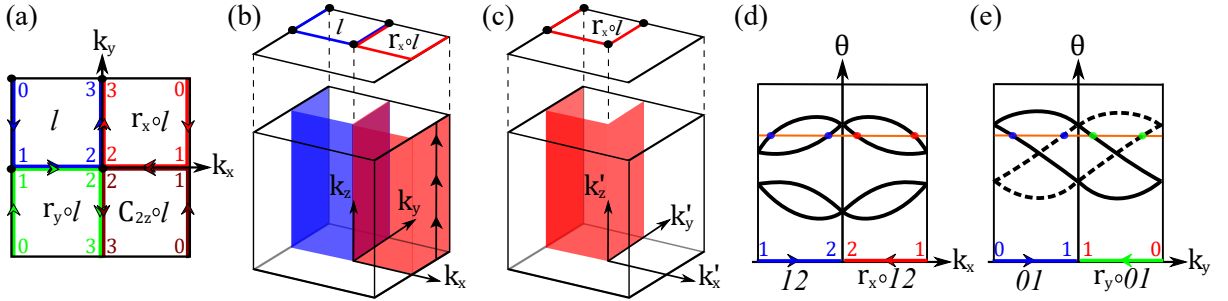}
\caption{(a) Illustration of $p \circ l$ in the surface Brillouin torus. (b) Bent subregions on which the $\Z_4$ invariants $\chi[p\circ l]$ are defined. (c) Bent subregion in which $r_x\circ \chi^{\pm}[l]$ is defined, in the reflected coordinates $(k_x',k_y',k_z')=(-k_x,k_y,k_z).$ 
%add
(d) Representative Zak-phase dispersion along $12$ and $r_x\circ 12$. (e) Representative Zak-phase dispersion along $01$ and $r_y\circ 01$. }\label{fig:coordinate_invariance}
\end{figure}

The rest of this appendix is devoted to proving \q{gchi}. Let $l$ be the oriented path in $(k_x,k_y)$-space on which $\chi^{\pm}[l]$ is defined through \q{wilsonex}. $l$ is illustrated in \fig{fig:coordinate_invariance}(a), in conjunction with the three other point-group mapped $p\circ l$; we remind the reader that $p$ is not necessarily a symmetry of the solid. A word of caution: $l$ was also used in the previous section to define a loop illustrated in \fig{fig:coordinate_invariance_chern}; in this section we use the same symbol $l$ for an open segment of the loop in  \fig{fig:coordinate_invariance_chern}.

For each of $p\circ l$ illustrated in \fig{fig:coordinate_invariance}(a), we define the quantities $\chi^{\pm}[p \circ l]$ which simply generalize our original definition in \q{wilsonex}:
 \e{ \chi^{\pm}[p\circ l] \eq \f1{\pi} \sum_{j=1}^{\noc/2} \left[ \theta_j^{\pm}[p\circ l(0)] -\theta_j^{\pm}[p\circ l(3)] + \int_{[p\circ l(0)]}^{[p\circ l(1)]}  d\theta_j^{\pm} + \int^{[p\circ l(3)]}_{[p\circ l(2)]} d\theta_j^{\pm} \right] +\f1{2\pi}\sum_{j=1}^{\noc} \int_{[p\circ l(1)]}^{[p\circ l(2)]} d\theta_j. \la{wilsonex2}}
\q{wilsonex} is a particularization of  $\chi^{\pm}[p\circ l]$  for $p$ being the identity operation. Here, we have parametrized $p\circ l(t)$ by $t\in [0,3]$ such that $t\in \{0,1,2,3\}$ lie on the high-symmetry wavevectors in the $k_x-k_y$ plane, as illustrated in \fig{fig:coordinate_invariance}(a).   
%remove $\theta_j[p\circ l(t)]$ Wilson loop represents the holonomy 
%add
$\{e^{i\theta_j[p\circ l(t)]}\}$ are eigenvalues of the Wilson loop -- 
for an oriented quasimomentum 
%remove circle 
%add
loop
which projects in the z direction to
% add
the wavevector
 $p\circ l(t)$,
%add
as illustrated by the triple arrows in \fig{fig:coordinate_invariance}(b); by definition,
the orientation of each 
%remove circle 
%add
loop is always in the direction of increasing $k_z$. 
%remove as indicated by the triple arrows in \fig{fig:coordinate_invariance}(b). 
%add

In congruence with our previous definitions,  $\chi^{\pm}[ p_1\circ l]$ is defined respect to a reference coordinate $\bk$, and we define $p_2\circ \chi^{\pm}[p_1\circ l]$ with respect to $\bk'=p_2\circ \bk$, with $p_1$ not necessarily equal to $p_2$.  
We caution that $\chi^{\pm}[p\circ l]$ and $p\circ \chi^{\pm}[l]$  are not necessarily equal, as will be seen in \q{gchiexplained}. 

\subsubsection{Proposition 1}

Let us prove an intermediate proposition: 
\e{ \chi^{\pm}[l] \equiv \chi^{\pm}[r_x \circ l]  \equiv \chi^{\mp}[r_y \circ l] \equiv \chi^{\mp}[C_{2z} \circ l], \la{inv2d}}
where $\equiv$ is an equivalence modulo four. \\

\noindent Let us introduce the shorthand $p \circ j (j+1)$, for $j \in \{0,1,2\}$, as the subset of $p\circ l(t)$ in which $t \in [j,j+1]$. That is, $l$ is the 
%remove concatenation 
%add
union 
of intervals $01$, $12$ and $23$, and so similarly we define $p \circ 01$, $p \circ 12$ and $p \circ 23$ for $p \circ l$. The relation in \q{z4inv} simply generalizes to
\e{ \chi^{\pm}[p \circ l] \equiv 2\cals^{\pm}_{p \circ 01}(\bar{\theta})+\cals_{p \circ 12}(\bar{\theta})+2\cals^{\pm}_{p \circ 23}(\bar{\theta}). \la{rhs}}
where $\cals_{p \circ ij}$ is defined analogously to $\cals_{ij}$, as introduced in the main text. We write it down for clarity: draw a constant-$\bar{\theta}$ reference line (for an arbitrarily chosen Zak phase $\bar{\theta}$) and consider its intersections with Zak bands along $p \circ l$.  For each intersection between $p \circ 12$, we calculate the sign of the velocity $d\theta/dt$, and sum this quantity over all intersections to obtain $\cals_{p \circ 12}(\bar{\theta})$; for $p \circ 01$ and $p \circ 23$, we consider only intersections with Zak bands in the $\Delta_{\pm}$ representation, and we similary sum over sgn[$d\theta/dt$] to obtain $\cals^{\pm}_{p \circ 01}(\bar{\theta})$ and $\cals^{\pm}_{p \circ 23}(\bar{\theta})$ respectively.  \\

\noindent \emph{Proof of $\chi^{\pm}[l] \equiv \chi^{\pm}[r_x \circ l]$}

\noindent Along the glide-invariant lines, $r_x \circ 01 = 01$ and $r_x \circ 23 = 23$, and therefore $\cals^{\pm}_{r_x \circ 01} = \cals^{\pm}_{01}$ and $\cals^{\pm}_{r_x \circ 23} = \cals^{\pm}_{23}$.  However, $r_x \circ 12 {\neq} 12$ lie on distinct lines which are related by time-reversal symmetry [which maps $(k_x,k_y)\rightarrow (-k_x,-k_y)$], as illustrated in \fig{fig:coordinate_invariance}(a). This symmetry imposes  $\cals_{12} = \cals_{r_x \circ 12}$, as we now explain. Suppose a Zak band over $12$ intersects our constant-$\bar{\theta}$ line with velocity $v$, then its time-reversed partner is a Zak band over $r_x\circ 12$, which intersects the $\bar{\theta}$ line with velocity $-v$. By $v$ and $-v$, we refer to velocities defined by varying the Zak phase of a Zak band with respect to 
%remove $k_y$.
%add
$k_x$. 
However, our definition of $\cals_{p\circ ij}$ involved velocities defined by varying the Zak phase with respect to a parameter that is specific to $p\circ ij$: the parameter for $12$ increases in the same direction as 
%remove $k_y$.
%add
$k_x$,
but the parameter for $r_x\circ 12$ increases in the opposite direction, as illustrated in \fig{fig:coordinate_invariance}(a) and (d). Therefore, each pair of time-reversed Zak bands contribute equally to  $\cals_{12}$ and $\cals_{r_x \circ 12}$, leading to $\cals_{12} = \cals_{r_x \circ 12}$. 
%add
For example, consider a representative Zak-band dispersion in \fig{fig:coordinate_invariance}(d), where $\cals_{12} = \cals_{r_x \circ 12}=2$ for the chosen reference line (colored orange). $\blacksquare$\\

\noindent \emph{Proof of $\chi^{\pm}[l] \equiv \chi^{\mp}[r_y \circ l]$}

\noindent Since $12=r_y\circ 12$, 
\e{\cals_{12} = \cals_{r_y\circ 12}.\la{rel1}} 
Time reversal relates $01$ and $r_y \circ 01$, are therefore imposes a relation between $\cals^{\pm}_{01}$ and $\cals^{\pm}_{r_y\circ 01}$, as we now derive. Recall from \s{sec:symm} that time-reversed partner states at $\pm k_y$ belong to opposite
%remove   glide 
representations $\Delta_{\pm}$
%add
of the glide $g_x$. 
This implies that (a) a Zak band in the $\Delta_{\pm}$ representation at $01$ has a time-reversed partner at $r_y \circ 01$ in the $\Delta_{\mp}$ representation; note that $01$ and $r_y\circ 01$ are distinct lines in $\bk$-space. (b) Moreover, as representatively illustrated in \fig{fig:coordinate_invariance}(e), time-reversed partners have opposite-sign velocities with respect to variation of $k_y$, but equal velocities with respect to varying the parameters of $01$ and $r_y\circ 01$ respectively. (a) and (b) together imply 
\e{\cals_{01}^{\pm} = \cals_{r_y\circ 01}^{\mp} \la{rel2}.} 
By cosmetic substitution of $12 \rightarrow 23$ in the above demonstration, we would show that 
\e{\cals_{23}^{\pm} = \cals_{r_y\circ 23}^{\mp}.\la{rel3}}
\q{rel1}, \q{rel2}, \q{rel3} and \q{rhs} together imply our claim. $\blacksquare$\\

\noindent Finally, $\chi^{\pm}[l] \equiv \chi^{\mp}[C_{2z} \circ l]$ may be proven from
\e{\cals_{12} = \cals_{C_{2z}\circ 12},\as \cals_{01}^{\pm} =\cals_{C_{2z}\circ 01}^{\mp},\as \cals_{23}^{\pm} =\cals_{C_{2z}\circ 23}^{\mp}.}

\subsubsection{Dependence on proper coordinate transformations}

Let the $p$ be a proper point-group transformation that preserves handedness of the coordinate system.  $p$ can
%remove be decomposed as $p_{\perp}$ (acting in the $k_x-k_y$ plane) and $p_{\parallel}\in \pm 1$ (acting in the $k_z$ line). 
%add
always be viewed as the composition of  a two-dimensional point-group operation ($p_{\perp}$) acting in the $k_x-k_y$ plane, and a one-dimensional point group operation acting in the $k_z$ line:
\e{ p \circ \bk = (p_{\perp}\circ (k_x,k_y),p_{\parallel} k_z), \as p_{\parallel}\in \pm 1.}
This gives a correspondence $p \leftrightarrow (p_{\perp},p_{\parallel})$.
We are particularly interested in 
\e{ C_{2x} \leftrightarrow (r_y,-1), \as C_{2y} \leftrightarrow (r_x,-1), \as C_{2z} \leftrightarrow (C_{2z},+1). \la{decomposec2}} 
For two coordinate parametrizations ($\bk$ and $\bk'=p\circ \bk$) of the same Hamiltonian, we argue that
\e{ p\circ \chi^{\pm}[l,k_z] = \chi^{\pm pr_xp^{-1}r_x^{-1}}[p_{\perp}\circ l,p_{\parallel}k_z], \la{gchiexplained}}
where $\chi^{\pm}[p\circ l,k_z]:=\chi^{\pm}[p\circ l]$ as defined in \q{wilsonex2}, and $\chi^{\pm}[p\circ l,k_z]$ is identical to $\chi^{\pm}[p\circ l]$ except that the orientation of each Wilson loop is reversed (from increasing $k_z$ to decreasing $k_z$). The above equation has the following justification: \\

\noi{i} A coordinate transformation effectively changes the bent quasimomentum region on which $\chi$ is calculated; this is reflected in a change in
%add
the
argument of $\chi$. For example, $r_x \circ \chi^{\pm}[l,k_z]$ is defined over the bent quasimomentum subregion $a'b'c'=r_x\circ abc$ that we illustrate in the primed coordinates [red sheet in \fig{fig:coordinate_invariance}(c)] and reference coordinates [red sheet in \fig{fig:coordinate_invariance}(b)]; $r_x\circ abc$ projects in the z direction to $r_x\circ l$. \\

%Implicit in the above claims is a consistent definition of glide that applies to any coordinate system.
%that is, we maintain the same sense for the $\pi$ rotation about $\vec{\bx}'$ axis as we had for $\vec{\bx}$; note that the sense of any $\pi$-rotation is important in half-integer-spin representations, in which a $2\pi$-rotation produces a nontrivial phase factor.\\

\noi{ii} Whether the glide representation changes under a coordinate transformation $(x,y,z) \rightarrow (x',y',z')=p\circ (x,y,z)$ depends on $p$. 
%remove Precisely, we mean that a Bloch state in the $\Delta'_{\pm}$ glide representation (with $\Delta'_{\pm}$ defined as the eigenvalue of 
%\e{g_{x'} = t(\vec{\by}'/2)\cali \exp\left[-i (L_{x'}+S_{x'})\pi/\hbar \right] \;\bigg) }
%does not necessarily transform in the $\Delta_{\pm}$ representation (with $\Delta_{\pm}$ defined as the eigenvalue of 
%\e{ g_x = t(\vec{\by}/2)\cali \exp\left[-i (L_x+S_x)\pi/\hbar \right].\;\bigg) \la{defineglideconsistently}}
%$\cali$ above is the spatial inversion operator; $L_j$ and $S_j$ are orbital and spin angular momentum operators respectively. The correspondence is 
%\e{ \Delta'_{\pm} \leftrightarrow \Delta_{\pm pr_xp^{-1}r_x^{-1}} \la{correspond}}
%where $pr_xp^{-1}r_x^{-1}=\pm 1$ indicates whether $p$ has inverted the orientation of the reflection component ($r_x$) of $g_x$.  What is meant by the orientation of $r_x$? $r_x$ can be viewed as the composition of $\cali$ with the two-fold rotation $C_{2x}$ (explicitly expressed with angular momentum generators in \q{defineglideconsistently}); $C_{2x}$ encodes an orientation, because a $\pi$ clockwise rotation differs from a $\pi$ anticlockwise rotation (by a $-1$ phase factor) in a spinor representation. For example, if $p=r_y$,  $g_{x'}$ and  $g_x$ have opposite orientations, because $r_x$ and $r_y$ anticommute in the spinor representation. \q{correspond} justifies the the change in the superscript of $\chi$ in \q{gchiexplained}. \\
%add
To appreciate this, let us recall that the reflection component ($r_x$) of glide $g_x$ has an associated orientation. Indeed, $r_x$ may be viewed as the composition of a spatial inversion  ($\cali$) with the two-fold rotation ($C_{2x}$) about the $x$-axis, and, for \emph{half-integer-spin representations}, we need to specify if this rotation is clockwise- or anticlockwise-oriented. That is to say, a $\pi$ clockwise rotation differs from a $\pi$ anticlockwise rotation by a $-1$ phase factor. Consequently, the \emph{same} glide-invariant state has glide eigenvalues with \emph{opposite} signs -- with respect to two glide operations which differ only in orientation. For a coordinate system $(x,y,z)$, we always define $g_x$ with a clockwise rotation about the $x$-axis; this was implicit in our previous definitions of $\Delta_{\pm}$ and $\chi^{\pm}$. Suppose a Bloch state transforms under $g_x$ with  eigenvalue $\Delta_{\pm}=\pm i e^{-ik_y/2}$;  the \emph{same} state may (or may not) transform with the inverted eigenvalue $\mp ie^{-ik_y/2}$ under the glide $g_{x'}$, which is defined with a clockwise orientation about the $x'$-axis [recall  $(x',y',z')=p\circ (x,y,z)$]. The glide eigenvalue is inverted if and only if the coordinate transformation $p$ inverts the orientation of a rotation about the $x$-axis, i.e., it depends on $pr_xp^{-1}r_x^{-1}=\pm 1$ (with $-1$ indicating an inversion). For example, if $p=C_{2x}$, $g_{x'}$ and  $g_x$ have the same  orientations;  if  $p=C_{2y}$,  $g_{x'}$ and  $g_x$ have opposite orientations, because $r_x$ and $C_{2y}$ anticommute in the half-integer-spin representation. This possible change in the glide representation is accounted for in \q{gchiexplained} by the superscript of $\chi$.\\

Beginning from \q{gchiexplained}, the next step is to express
\e{ p\circ \chi^{\pm}[l,k_z] \equiv  p_{\parallel}\chi^{\pm pr_xp^{-1}r_x^{-1}}[p_{\perp}\circ l,k_z]. \la{gchiexplained2}}
To justify this, $p_{\parallel}=-1$ implies that the orientation of the Wilson loop flips, thus $e^{i\theta}(t) \rightarrow e^{-i\theta}(t)$, and  the velocities at the reference Zak phase are likewise inverted; cf.\ \q{rhs}. \\

Finally, inserting \q{decomposec2} and \q{inv2d} [which should be understood as relating $\chi$ with constant $k_z$ arguments] into \q{gchiexplained2}, we obtain 
\e{ &C_{2x}\circ\chi^{\pm}[l,k_z]\equiv -\chi^{\pm}[r_y\circ l,k_z] \equiv -\chi^{\mp}[l,k_z],\lin
&C_{2y}\circ\chi^{\pm}[l,k_z]\equiv -\chi^{\mp}[r_x\circ l,k_z] \equiv -\chi^{\mp}[l,k_z],\lin
&C_{2z}\circ\chi^{\pm}[l,k_z]\equiv \chi^{\mp}[C_{2z}\circ l,k_z] \equiv \chi^{\pm}[l,k_z],}
from which \q{gchi} follows directly.

\section{Consideration of light sources for photoemission}\la{app:lightsource}

To exploit the selection rule  developed in \s{sec:pes}, we would like that the electron-photon coupling $H_{int}$ transforms in a one-dimensional representation of glide reflection [cf.\ \q{onedimrepHint}]. As we will show in this appendix, this transformation holds for a linearly-polarized light source, with photon wavevector  parallel to the glide-invariant yz plane, and with the polarization vector $\vec{\epsilon}$  either orthogonal [see \fig{fig:pes}(d)] or parallel [\fig{fig:pes}(c)] to the glide-invariant plane. To orient $\vec{\epsilon}$ relative to the glide plane, we would need to know the sample's crystallographic orientation; this may be obtained by independent experiments (e.g., X-ray diffraction), or by comparison of the angle-resolved photoemission data to a first-principles calculation (where the glide plane is known).

For the purpose of  demonstrating \q{onedimrepHint}, it is useful to distinguish between normally  and obliquely  incident light. With oblique incidence, we identify (by standard convention) the parallel alignment as $p$ polarization, and the orthgonal alignment as $s$ polarization, e.g., compare \fig{fig:pes}(c) and (d). For normal incidence, the two types of polarization are indistinguishable.

The cases of normal incidence (both parallel and orthogonal alignments) and oblique incidence (orthogonal alignment) will be dealt with in \s{sec:classicalmaxwell}, where we prove  \q{onedimrepHint} within the classical approximation\cite{feibelman_surfacereflectivity,feibelman_review,goldmann_onestep_Cu,feder_review} of light within the solid. This classical approximation is invalid (for surface photoemission) in the case of oblique incidence (parallel alignment); nevertheless, so far as nonlinearities in the optical response (of the medium) can be neglected, we will find in \s{sec:linearresponse} that  \q{onedimrepHint} still holds. 

\subsection{Normal incidence (parallel and orthogonal alignments) and oblique incidence (orthogonal alignment)}\la{sec:classicalmaxwell} 

For such incidence angles and polarizations, the incident electric field is parallel to the surface, allowing for a classical, Maxwell-based approximation of the electromagnetic field (within the solid). We briefly review why: corrections to the classical approximation are known as local fields, which are believed to be only significant near the surfaces of solids,\footnote{Precisely, they are significant near the surfaces of metals, or of insulators with surface states (including a wide class of topological insulators).} where surface plasmons and electron-hole pairs may be excited by the incident radiation.\cite{feibelman_surfacereflectivity,levinson_surfacephotonfield,feibelman_review} Consequently, local-field effects are especially relevant to  surface photoemission, which is the main application in \s{sec:pesZ4}. It is known that local-field effects are negligible if  the incident electric field is aligned parallel to the surface (i.e., $\vec{\epsilon}$ lies in the xy-plane).\cite{feibelman_surfacereflectivity,levinson_surfacephotonfield,feibelman_review,goldmann_onestep_Cu,feder_review}
The reason is that surface-parallel electric-field components vary smoothly across the surface, while surface-normal electric-field components can vary rapidly on the order of atomic distances (thus invalidating the dipole approximation). Even within the classical, Maxwell-based approximation, it is known that surface-normal field components are discontinous across the interface of two distinct media due to the presence of a surface charge;\cite{jackson_classicalem} this surface charge is an idealization, and its proper, quantum description is given by the aforementioned surface plasmons and electron-hole pairs.\cite{feibelman_review} \\

Within the classical approximation, and for the above-stated conditions on the light source, Fresnel's equations\cite{jackson_classicalem} inform us that the photon field within the solid remains linearly polarized, with a polarization vector $\vec{\epsilon}$  (within the solid) that is identical to the polarization vector of the light source.\\

 In the temporal gauge, the electric field and vector potential are parallel, hence $\ba$ (the screened vector potential within the solid) is  proportional to $\vec{\epsilon}$. So far as we are concerned only with the absorption of photons,  $\ba$ (occurring in the electron-photon coupling $H_{int}$) may be equated with $a_0\vec{\epsilon}e^{i\bq\cdot \br}$, where $a_0$ is a spatially-independent constant, and  $\bq$ is the wavevector of the photon within the solid.  \\

For normally-incident light ($\bq{=}{-}\omega \vec{z}/c$) with the polarization vector  parallel to the glide plane ($\epsilon{=}\vec{y}$), $H_{int}$ commutes with the glide operation $\hatbmx$.\\

 If the polarization vector is orthogonal to the glide plane ($\vec{\epsilon}{=}\vec{x}$),  $H_{int}$ anticommutes with $\hatbmx$ in the case of normal incidence.\\

 For non-normal incidence and $\vec{\epsilon}{=}\vec{x}$, $\hatbmx H_{int}\hatbmx^{-1}{=}{-}e^{-iq_yR_2/2}H_{int}$; the $q_y$-dependent phase factor originates from the half-lattice translation ($y{\rightarrow}y{-}R_2/2$) in $\hatbmx$.

\subsection{Oblique incidence (parallel alignment)}\la{sec:linearresponse}

As explained in the previous \s{sec:classicalmaxwell}, the classical approximation is \emph{not} satisfied if the incident electric field has a component normal to the surface -- as would be the case for  $p$-polarized radiation at oblique incidence.\cite{feibelman_surfacereflectivity,levinson_surfacephotonfield,feibelman_review} 

Nevertheless, so long as the optical response of the medium is linear (though not necessarily local\cite{feibelman_review}), the electron coupling to the medium-induced electromagnetic field (given by vector potential $\ba^{ind}$) transforms in the same glide representation as the electron coupling to the externally applied field (given by $\ba^{ext}$).\footnote{We thank Ji Hoon Ryoo for alerting us to this argument.} That is to say, if $\hatbmx \bp\cdot \ba^{ext}\hatbmx^{-1} = e^{-iq_yR_2/2}\bp\cdot \ba^{ext}$, so must  $\hatbmx \bp\cdot \ba^{ind}\hatbmx^{-1} = e^{-iq_yR_2/2}\bp\cdot \ba^{ind}$. This follows from the assumed existence of a linear functional relating the two potentials:
\e{ a_i^{ind}(\br)=\sum_{j=x,y,z}\int  \chi_{ij}(\br,\br')a_j^{ext}(\br')d\br',}
with the susceptibility  satisfying the glide-symmetric constraint:
\bal
 \chi_{ij}(\br,\br') = \sum_{a,b}[\mir_x]_{ia}[\mir_x]_{jb}\chi_{ij}(g_x\circ\br,g_x\circ\br'), \;\; \mir_x:= \begin{pmatrix} -1& 0&0 \\0&1&0\\0&0&1 \end{pmatrix}, \;\;g_x\circ(x,y,z):=(-x,y-R_2/2,z).
\end{align}
Consequently, the electron coupling to the \emph{total} photon field transforms as $\hatbmx H_{int}\hatbmx^{-1} = e^{-iq_yR_2/2}H_{int}$.

%To appreciate why, translational symmetry ensures that the crystal wavenumber $k_y$  is preserved, which implies that the glide eigenvalues $(\Delta_{\pm}{=}{\pm}e^{-ik_ya_2/2})$ of the photoelectron and initial state  are equal up a minus sign.

% There are two limits for the photon wavelength where single-photon absorption is allowed in this scenario: 

\end{widetext}

\bibliography{bib_Apr2018}
%\bibliography{bib_June2017}

\end{document}